\documentstyle[sprocl,epsf]{article}

\newif\ifBigFig	\BigFigtrue

\newdimen\rotdimen
\def\vspec#1{\special{ps:#1}}
\def\rotstart#1{\vspec{gsave currentpoint currentpoint translate
   #1 neg exch neg exch translate}}
\def\rotfinish{\vspec{currentpoint grestore moveto}}
\def\rotr#1{\rotdimen=\ht#1\advance\rotdimen by\dp#1%
   \hbox to\rotdimen{\hskip\ht#1\vbox to\wd#1{\rotstart{90 rotate}%
   \box#1\vss}\hss}\rotfinish}
\def\rotl#1{\rotdimen=\ht#1\advance\rotdimen by\dp#1%
   \hbox to\rotdimen{\vbox to\wd#1{\vskip\wd#1\rotstart{270 rotate}%
   \box#1\vss}\hss}\rotfinish}%
\def\rotu#1{\rotdimen=\ht#1\advance\rotdimen by\dp#1%
   \hbox to\wd#1{\hskip\wd#1\vbox to\rotdimen{\vskip\rotdimen
   \rotstart{-1 dup scale}\box#1\vss}\hss}\rotfinish}%
\def\rotf#1{\hbox to\wd#1{\hskip\wd#1\rotstart{-1 1 scale}%
   \box#1\hss}\rotfinish}%

\def\dofig#1#2{\epsfxsize=#1\centerline{\epsfbox{#2}}}
\def\dofigs#1#2#3{\centerline{\epsfxsize=#1\epsfbox{#2}%
   \epsfxsize=#1\epsfbox{#3}}}

\def\ellell{\ell^+\ell^-}
\def\etmiss{\slashchar{E}_T}
\def\sgn{\mathop{\rm sgn}}
\def\fb{{\rm fb}}
\def\calL{{\cal L}}
\def\lsp{{\tilde\chi_1^0}}
\def\tchi{{\tilde\chi}}
\def\tq{{\tilde q}}
\def\tg{{\tilde g}}
\def\tell{{\tilde\ell}}
\def\ttau{{\tilde\tau}}
\def\tb{{\tilde b}}

\def\Meff{M_{\rm eff}}
\def\mhalf{m_{1/2}}
\def\calL{{\cal L}}
\def\calO{{\cal O}}
\def\jet{{\rm jet}}
\def\jets{{\rm jets}}
\def\Msusy{M_{\rm SUSY}}
\def\pawplot{\vskip-10pt}
\def\MeV{{\rm MeV}}
\def\GeV{{\rm GeV}}
\def\TeV{{\rm TeV}}
\def\cmsec{{\rm cm^{-2}s^{-1}}}
\def\mb{{\rm mb}}
\def\pb{{\rm pb}}
\def\fb{{\rm fb}}
\def\hb{\hfil\break}

\catcode`@=11
\def\@cite#1#2{{$^{#1}$\if@tempswa\typeout
        {IJCGA warning: optional citation argument
        ignored: `#2'} \fi}}
\catcode`@=12

\def\simge{
    \mathrel{\rlap{\raise 0.511ex
        \hbox{$>$}}{\lower 0.511ex \hbox{$\sim$}}}}
\def\simle{
    \mathrel{\rlap{\raise 0.511ex 
        \hbox{$<$}}{\lower 0.511ex \hbox{$\sim$}}}}

\def\slashchar#1{\setbox0=\hbox{$#1$}           
   \dimen0=\wd0                                 
   \setbox1=\hbox{/} \dimen1=\wd1               
   \ifdim\dimen0>\dimen1                        
      \rlap{\hbox to \dimen0{\hfil/\hfil}}      
      #1                                        
   \else                                        
      \rlap{\hbox to \dimen1{\hfil$#1$\hfil}}   
      /                                         
   \fi}                                         %
\def\etmiss{\slashchar{E}_T}

\def\mod{\mathop{\rm mod}}
\def\half{\frac12}

\font\twelvess=cmss10 scaled \magstep1

\begin{document}

\begingroup
\parindent=20pt
\thispagestyle{empty}
\vbox to 0pt{
\vskip-1.4in
\moveleft.75in\vbox to 8.9in{\hsize=6.5in
{\large
\centerline{\twelvess BROOKHAVEN NATIONAL LABORATORY}
\vskip6pt
\hrule
\vskip1pt
\hrule
\vskip4pt
\hbox to \hsize{December, 1997 \hfil BNL-HET-98/1}
\vskip3pt
\hrule
\vskip1pt
\hrule
\vskip3pt

\vskip1in
\centerline{\LARGE\bf Supersymmetry Signatures at the CERN LHC}
\vskip.5in
\centerline{\bf Frank E. Paige}
\vskip4pt
\centerline{Physics Department}
\centerline{Brookhaven National Laboratory}
\centerline{Upton, NY 11973 USA}

\vskip.75in

\centerline{ABSTRACT}
\vskip8pt
\narrower\narrower
	These lectures, given at the 1997 TASI Summer School,
describe the prospects for discovering supersymmetry (SUSY) and for
studying its properties at the Large Hadron Collider (LHC) at CERN.
If SUSY exists at a mass scale less than 1--$2\,\TeV$, then it should
be easy to observe characteristic deviations from the Standard Model
at the LHC. It is more difficult to determine SUSY masses because in
most models there are two missing particles $\lsp$ in every event.
However, it is possible to use various kinematic distributions to make
precision measurements of combinations of SUSY masses and other
quantities related to SUSY physics. In favorable cases such
measurements at the LHC can determine the parameters of the underlying
SUSY model with good accuracy.

\vskip.75in

	To appear in {\sl TASI 97: Supersymmetry, Supergravity and
Supercolliders} (Boulder, CO, 1997).

\vskip0pt
}
\vfil
	This manuscript has been authored under contract number
DE-AC02-76CH00016 with the U.S. Department of Energy.  Accordingly,
the U.S.  Government retains a non-exclusive, royalty-free license to
publish or reproduce the published form of this contribution, or allow
others to do so, for U.S. Government purposes.
}
\vss}
\newpage

\thispagestyle{empty}
\tableofcontents
\newpage

\endgroup

\setcounter{page}{1}

\title{Supersymmetry Signatures at the CERN LHC}
\author{Frank E. Paige}
\address{Physics Department\\
Brookhaven National Laboratory\\
Upton, NY 11973 USA}
\maketitle

\abstracts{These lectures, given at the 1997 TASI Summer School,
describe the prospects for discovering supersymmetry (SUSY) and for
studying its properties at the Large Hadron Collider (LHC) at CERN.
If SUSY exists at a mass scale less than 1--$2\,\TeV$, then it should
be easy to observe characteristic deviations from the Standard Model
at the LHC. It is more difficult to determine SUSY masses because in
most models there are two missing particles $\lsp$ in every event.
However, it is possible to use various kinematic distributions to make
precision measurements of combinations of SUSY masses and other
quantities related to SUSY physics. In favorable cases such
measurements at the LHC can determine the parameters of the underlying
SUSY model with good accuracy.}


\section{Introduction\label{sec:intro}}

	The theoretical attractiveness of having supersymmetry (SUSY)
at the electroweak scale has been discussed by many
authors.\cite{SUSY,Sally97} But while SUSY is perhaps the most
promising ideas for physics beyond the Standard Model, we will not
know if it is a correct idea until SUSY particles are discovered
experimentally. LEP might still discover a SUSY particle, but it has
already run at $\sqrt{s} = 183\,\GeV$, and its reach will be limited
by its maximum energy, probably $\sqrt{s} = 193\,\GeV$. It is less
unlikely that LEP might discover a light Higgs boson:  the current
bound\cite{Janot} of $\sim77\,\GeV$ is expected to be improved to
$\simle95\,\GeV$,\cite{LEPhiggs} whereas the upper limit on the mass is
$\sim130\,\GeV$ in the Minimal Supersymmetric Standard
Model\cite{Sher96} and $\sim150\,\GeV$ more generally.\cite{Kane93}
Finding a light Higgs would not prove the existence of SUSY, but it
would certainly be a strong hint.

	The Tevatron has a better chance of finding SUSY particles,
particularly from the process\cite{DPF95,D03l}
$$
\bar p p \to \tchi_2^0 \tchi_1^\pm \to \ellell \ell^\pm \etmiss\,.
$$
This can be sensitive to $M(\tchi_2^0) \approx M(\tchi_1^\pm) \simle
200\,\GeV$ for some choices of the other parameters, e.g., small
$\tan\beta$, given an integrated luminosity of $2\,\fb^{-1}$ in Run~2
and more in future runs.\cite{BCPT2} But, like LEP, the Tevatron
cannot exclude SUSY at the weak scale.

	The decisive test of weak scale SUSY, therefore, must await
the Large Hadron Collider (LHC) at CERN. The LHC can detect gluinos
and squarks in the MSSM up to $\sim2\,\TeV$ with only 10\% of its
design integrated luminosity per year. Discovering gluinos and squarks
in the expected mass range, $\simle1\,\TeV$, seems straightforward,
since the rates are large and the signals are easy to separate from
Standard Model backgrounds. Other SUSY particles can be found from the
decays of gluinos and squarks. The difficult problem is not
discovering SUSY if it exists but verifying that the new physics is
indeed SUSY, separating the various SUSY signals, and interpreting
them in terms of the parameters of an underlying SUSY model. This is
more difficult for the LHC than for an $e^+e^-$ machine of sufficient
energy, but some progress has been made recently.

	The first few sections of these lectures are mainly review.
Section~\ref{sec:sigma} reviews the SUSY production cross sections and
some basic facts about QCD perturbation theory. Section~\ref{sec:sim}
discusses event generators, which are used to translate production
cross sections into experimental signals. Section~\ref{sec:det}
summarizes the capabilities of ATLAS and CMS, the two main LHC
detectors, to detect these signals.

	The next sections concentrate on SUSY measurements at the LHC
in the context of the minimal supergravity (SUGRA) model,\cite{SUGRA}
although the general results should apply to other models, at least
those in which the lightest SUSY particle escapes the detector.
Section~\ref{sec:reach} shows the reach in SUGRA parameter space for
various signals and describes how to make a first estimate of the SUSY
mass scale.  Section~\ref{sec:precise} describes examples of a recently
developed approach to extracting information about SUSY masses and
other parameters from LHC data for five specific SUGRA points.
Section~\ref{sec:fits} shows what the resulting errors on the SUGRA
parameters would be at these points.  Section~\ref{sec:large}
discusses preliminary results at a SUGRA point with large $\tan\beta$
that has very different properties. 

	Section~\ref{sec:higgs} discusses the LHC discovery potential
for Higgs bosons both in the Standard Model and in the MSSM. Finally,
Section~\ref{sec:last} draws some conclusions.

\section{SUSY Cross Sections\label{sec:sigma}}%

	The LHC is a $pp$ collider to be built in the existing LEP
tunnel at CERN (the European Laboratory for Particle Physics, located
near Geneva, Switzerland) with a center-of-mass energy $\sqrt s =
14\,\TeV$ and a luminosity $\calL = 10^{33}$--$10^{34}\,\cmsec$. It
will have two major experiments, ATLAS\cite{ATLAS} and CMS,\cite{CMS}
for studying high-$p_T$ physics like SUSY. Two smaller experiments ---
LHC-B for $B$ physics and ALICE and for heavy ion physics --- have also
been proposed but will not be discussed further here. Construction
of both the accelerator and the experiments is expected to be
completed in 2005.\cite{LHCsched}

	Fine tuning arguments\cite{Anderson} suggest that the SUSY
masses should be below about $1\,\TeV$ if SUSY is relevant to
electroweak physics. Then SUSY production at the LHC is dominated by
the production of gluinos and squarks. The elementary $\tg$ and $\tq$
cross sections only depend on the color representations and spins of
these particles --- which of course are fixed by supersymmetry --- and
on their masses. Thus they are less model dependent than the cross
sections for gaugino production, which also depend on couplings
determined by the mixing matrices.

	Perturbative QCD tells us that inclusive production cross
sections can be computed as a power series in the strong coupling
coupling $\alpha_s(Q)$ evaluated at a scale $Q$ of order the masses
involved. For example, the lowest order contribution to the elementary
process $gg \to \tg\tg$ is given by\cite{EHLQ}
\begin{eqnarray*}
{d\hat\sigma \over d\hat t} &=& {9\pi\alpha_s^2\over4\hat s^2} 
\Biggl\{{2(\hat t-M^2)(\hat u-M^2)\over \hat s^2} + \\
&&\qquad\qquad\left[{(\hat t-M^2)(\hat u-M^2)-2M^2(\hat
t+M^2)\over(\hat t-M^2)^2} + (\hat t\leftrightarrow \hat u)\right] + \\
&&\qquad\qquad{M^2(\hat s-4M^2)\over(\hat t-M^2)(\hat u-M^2)} \Biggr\}\,,
\end{eqnarray*}
where $\hat s$, $\hat t$, $\hat u$ are usual parton process invariants
and $M = M_\tg$ is the only relevant mass in this case. The lowest
order cross section for $gg \to \tq \bar{\tq}$ depends on both $M_\tq$
and $M_\tg$. The cross sections for gaugino pair production and
associated production depend on the masses and also on couplings which
are determined by the chargino and neutralino mixing matrices.

	The elementary cross sections are then related to $pp$ cross
sections by the QCD-improved parton model, which is based on the
impulse approximation for processes with large $Q^2$. In the parton
model the $pp$ cross section is given by a convolution of the
elementary parton-parton cross sections and the appropriate parton
distributions, i.e., the probabilities of finding quarks or gluons
with given momentum fractions $x_i$ in the incoming protons:
$$
\sigma = \sum_{ij} \int dx_1dx_2\, \hat\sigma_{ij} f_i(x_1,Q^2)
f_j(x_2,Q^2)\,.
$$
Here $Q^2$ is a measure of the scale, e.g.\ $Q^2 = p_T^2 + M^2$,
and $x_i$ are the momentum fractions of the incoming partons. By
elementary kinematics they satisfy
$$
\hat s = x_1 x_2 s, \quad x_i = \sqrt{\hat s \over s} e^{\pm \hat y}\,,
$$
where $\hat y$ is the rapidity of the center of mass of the produced
system.

\begin{figure}[t]
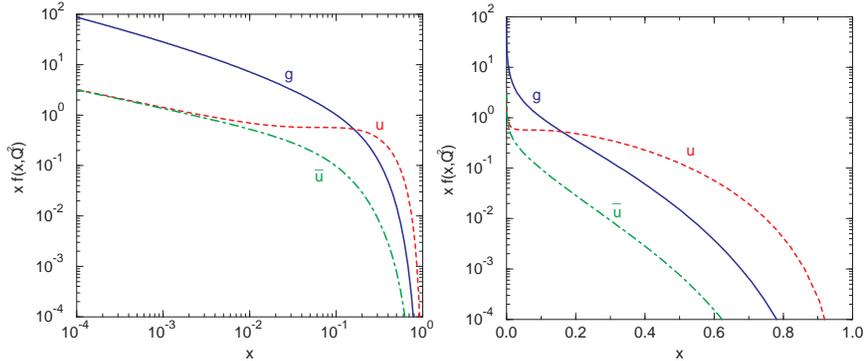

\dofigs{2.25in}{pdf.ai}{pdf2.ai}
\caption{$g$, $u$, and $\bar u$ parton distributions $xf(x,Q^2)$ in
the proton vs.\ $x$ for $Q=100\,\GeV$ using the
CTEQ3L\protect\cite{CTEQ3} parameterization.\label{pdf}}
\end{figure}

	Representative parton distributions for $Q=100\,\GeV$ are
shown in Figure~\ref{pdf}. Note that
$$
x \sim \sqrt{\hat s \over s} \simge {2 M_\tg \over \sqrt s}\,
$$
so that $g(x)$ is large and $gg$ processes dominate the production of
squarks and gluinos for most masses of interest.

\begin{figure}[t]
\dofig{1.25in}{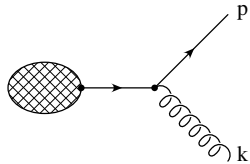}
\caption{Soft and collinear singularities both come from gluons
attached to external lines, only one of which is shown. Soft
singularities come from $p^2=m^2$ and $k\to0$; collinear ones come
from $p^2=k^2=0$ and $\theta_{pk}\to0$. The gluon can be either real
or virtual.\label{feyn1}}
\end{figure}

	The QCD-improved parton model is intuitively plausible, but we
must require that it is consistent with higher order perturbation
theory. Consider adding one gluon emission or loop. Then after
renormalization of ultraviolet divergences in the usual way,
individual graphs are found to give a series not in $\alpha_s$ but in
$\alpha_s\ln^2Q^2$. These logarithms must cancel if perturbation
theory is to be usable. There is only one large $Q^2$ in the problem,
so the logarithms must reflect singularities as the external particles
are put on mass shell. There are actually two distinct types of
singularities, both well known from QED:

	{\it Soft or Infrared Singularities:} These arise from a gluon
with $k \to 0$ attached to an external line, Figure~\ref{feyn1}. Then
as the external line goes on shell, so does the internal propagator.
Soft singularities arise from the divergent multiplicity of soft
gluons; the probability of radiating no gluons from an accelerated
color charge is zero. The total cross section must be finite ---
something must happen. Hence the singularities must cancel between
processes with different numbers of gluons, i.e., between real and
virtual graphs. This can be proven to all orders in perturbation
theory for QED.\cite{KLN,Weinberg65} The situation in QCD is more
complicated because the gluons themselves radiate, but the
cancelation certainly works in all cases that have been tried.

	{\it Collinear or Mass Singularities:} These arise from the
emission of a hard gluon parallel to a massless parton in the initial
state, i.e., $p^2 = k^2 =0$ and $k \| p$ in Figure~\ref{feyn1} .
(There also can be collinear singularities in the final state for
differential cross sections.) They do not cancel for initial states
like hadrons with limited transverse momenta, but they are universal
because they come from on-shell poles.\cite{Politzer77} Hence they
cancel if you calculate one physical process, e.g., $\tg\tg$
production, in terms of another, e.g., deep inelastic scattering.
Equivalently, they can be absorbed into universal parton distributions
defined by deep inelastic scattering. It is straightforward to verify
this at one loop.  The general proof is complex because soft and
collinear singularities get tangled,\cite{Factor} but the result is
believed to be true.

	The collinear singularities in the parton distributions lead
to a series in $(\alpha_s(Q^2)\ln Q^2) = \calO(1)$. These leading
logarithms can be summed to all orders in perturbation theory:
Altarelli and Parisi\cite{Altarelli77} matched the operator product
expansion to the 1-loop graphs, while Gribov and
Lipatov\cite{Gribov72} and Dokshitser\cite{Dokshitser} studied the
origin of the logarithms in perturbation theory directly. The result
is that the parton distributions satisfy the DGLAP equations,
\begin{eqnarray*}
{\partial q_i(x,Q^2) \over \partial Q^2} &=& \displaystyle
{\alpha_s(Q^2) \over 2 \pi Q^2} \displaystyle \int_x^1 {dx' \over x'}
   \left[ q_i(x',Q^2) P_{qq}({x \over x'})
+ g(x',Q^2) P_{qg}({x \over x'}) \right]\,,\\ 
{\partial g(x,Q^2) \over \partial Q^2} &=& \displaystyle
{\alpha_s(Q^2) \over 2 \pi Q^2} \displaystyle \int_x^1 {dx' \over x'}
   \Bigg\lbrack \sum_j q_j(x',Q^2) P_{gq}({x \over x'})
+ g(x',Q^2) P_{gg}({x \over x'}) \Bigg\rbrack\,,
\end{eqnarray*}
where the DGLAP functions
\begin{eqnarray*}
P_{qq}(x) &=& c_F \left({1+x^2 \over 1-x}\right)_+
+c_{qq}\delta(1-x)\,,\\
P_{gq}(x) &=& c_F \left({1+(1-x)^2 \over x}\right)\,,\\
P_{qg}(x) &=& \frac12 \left(x^2 + (1-x)^2\right)\,,\\
P_{gg}(x) &=& 2 c_A \left({{x \over 1 - x} + {1-x \over x} + x(1 - x)}
\right)_+ +c_{gg}\delta(1-x)\,,
\end{eqnarray*}
reflect the various QCD couplings. The $+$ subscript in these
functions indicates that the singularity is to be treated as a
distribution,
$$
\int dx\, {f(x) \over (1-x)_+} \equiv \int dx\, {f(x)-f(1) \over
(1-x)}\,.
$$
The $1/(1-x)$ singularities come from the radiation of soft gluons,
while the $\delta(1-x)$ terms come from virtual graphs. A little
thought will show that using such a distribution implements the
real-virtual cancellation just like the usual perturbative calculation.
It is not actually necessary to calculate the virtual graphs: the
coefficients of the delta functions can be determined by momentum
conservation.\cite{Altarelli77}

	The DGLAP equations correspond to a simple picture of the
$Q^2$ evolution of parton distributions. As $Q^2$ increases, more
gluons are radiated, so the distributions soften at large $x$ and
increase at small $x$. This radiation is responsible for the rise at
small $x$ in Figure~\ref{pdf}.

\begin{figure}[t]
\dofig{2.5in}{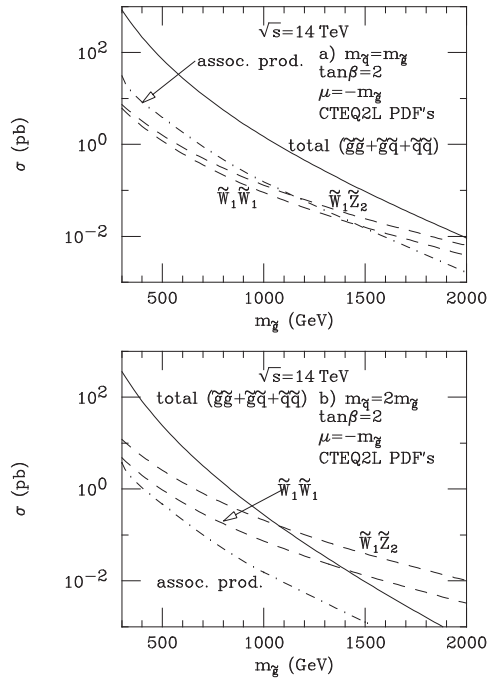}
\caption{Total SUSY MSSM production cross sections\protect\cite{DPF95}
at the LHC using $M_i = \alpha_i M_\tg / \alpha_s$ to determine the
gaugino masses.\label{lhcsigma}} 
\end{figure}

	The lowest-order QCD cross sections\cite{BCPT1,DPF95} for SUSY
particle production in the MSSM are shown in Figure~\ref{lhcsigma}.
These cross sections are similar but not identical to those in the
SUGRA model; the gaugino masses are scaled from $\tg$ like
$$
{M_1 \over \alpha_1} = {M_2 \over \alpha_2} = {M_\tg \over  \alpha_s}
$$
instead of being calculated from the renormalization group equations.
However, the gluino and squark cross sections are model independent,
and these dominate except for very high masses and heavy squarks.

\begin{figure}[t]
\epsfysize=2in
\dofig{2.5in}{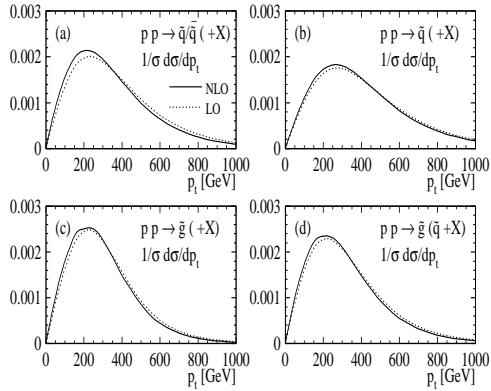}
\caption{Normalized LO and NLO $p_T$ distributions\protect\cite{BHSZ}
for squarks and gluinos at the LHC with $M_\tq=600\,\GeV$,
$M_\tg=500\,\GeV$.\label{lhcpt}}
\end{figure}

	In addition to their effects on parton distributions, higher
order QCD corrections also give finite $\calO(\alpha_s)$ corrections
to $pp$ cross sections. These corrections typically give significant
corrections to the overall normalization. That is, they give
$$
\sigma \approx K \sigma_{LO}
$$
with a ``$K$-factor''
$$
K = 1 + C {\alpha_s \over \pi}
$$
that is typically substantially larger than one but less than two for
the natural lowest-order scale choice because $C \sim \pi^2$ rather
than $C\sim1$.  While the effect on the normalization is significant,
the effect on the shape of inclusive distributions is typically small.
This is illustrated in Figure~\ref{lhcpt}, which shows the $p_T$
distributions for the leading-order (LO) and next-to-leading-order
(NLO) calculations for SUSY production at the LHC.\cite{BHSZ} The two
shapes are clearly almost identical.

	The similarity in the shapes of the LO and NLO distributions
seems to be a general feature of perturbative QCD calculations for a
wide variety of $pp$ processes. It is partly understood, at least for
the simple case of Drell-Yan.\cite{KP,AEM} The NLO Drell-Yan
calculation contains a $\ln^2(-Q^2)$ factor from the overlapping soft
and collinear singularities. The logarithm is canceled by a
$\ln^2(Q^2)$ factor for deep inelastic scattering, but there is a
$\pi^2$ from the difference. This $\pi^2$ multiplies the natural scale
of $(4\alpha_s/3\pi)\sigma_0$, where $\sigma_0$ is the lowest order
cross section, so it produces an overall normalization factor but no
change in shape. It is not clear, however, that this $\pi^2$ factor is
the dominant effect even in the simple case of Drell-Yan.

	The cross sections shown in Figure~\ref{lhcsigma} correspond
to large rates. ``Low'' luminosity at the LHC is $10^{33}\,\cmsec$
(about $10^2$ times that currently achieved at the Tevatron) while the
design luminosity is $10^{34}\,\cmsec$. A ``Snowmass year'' is defined
to be $10^7\,{\rm s}$, not $3.1536 \times 10^7\,{\rm s}$; it
represents the typical effective running time per year for an
accelerator. For example, in the recently completed Tevatron Run~I, an
integrated luminosity of about $100\,\pb^{-1}$ was obtained in about
two years with peak luminosity of about $10^{31}\,\cmsec$.  Given a
luminosity of $10^4-10^5\,\pb^{-1}$ per year, a large number of SUSY
events will be produced if the masses are below $\sim1\,\TeV$, as can
be seen from Table~\ref{tbl:rate}.

\begin{table}[t]
\caption{Events/year for various SUSY masses at LHC at low and high
luminosity using the cross sections shown in
Figure~\protect\ref{lhcsigma}.\label{tbl:rate}}
\medskip
\begin{center}
\begin{tabular}{ccc}
\hline\hline
$M_\tg = M_\tq$ (GeV)	& $\sigma$ (pb)	& Events\\
\hline
500\strut		& 100		& $10^6-10^7$\\
1000			& 1		& $10^4-10^5$\\
2000			& 0.01		& $10^2-10^3$\\
\hline\hline
\end{tabular}
\end{center}
\end{table}

	The gluino and squark $p_T$ distributions must obviously peak
at $p_T \sim M$, since smaller $p_T$'s are suppressed by phase space,
while larger ones are suppressed by the requirement of additional
energy for the initial partons. This peaking can be seen in
Figure~\ref{lhcpt} both for the LO and for the NLO calculations.  This
means that the gluino or squark decay rest frame and the lab frame are
similar, and the decay products in the lab frame are spread out over
phase space. Gluinos or squarks decay via several steps to the lightest
SUSY particle (LSP), taken to be the $\lsp$. Thus a typical event
might be:
\begin{eqnarray*}
&g + g \to \tg + \tg& \\
&\tg \to \tq_L \bar q;\quad \tq_L \to \tchi_2^0 q;\quad \tchi_2^0 \to
\lsp \ell^+\ell^- (\lsp q \bar q)&\\
&\tg \to \tq_R \bar q;\quad \tq_R \to \lsp q\,.&
\end{eqnarray*}
This example event contains five new particles(!). The $\lsp$ has
electroweak couplings and must produce virtual sleptons or squarks
when interacting with matter, so interacts weakly and escapes from the
detector, giving $\etmiss$.  Signatures for such events involve
multiple jets, large $\etmiss$ from the two weakly-interacting
$\lsp$'s, possible leptons from decays of gauginos or sleptons in the
cascade decay process, and $b$ jets that are either democratically
produced or enhanced by smaller third-generation squark masses and
large Yukawa couplings.

	All of these features provide handles that can be used to
separate the SUSY signals from Standard Model backgrounds. As will be
discussed in Section~\ref{sec:det} below, the LHC detectors can
measure the $E_T$ of all jets and leptons, so they can determine
$\etmiss$ (with possible problems from cracks and resolution tails).
They can identify and measure electrons and muons well; hadronically
decaying $\tau$'s can also be identified, although with more
background. Finally, $b$ jets can be tagged using vertex detectors to
observe displaced vertices.

	The single most important experimental signature for SUSY at
the LHC is probably missing transverse energy $\etmiss$. The Standard
Model backgrounds for $\etmiss$ come from all the possible ways to
produce high-$p_T$ $\nu$'s:
\begin{itemize}
\item	$W \to \ell\nu$, $\tau\nu$
\item $Z \to \nu\bar\nu$, $\tau\tau$
\item	$Q \bar Q \to \ell\nu X$, $Q=c,b,t$.
\item	$g \to Q \bar Q$, $Q \to \ell\nu X$
\end{itemize}
The electroweak cross sections are suppressed by powers of $\alpha$,
while the heavy quark ones are suppressed by color and spin factors.
Both are suppressed by leptonic branching ratios and by the fact that
the missing $\nu$ typically has smaller $p_T$ than its parent. There
are also backgrounds to $\etmiss$ from cracks and resolution tails in
the detector. These are difficult to estimate in any simple way, but
detailed calculations indicate that real neutrinos dominate for events
with multiple jets and/or leptons and $\etmiss \simge 100\,\GeV$. 

\section{Event Simulation\label{sec:sim}}%

	SUSY signatures typically involve multiple jets and/or leptons
plus missing transverse energy $\etmiss$ arising from complex cascade
decays. The Standard Model backgrounds for these signatures arise only
from high orders of perturbation theory, and the dominant
contributions come from those regions of phase space that give soft or
collinear singularities. Thus it is appropriate to use Monte Carlo
event generators that simulate complete events and include the most
important effects of QCD radiation to calculate both the signals and
the backgrounds. Such event generators also provide complete events
that can be used to estimate detector-induced backgrounds.

	Three general-purpose event generators are in common use:
HERWIG\cite{Herwig}, ISAJET\cite{Isajet}, and PYTHIA.\cite{Pythia} Of
these, HERWIG provides the best theoretical treatment of QCD but
presently contains no SUSY processes. ISAJET has the most detailed
treatment of SUSY. PYTHIA provides a well-tuned description of QCD and
hadronization plus many SUSY processes. While a precise description of
QCD and hadronization may eventually be important, it it probably not
essential for the exploratory studies now being done.

	Any Monte Carlo event generator requires a random number
generator. If the numbers were truly random, then each run would
produce different results, and debugging would be nearly impossible.
Hence we require a pseudo-random number generator that produces the
same sequence of random numbers given the same initial seed(s). A
commonly used algorithm is the congruential generator.\cite{Knuth} It
is surprisingly simple: 
\begin{eqnarray*}
&s_0 > 0\,,&\\
&s_{n+1} = (a s_n + c) \mod m\,.&
\end{eqnarray*}
For careful choices of $a$ and $c$ and a large value of $m$, e.g.{}
$a=7^5$, $c=0$, $m=2^{31}-1$ or $a=6909031764870$, $c=0$, $m=2^{48}$,
this algorithm yields a uniformly distributed sequence of integers
satisfying most tests of randomness. (It is obvious that there are
also extremely bad choices.) Then $x_n = s_n/m$ provides a uniformly
distributed real number in $(0,1)$. The congruential generator does
have the limitation that $k$-tuples of $x_n$ lie on at most $m^{1/k}$
planes in $k$-space.  The random number algorithm does not at this
time seem to be the critical issue for LHC SUSY physics, so we will
not discuss it further here.

\subsection{Hard Processes and Perturbative QCD\label{sec:hard}}

	Given a random number generator, one can construct an
algorithm for generating events according to any desired parton
processes. This section outlines the basic steps, but since this
Summer School is devoted to SUSY rather than to QCD, it opts for
simplicity rather than attempting to review the state of the
art.\cite{Webber}

	{\bf Step~1:} Generate the parton process. i.e., pick the
kinematic variables according to the appropriate lowest order parton
cross section. Using a lowest order cross section is important: higher
order cross sections would involve singular distributions. These can be
treated in principle using a cutoff, but the resulting distributions
have very large positive/negative weight fluctuations.

	This step is simple in principle, although it may be
complicated in practice. From the definitions of cumulative
probability distributions and random variables, if $x$ is distributed
according to a (normalized) probability distribution $f(x)$, $0<x<1$,
then 
$$
\int_0^x dx'\,f(x') = \zeta,\quad 0<\zeta<1\,,
$$
where $\zeta$ is a uniformly distributed random variable that can be
generated using a congruential or other generator. In a few cases,
e.g.\ $f(x) = (n+1)x^n$, this relation can be solved analytically.
Generally, an analytic solution is not possible, but one can find a
bound $f(x) < f_{\rm max}$ and use a rejection algorithm, accepting
$x=\zeta_1$ if $f(x) > \zeta_2 f_{\rm max}$.  It is possible to
combine these methods and/or change variables to improve the
efficiency.

	To generate a parton process like $gg \to \tg\tg$, one should
choose a set of variables like $p_T, y_1, y_2, \phi$ to isolate the
rapid cross section variation in a just a few variables, e.g., $p_T$.
For processes like gluon and light quark jets that do not involve a
mass scale, some form of importance sampling or weighting should be
used to generate $p_T$ efficiently. For process like $Z$ or $H$ with a
narrow $s$-channel resonance, one should instead choose a set of
variables such as $\hat s, \hat t, \hat y, \phi$. Generating the hard
scattering is basically straightforward, although it can involve
rather complex computer code.

\bigskip

	{\bf Step~2:} Add QCD radiation. Perturbation theory describes
inclusive cross sections, but it does not necessarily give a good
description of event structure. QCD is approximately scale invariant,
with a dimensionless coupling $\alpha_s$ whose running with $Q^2$ is
only logarithmic. Hence, radiation at all $Q^2$ scales is important;
it produces fixed angle jets with $0 \ll \langle M^2_\jet \rangle \sim
\alpha_s(p_T^2) p_T^2 \ll p_T^2$.

	The most important QCD radiation is the emission of gluons
collinear with initial or final partons. But we know from the
factorization theorems of perturbative QCD that collinear
singularities factorize, not for amplitudes but for cross sections.
That is,
$$
\sigma_{n+1} \sim \sigma_n \otimes {\alpha_s(p^2) \over 2\pi p^2}
P_{ij}(z),\quad p^2 \to 0\,,
$$
where $P_{ij}$ is the appropriate Altarelli-Parisi (DGLAP) function
and $z \sim p_i/p$. The starting point for QCD-based event generators
like HERWIG, ISAJET, and PYTHIA is to use this collinear approximation
for gluon radiation together with exact, non-collinear kinematics.
This is known as the branching approximation.\cite{FoxWolfram} The
branching approximation correctly describes the dominant, leading-log
QCD effects, but because it uses exact kinematics, it also gives a
fairly good approximation for real higher-order QCD effects such as
multiple jet production.

	The discussion that follows is limited only to the simplest
case of leading-log gluon radiation from an outgoing quark line. This
does not represent the state of the art, but it is simple to present
and illustrates the main point, namely that a classical branching
process can provided a reasonable approximation to QCD to all orders
in perturbation theory. The cross section for the radiation of $n$
gluons from an outgoing quark is given in the collinear limit by
$$
\frac{1}{\sigma} {d\sigma \over dp_1^2 dz_1 \ldots dp_n^2 dz_n}
= \left[{\alpha_s(p_1^2) \over 2 \pi p_1^2} P(z_1)\right] \ldots
\left[{\alpha_s(p_n^2) \over 2 \pi p_n^2} P(z_n)\right]\,.
$$
where $z$ is the momentum fraction. While $z$ is well defined in the
collinear limit, its non-collinear extension is model dependent.  A
good choice is
$$
z_i = {E_{i+1} + |\vec p_{i+1}| \over E_{i} + |\vec p_{i}|}\,.
$$
With this choice, radiation from the quark and antiquark in $e^+e^-
\to q \bar q$ is treated symmetrically.

	Introduce a minimum mass $t_c$ to regulate both the soft and
the collinear singularities. For a branching $t_0 \to t_1 + t_2$ in
the initial $t_0$ rest frame, the final energy and momenta are given
by simple two-body kinematics:
$$
E_1^* = {t_0+t_1-t_2 \over 2\sqrt{t_0}}\,, \qquad
p_1^* = {[(t_0-t_1-t_2)^2 - 4t_1t_2]^{1/2} \over 2\sqrt{t_0}}\,.
$$
The minimum energy and momentum of decay product 1 depend on the
relation between the mass and velocity of the decaying system, but the
maximum energy and momentum always correspond to $\cos\theta=1$.
Boosting the center of mass momenta back to the lab frame, we find
$$
z_{\rm max} = 1 - z_c = {t_0+t_1-t_2 \over 2t_0} 
+ {[(t_0-t_1-t_2)^2-4t_1t_2]^{1/2} \over 2t_0}\,.
$$
Setting $t_1=t_2=t_c$, the cutoff mass, gives
$$
z_c = \half \left[1 - \sqrt{1-{4t_c \over t_0}}\right]\,.
$$
This cutoff will be used in what follows.

	The next step is to sum up the radiation softer than the
cutoff into a Sudakov form factor, i.e., into the probability of
evolving from an initial mass $t$ to a final mass $t'$ emitting no
resolved radiation greater than the cutoff. The calculation of this
Sudakov form factor is particularly simple if one takes the cutoff to
be not a fixed mass $t_c$ but a fixed smallest $z_c$ defined by the
initial mass and $t_c$ cutoff. Kinematics requires
$$
t \ge t_1 \ge \ldots \ge t'\,.
$$
By definition of a $z$ cutoff, all the $z_i$ are in the interval$$
{\cal R} = \{ z < z_c  \} \cup \{ z > 1-z_c \}\,.
$$
Thus, using the factorized form for gluon emission, the Sudakov form
factor $S$ is given by
\begin{eqnarray*}
S_i(t,t') &= \displaystyle \sum_n \int_{t'}^{t} dt_1 \, {\alpha_s(t_1) 
   \over 2 \pi t_1}
\int_{t'}^{t_1} dt_2 \, {\alpha_s(t_2) \over 2 \pi t_2} \ldots
\int_{t'}^{t_{n-1}} dt_n \, {\alpha_s(t_n) \over 2 \pi t_n} \\
&\quad\times \displaystyle \int_{\cal R} dz_1 P_i(z_1) \ldots 
   \int_{\cal R} dz_1 P_i(z_1)\,.
\end{eqnarray*}
The form of this expression is simple because using a $z$ cutoff
decouples the $z$ integrals from the $t$ ones.

	Virtual graphs contribute $\delta(1-z)$ terms to the DGLAP
functions. Momentum conservation requires that the $x$-weighted
integral of the gluon and quark distributions give the total momentum
fraction, unity:
$$
\int_0^1 dx\,\left[xg(x,Q^2) + \sum_q xq(x,Q^2)\right] = 1\,.
$$
This fixes the constants $c_{qq}$ and $c_{gg}$ in the DGLAP functions 
to satisfy
$$
\int_0^1 dz\, \left[{P_{ii}(z) + c_{ii}\delta(1-z)}\right] = 0\,,
$$
where the $1/(1-z)$ singularities in the functions $P(z)$ are
regulated using "$+$" distributions:
$$
\int dz\,{f(z) \over (1-z)_+} \equiv \int dz\,{f(z)-f(1) \over
(1-z)}\,.
$$
The physical meaning of these $+$ distributions is that the soft
singularities cancel between the real and virtual graphs. Because the
complete integral over $P$ vanishes, the $z$ integrals in the
definition of $S$ over the unresolved soft radiation in the region
${\cal R}$ can be related to a well-defined integral over non-soft
values of $z$ in the complement of ${\cal R}$: 
$$
\int_{\cal R} dz \, P_i(z) = - \int_{z_c}^{1-z_c} dz \, P_i(z) 
\equiv - \gamma_i(z_c)\,.
$$
The integrals in the definition of $S$ over the $t_i$ are elementary,
and the nesting gives a factor of $1/n!$. The final result has the
simple form
$$
S_i(t,t') = \left[{\alpha_s(t) \over \alpha_s(t')}\right]
^{\frac{2}{b_0}\gamma_i(z_c)}\,.
$$

	Knowing the cumulative distribution $S$ is enough to generate
the next mass and hence the complete shower. Given $S$, the
probability $\Xi(t)$ that the first resolved radiation occurs at $t$
is given by
$$
\int_{t_c}^{t_0} dt\,\Xi(t) = 1 - S_i(t_0,t_c)\,.
$$
The mass-squared $t$ of the quark at the next branching is therefore
generated by solving the equation
$$
{1-S_i(t,t_c) \over 1-S_i(t_0,t_c)} = \zeta\,,
$$
where $\zeta$ is a uniformly distributed random number in $(0,1)$.
Given the form for $S$, this equation can be easily solved in terms of
elementary functions. This solution can then be used to build the
complete shower:
\begin{enumerate}
\item	Generate $t$.
\item	Generate $z_c < z < 1-z_c$ according to $P_i$.
\item	Check $z$ against $t_c$ limits.
\item	Generate $t_1$, $t_2$ starting at $z$, $(1-z)t$.
\item	Solve 2-body kinematics.
\item	Iterate.
\end{enumerate}
This simple Monte Carlo algorithm generates all the leading-log
effects of QCD radiation and provides a reasonable approximation to
non-leading effects such as multi-jet production.

	The algorithm just described can be extended to include $g \to
gg, q \bar q$ and initial state radiation. It can also be refined in
various ways, in particular to include in an approximate way coherent
interference effects that reduce the radiation of soft gluons. For a
recent review see Webber.\cite{Webber}

\bigskip

	{\bf Step~3:} Generate decays of the SUSY particles. This
involves selecting the decay mode using branching ratios calculated
from the SUSY masses and couplings and then then generating the decay
using 2-body or 3-body phase space. Parton showers from any outgoing
quarks or gluons are added using the branching approximation with an
initial scale equal to the parent particle mass.

	The use of phase space for a 3-body decay, say $\tq \to \lsp q
\bar q$, is not a good approximation if the squark $\tq$ is just
slightly heavier than the gluino. The squark poles should cause the
matrix element to peak where the gaugino and one quark are hard, while
the other quark is soft. All distributions should be continuous as the
squark mass varies from below to above the gluino mass, but the phase
space approximation is not. This problem does not affect any of the
points to be studied here. The approximations also lose information on
spin correlations between particles, but this is probably not very
important, at least for LHC studies.

\bigskip

	{\bf Step~4:} Hadronize the partons, i.e., generate a jet of
hadrons for each parton. This involves soft, nonperturbative QCD
physics, so one must rely on models tuned to experimental
fragmentation data. The physical picture is that the hard scattering
creates separated color charges connected by ``strings'' of gluons. As
the strings stretch, they break, pulling a $q \bar q$ pair out of the
vacuum. Since QCD is strong only at low $Q^2$, this breaking is a soft
process which only occurs with low $p_T$ and small $\Delta y$. Thus it
produces a jet of hadrons with limited $p_T$ relative to the parton
direction. Hence, the event structure is mainly controlled by
perturbation theory, not by the soft physics model. 

	The hard part of a $pp$ interaction never accounts for more
than a small fraction of the total energy. The rest of the energy is
carried off by spectators, which produce low-$p_T$ hadrons more or
less uniformly in rapidity. Typically the leading particle in the beam
jet carries half of the beam energy and just a few hundred MeV of
transverse momentum. As a result, much of the energy escapes down the
beam pipe: one can measure missing transverse energy $\etmiss$ but not
missing total energy. The models for beam jets are essentially
parameterizations of experimental data. Fortunately, the beam jets are
not very important for the experimental signatures, although they do
have some effect on the isolation requirements for leptons and
photons.  

\subsection{Soft or $\ln s$ Physics\label{sec:soft}}%

	The production of SUSY particles and of all the significant
Standard Model backgrounds for them are hard processes that can be
calculated in QCD perturbation theory. But most of the events at the
LHC are soft interactions, for which there is no good theoretical
description. These events are not important as backgrounds for new
physics, but they dominate the rate and so are important for the
design of the detectors. Fortunately, soft physics varies slowly with
$\ln s$, so reliable extrapolations from lower energy can be made even
without a good theoretical model.

	The geometrical size of the proton is set by $1/M_\pi$,
implying a natural geometrical cross section
$$
\sigma = \pi \left({\hbar c \over M_\pi}\right)^2 \approx 62\,\mb
$$
since $\hbar c = 197\,\hbox{MeV-f}$. This should be compared with
SUSY cross sections ranging from $0.01\,\pb$ to $100\,\pb$ depending
on the masses. The total cross section is of course given by the
imaginary part of the forward scattering amplitude,
$$
\sigma_T = {1\over s} \Im f(s,0)\,.
$$
In the high energy limit with fixed momentum transfer ($s \to \infty$,
$t$ fixed) the exchange of an elementary particle of spin $J$ in the
$t$ channel leads to
$$
f(s,t) \sim P_J(z_t) \sim s^J\,.
$$
The Froissart bound, based only on analyticity properties that can be
proven in field theory and on unitarity of the $S$-matrix, limits the
growth of the cross section to
$$
\sigma_T \le \hbox{const} \ln^2 s
$$
so we must have $J\le1$. 

\begin{figure}[t]
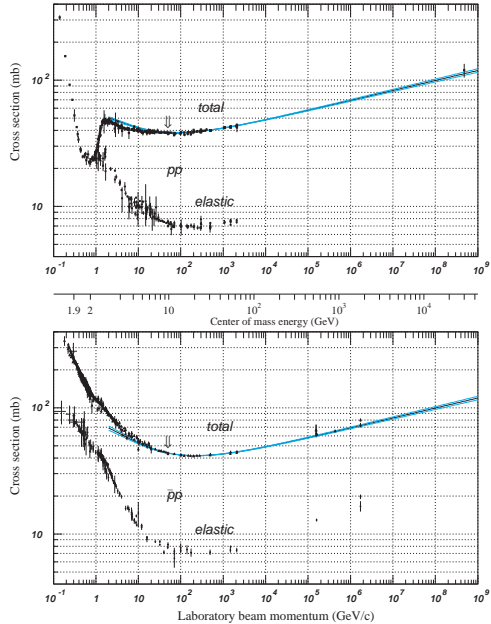

\dofig{2.5in}{sigmapp.ai}
\caption{Compilation of $pp$ and $\bar pp$ cross
sections.\protect\cite{PDG}\label{sigmapp}} 
\end{figure}

	The experimental data on the $pp$ and $\bar pp$ total cross
sections, Figure~\ref{sigmapp}, show rapid variation at low energy and
large differences between the $pp$ and $\bar pp$. This difference is
not surprising, since $\bar pp$ has a large cross section to
annihilate into mesons, while $pp$ cannot do this. At high energy,
however, the $pp$ and $\bar pp$ cross sections become equal and vary
slowly with $\ln s$, in a way quite consistent with the Froissart
bound. 

\begin{figure}[t]
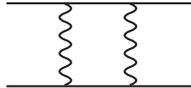

\dofig{1in}{feynbox.ai}
\caption{The discontinuity of the box graph gives a constant cross
section for $s\to\infty$.\label{feynbox}}
\end{figure}

	While the total cross section is certainly nonperturbative, it
is useful to consider how multiparticle production works in
perturbation theory. Because the gluon has spin one, the box graph,
Figure~\ref{feynbox}, gives a constant cross section for large $s$.
Of course the cross section is infinite for a massless gluon exchanged
between unconfined quarks, so an infrared cutoff is needed.  Hence
this whole discussion is only qualitative.  However, it is possible to
do a rigorous calculation of the behavior of the high-$p_T$ jet cross
section for $s \gg p_T^2$ in QCD perturbation theory.\cite{Mueller87}

\begin{figure}[t]
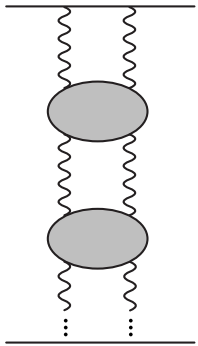

\dofig{1in}{feynladder.ai}
\caption{The discontinuity of the (generalized) ladder graph gives a
factor of $\ln s$ from each longitudinal phase space
integral.\label{feynladder}}
\end{figure}

	The generalized ladder graph in Figure~\ref{feynladder}
similarly remains constant as the generalized rungs, the shaded blobs
in the figure, are moved in rapidity. Hence, this graph gives a factor
of $\ln s$ from each longitudinal phase space integral. It can be
shown that graphs like this give the leading powers of $\ln s$ for
each order in perturbation theory. The $n$ longitudinal integrals are
ordered in rapidity and so give
$$
\int_0^{\ln s} dy_1 \dots \int_0^{y_{n-1}} dy_n = {1\over n!}\ln^n s\,.
$$
The discontinuity across the generalized rungs produces particles in
the final state, and the distribution of these particles, like that of
the rungs, is flat in rapidity. Furthermore, after many steps in
rapidity, any quantum number exchange will be washed out, so the
distribution in the central region is universal. All of these
properties of this toy model are in qualitative agreement with
experiment. However, the growth of Figure~\ref{feynladder} with $\ln
s$ violates the Froissart bound, so more complicated graphs must become
important.

\begin{figure}[t]
\dofig{2.5in}{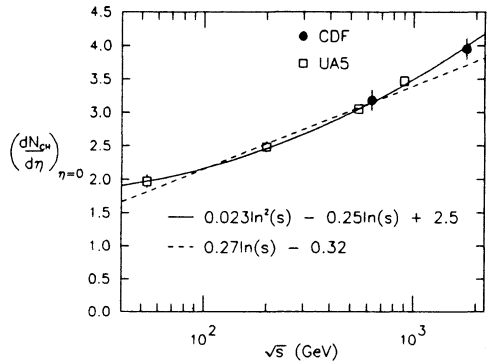}
\caption{Compilation of $dN/d\eta \approx dN/dy$ for $pp$ and $\bar
pp$ interactions.\protect\cite{CDFmult}\label{dndeta}}
\end{figure}

	Data on the charged multiplicity $dN/d\eta$, where
$$
\eta \equiv -\ln\tan{\theta\over2} \approx y \equiv \ln{E+p_L\over
E-p_L}\quad {\rm for}\quad p_T \gg m\,,
$$
are shown in Figure~\ref{dndeta} for $pp$ and $\bar pp$ interactions.
The rise is consistent with either a power of $\ln s$ or a small power
of $s$. A smooth extrapolation to $14\,\TeV$ gives
$$
{dN_{\rm ch} \over d\eta} \approx 6\,.
$$
The mean $p_T$ also grows slowly with $\sqrt s$, rising from about
$0.35\,\GeV$ at low energy to about $0.5\,\GeV$ at $1.8\,\TeV$. Most
of this rise probably comes from QCD jets with transverse momenta of a
few GeV. Extrapolation suggests $\langle p_T\rangle \sim0.65\,\GeV$ at
LHC.

	Soft interactions do not produce backgrounds for SUSY or other
interesting processes, but they do give a very high interaction rate.
The LHC bunch crossing rate is $40\,{\rm MHz}$, and there are $\sim2$
interactions/bunch at $10^{33}\,\cmsec$ and $\sim20$
interactions/bunch at $10^{34}\,\cmsec$. 

\section{LHC Detectors\label{sec:det}}%

	Two LHC large detectors, ATLAS\cite{ATLAS} and CMS,\cite{CMS}
are just beginning construction. Both ATLAS and CMS are designed to
identify and measure all the Standard Model quanta --- $\gamma$, $e$,
$\mu$, $\tau$, $g$ and $q$ jets, and $b$ jets --- over $|\eta| \simle
2.5$ ($\theta \simge 10^\circ$) and to measure energy flow over
$|\eta| \simle 5$ ($\theta\simge 1^\circ$) to determine the missing
transverse energy $\etmiss$. This broad coverage is essential for
complex signatures like SUSY. The cost of each detector by CERN
accounting rules is CHF~475M.\footnote{Both the U.S.{} and CERN do not
include physicists salaries. CERN also does not include most EDIA
(Engineering, Design, Inspection, and Administration) and labor costs.
This means that 1~CHF at CERN is roughly equivalent to \$2 in the
U.S.} Each collaboration currently includes about 1500 physicists and
senior engineers. Thus, both ATLAS and CMS are more like laboratories
than traditional experiments.

\begin{figure}[p]
\setbox0=\vbox to \textwidth{
\ifBigFig\centerline{\epsfbox{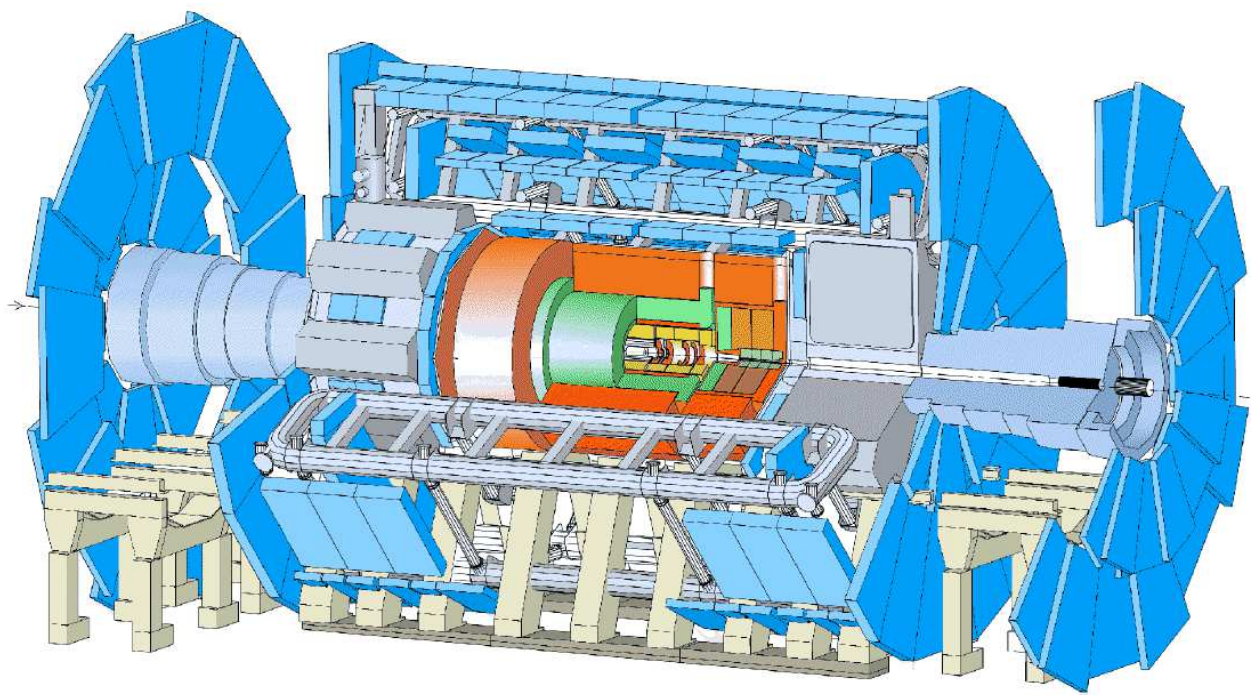}}%
\else\fi
\vfil
\caption{Cutaway view of the ATLAS detector\protect\cite{ATLAS} as of
1997, showing from inside out the central detector, 2~Tesla solenoidal
magnet coil, liquid argon electromagnetic calorimeter, hadron
calorimeter and muon system including the lumped toroidal magnet
coils and three layers of muon chambers.\label{ATLASdet}}
}
\centerline{\rotl0}
\vfill
\end{figure}

\begin{figure}[t]
\ifBigFig\dofig{\textwidth}{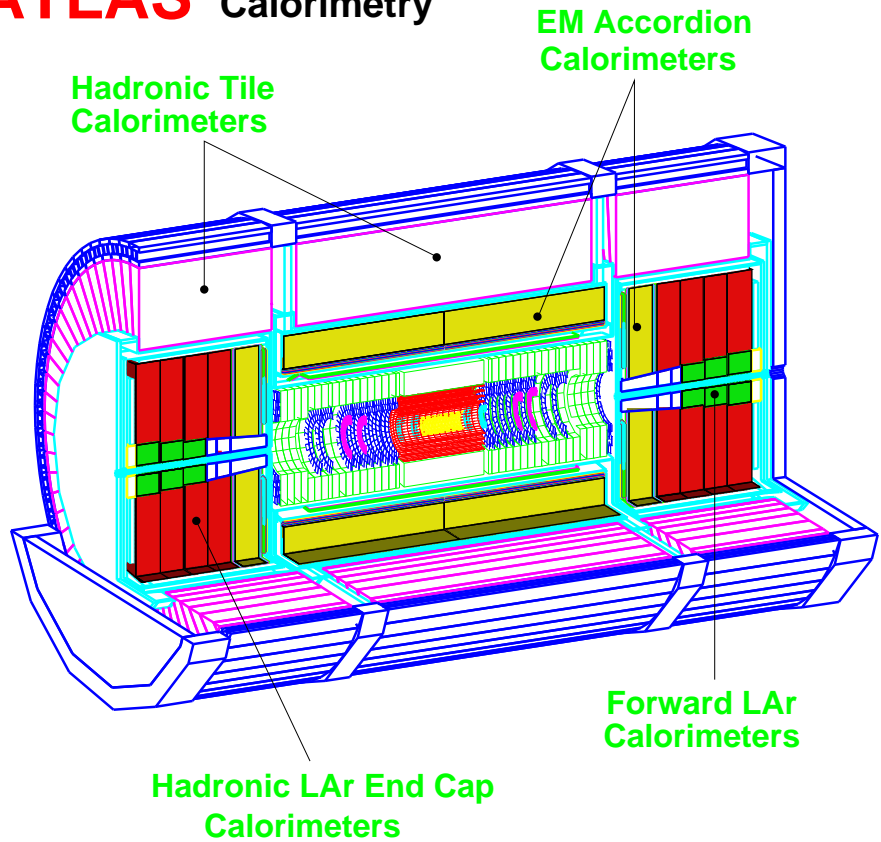}%
\else\vbox to \textwidth{\vfill}\fi
\medskip
\caption{Cutaway view of the ATLAS calorimeter and central tracker,
the central part of Figure~\protect\ref{ATLASdet}.\label{ATLAScal}}
\end{figure}

\begin{figure}[p]
\setbox0=\vbox to \textwidth{
\ifBigFig\centerline{\epsfbox{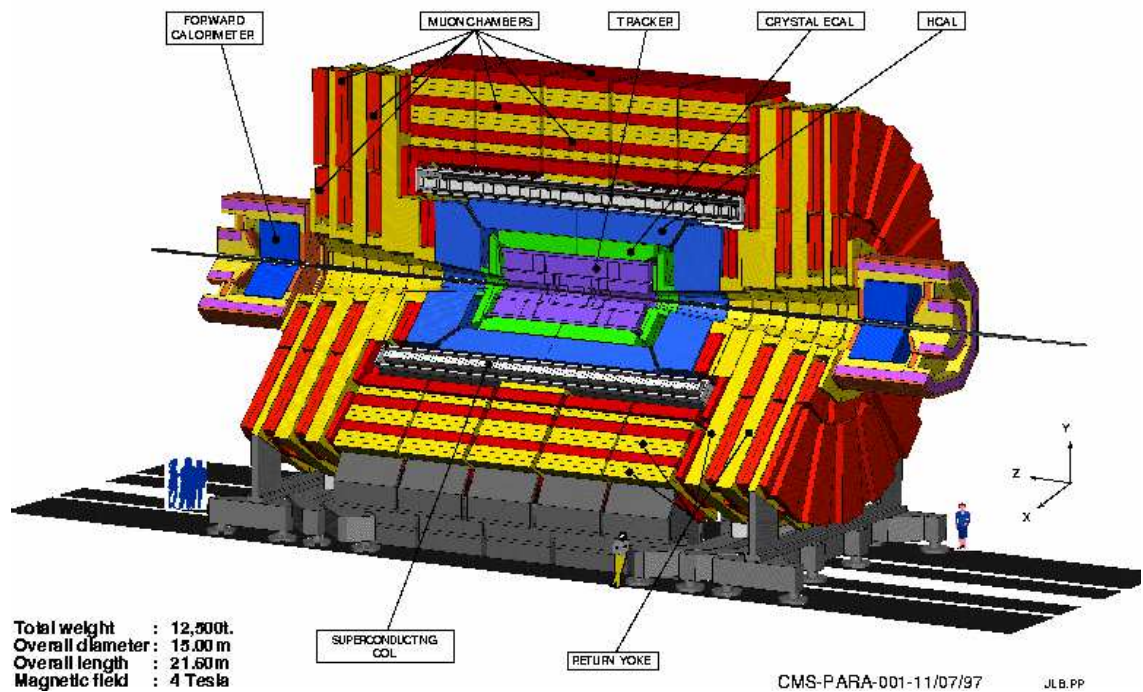}}%
\else\fi
\vfil
\caption{Cutaway view of the CMS detector\protect\cite{CMS} as of 1997,
showing from inside out the central tracker, crystal electromagnetic
calorimeter, hadron calorimeter, 4~Tesla solenoidal magnet coil, and
muon system.\label{CMSdet}} 
}
\centerline{\rotl0}
\vfill
\end{figure}

\begin{figure}[t]
\dofig{\textwidth}{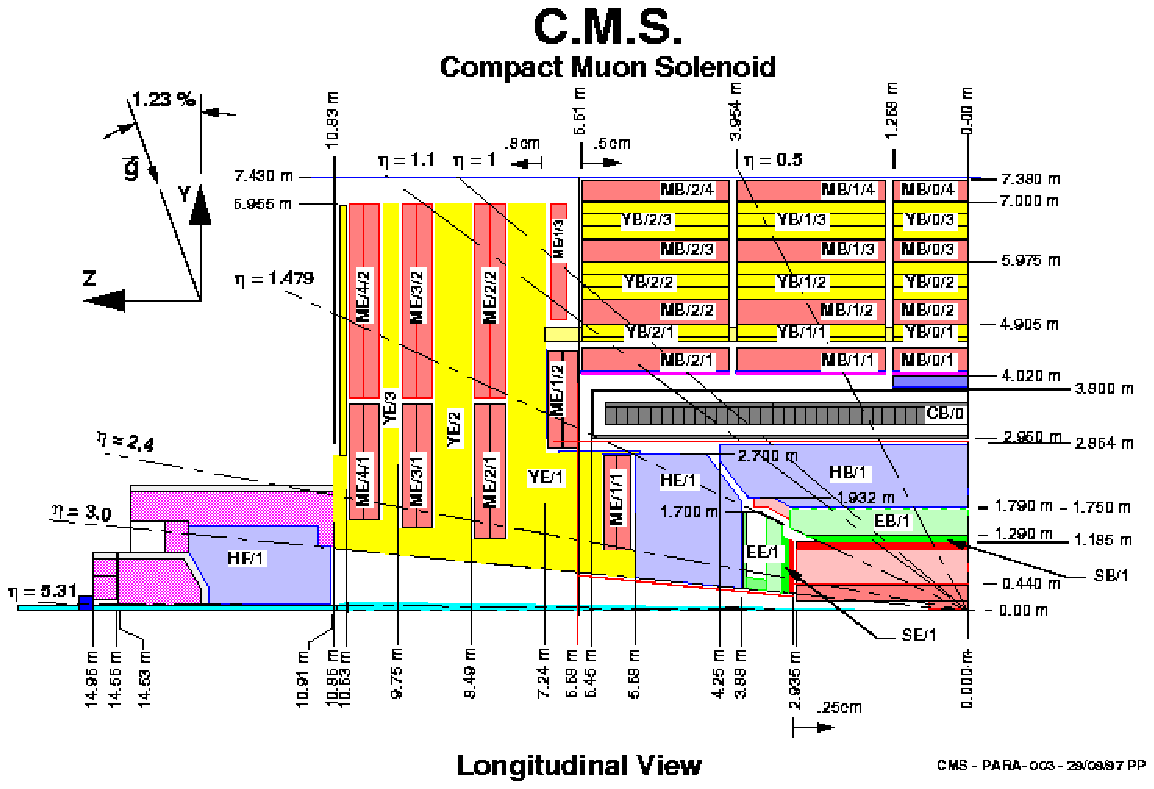}%
\medskip
\caption{Cross section of one quarter of the CMS
detector,\protect\cite{CMS} showing the same elements as
Figure~\protect\ref{CMSdet}.\label{CMSlong}}
\end{figure}

	The cost and scale of the detectors is driven both by the need
to measure the high energies and large range of possible interesting
processes such as SUSY production and by the need to cope with the
high rates caused by soft physics at the LHC. While the detectors are
quite different in detail, they both have the same basic elements.
These are from inside out: a silicon vertex detector intended
primarily for tagging $b$ jets, an inner tracker to measure the
momenta of charged particles in a magnetic field, an electromagnetic
calorimeter to measure the energies and directions of photons and
electrons, a hadron calorimeter to measure jets, a forward calorimeter
mainly to measure $\etmiss$ but also to tag forward jets, and a muon
system to identify and to measure muons. These parts can be seen in
Figures~\ref{ATLASdet} -- \ref{CMSlong}. Section~\ref{sec:detelem}
below gives a brief description of each of these elements and their
most important performance parameters. This is followed by
Section~\ref{sec:detperf}, which explains how the parts are used in
combination to detect physics signatures. The discussion given here is
necessarily superficial, but hopefully it will prove useful to
theorists interested in LHC physics. For details the reader should see
the ATLAS\cite{ATLAS} and CMS\cite{CMS} {\sl Technical Proposals} and
the {\sl Technical Design Reports} for the detector subsystems. The
Particle Data Group\cite{PDG} provides useful general information on
particle detectors.

\subsection{Detector Elements\label{sec:detelem}}

	{\it Silicon vertex detector:} Both ATLAS and CMS have silicon
microstrip detectors covering $|\eta|<2.5$. These are primarily
intended to tag $b$ jets by detecting the displaced vertices of $B$
hadrons, but they also contribute to the momentum measurement. The
detector elements are made out of silicon wafers similar to those used
to make computer chips. A charged particle passing through the silicon
causes ionization, and the resulting electrons are collected on strips
about $80\,\mu{\rm m}$ wide. An amplifier and discriminator on each
strip determines which strips were hit. For a uniform distribution of
a variable $x$ in an interval $(-a/2,a/2)$,
$$
\langle x^2 \rangle = {1\over a}\int_{-a/2}^{a/2} dx\,x^2 =
{a^2\over12}\,,
$$
so the nominal resolution orthogonal to the strip direction is about
$80\,\mu{\rm m}/\sqrt{12}$. The resolution in practice is very
similar. Measurement of the other coordinate is obtained by using
small-angle stereo, i.e., by placing the strips of adjacent layers at a
small angle. The innermost layers of the silicon detectors will use
pixels of about $50\,\mu{\rm m} \times 300\,\mu{\rm m}$ rather than
strips. These are still under development.

	{\it Central Tracker:} Given the high multiplicity at the LHC,
it is a difficult problem in pattern recognition to combine the right
hits to find the real tracks. Having many layers in the tracking
system helps, but it is prohibitive in terms of both cost and material
to make many silicon layers. In ATLAS the outer portion of the tracker
uses 70 layers of straw tubes. These have a thin conducting outer
shell and a fine wire in the center, with a high voltage between them.
Charged particles produce ionization in the gas, which is chosen to
give a constant drift velocity for the produced electrons. These
electrons drift to the central wire, where the high field produces an
avalanche with a typical gas gain of $\sim10^4$. The time of this
signal is measured, and the drift velocity is used to convert this to
a drift distance with a resolution of order $100\,\mu{\rm m}$. 

	The ATLAS tracker is a ``Transition Radiation Tracker.'' It
also incorporates a transition radiation detector, which detects the
$X$-rays emitted by a charged particle passing through a dielectric
interface. This radiation is proportional to $\gamma$ and is useful
for identifying low-$p_T$ electrons. 

	The CMS outer tracker uses gas microstrip detectors, which are
still being developed. The ionization is produced in a layer of gas,
but it is detected using closely spaced strips on a substrate read out
in a manner similar to silicon strips. There are fewer layers than in
the ATLAS tracker, but each layer has better position resolution and
lower occupancy.

	The purpose of the central tracker is to determine the momenta
of charged particles by measuring their curvature in the central
solenoidal magnetic field. As a consequence of the Lorentz force, $F=e
\vec v \times \vec B$, a charged particle in a uniform solenoidal
magnetic field $B$ follows a helix with a radius of curvature
$$
R={p_T \over 0.3B}\,,
$$
with $R$ in meters, $B$ in Tesla, and $p_T$ in GeV. Simple geometry
shows that the resulting sagitta\footnote{The sagitta, a standard term
in elementary geometry, is the maximum separation between the arc of a
circle and its chord, the straight line between its ends.} $s$ for a
radial length $L \ll R$ is given by
$$
s = {L^2 \over 8 R} = {0.3 BL^2 \over 8p_T}
$$
with $s$ and $L$ in meters and $p_T$ in GeV. For $p_T=100\,\GeV$,
$B=2\,{\rm T}$, and $L=1\,{\rm m}$, the sagitta is $s=750\,\mu{\rm
m}$. The resolution depends on the chamber layout and position
resolution and on the multiple scattering in the chamber material.
ATLAS has $B=2\,{\rm T}$, typical for most detectors, and $R = 1\,{\rm
m}$, giving
$$
{\Delta p_T \over p_T} \approx 0.7\left({p_T \over 1\,\TeV}\right)
\oplus 0.014 \,,
$$
where the constant term comes from multiple scattering and is added in
quad\-ra\-ture. The resolution for tracks beyond the corner of the
solenoid degrades like $1/\sin^2\theta$. CMS has a very high field,
$B=4\,{\rm T}$, and also a larger radius, giving
$$
{\Delta p_T \over p_T} \sim 0.1\left({p_T \over 1\,\TeV}\right)\,,
$$
again degrading beyond the corner of the solenoid. ATLAS has a
comparable resolution for muons only using its muon system.

	{\it Electromagnetic Calorimeter:} Precision electromagnetic
calorimetry has been used in several $e^+e^-$ detectors, but ATLAS and
CMS are the first hadron-collider detectors to have it. The demand for
very high resolution is driven by the search for $h \to \gamma\gamma$;
it will be seen in Section~\ref{sec:higgs} that this gives a narrow
peak on a large $\gamma\gamma$ continuum background. The ability to
separate $e$ and $\gamma$ depends on the tracker, so the useful
calorimeter coverage is also $|\eta|<2.5$.

	Any calorimeter creates a shower in some dense material and
uses the total charge or light output from this shower to determine
the energy of the initiating particle. The energy is divided among
more and more particles until it is completely absorbed. It is
important to realize that electromagnetic interactions with a dense
material like lead are much stronger than hadronic ones:
electromagnetic interactions scale like $Z^2$ while hadronic ones
scale like $A^{2/3}$. For lead, the radiation length $X_0$, the
distance in which a high energy electron loses all but $1/e$ of its
energy, is $0.56\,{\rm cm}$, while the inelastic hadronic interaction
length $\lambda$ is $17.1\,{\rm cm}$.  (For a light material like
aluminum $X_0=8.9\,{\rm cm}$ and $\lambda=39.4\,{\rm cm}$.) Thus, an
electromagnetic calorimeter $\sim 25X_0$ thick will contain almost all
of the shower from high energy electrons or photons while absorbing
little hadronic energy. Thus the electromagnetic calorimeter is always
in front of the hadronic one.

	The ATLAS electromagnetic calorimeter uses lead plates with
gaps filled with liquid argon. Electrons from the shower drift under
high voltage through the liquid argon and are collected on readout
pads. The energy is proportional to the total charge, with an energy
resolution
$$ 
{\Delta E \over E} \approx {10\% \over \sqrt{E}} \oplus 0.5\% 
$$
with $E$ measured in GeV. The first term here comes from the shower
multiplicity and the fluctuations in sampling it: the multiplicity $N$
of particles in the shower is proportional to the energy, and the
Poisson fluctuation in $N$ is $1/\sqrt{N}$. Note that this term gives
the same resolution for one particle or from several with the same
total energy, the errors being added in quadrature, as one would
expect for a calorimeter. The small constant term arises from many
sources and is added in quadrature. The ATLAS calorimeter can also use
the position of the shower as a function of depth to measure the
direction of a photon with an accuracy of about $50\,{\rm
mr}/\sqrt{E}$.

	CMS uses a dense, transparent crystal, ${\rm PbWO_4}$, both to
create the shower and to convert it into scintillation light that can
be detected by photodiodes. Because there are no inert lead plates,
the whole shower can be measured, the sampling fluctuations are
reduced, and the resolution is therefore better,
$$ 
{\Delta E \over E} \approx {2\% \over \sqrt{E}} \oplus 0.5\% \,,
$$
with $E$ measured in GeV. While crystals give better energy
resolution, they make it harder to achieve fine segmentation and good
pointing accuracy, and controlling the crystal quality is not trivial.
At low luminosity, CMS will rely on tracking to determine the vertex
and hence the photon direction. At high luminosity, it will add a
preshower detector to measure the starting position of the shower and
provide directional information at the cost of some energy resolution.

	{\it Hadron Calorimeter:} The hadron calorimeter follows the
electromagnetic one and measures the energy of both charged and
neutral hadrons. Hadronic showers have intrinsically larger
fluctuations than electromagnetic ones, and additional errors are
introduced for jets by the clustering algorithm. Both the ATLAS and
the CMS central hadron calorimeters use steel plates (which also act
as the magnetic flux return) and sample the hadronic showers with
scintillators read out by photomultiplier tubes. (The ATLAS endcap
hadron calorimeter uses copper plates and liquid argon.)  The
resulting jet resolutions when combined with the electromagnetic
calorimeters are roughly 
$$ 
{\Delta E \over E} \approx {60\% \over \sqrt{E}} \oplus 3\%\,,\qquad
|\eta|<3\,,
$$
with $E$ measured in GeV. The first term comes from the shower
multiplicity and the fluctuations in sampling it, as for the
electromagnetic calorimeter. The second term is added in quadrature
and comes, e.g., from the fact that there are fluctuations in the
fraction of the energy carried by $\pi^0$'s and by charged $\pi$'s,
and the calorimeter responds differently to these.\footnote{It is
possible to build calorimeters like that in the ZEUS detector at HERA
with nearly equal electromagnetic and hadronic responses, but they
generally have poorer electromagnetic resolution, the primary emphasis
for both ATLAS and CMS.} The forward calorimeters cover $3<|\eta|<5$
with cruder energy resolution since $E \gg p_T$ in this region.

	{\it Solenoid:} The central trackers require solenoids to
provide the magnetic field. The 2~Tesla solenoid in ATLAS is thin,
$<1X_0$, so it can be placed in front of the electromagnetic
calorimeter without degrading its resolution too much. The 4~Tesla
solenoid in CMS must be much thicker and so must be placed outside the
hadron calorimeter.

	{\it Muon System:} Muons radiate much less than electrons and
so penetrate the whole calorimeter with small energy losses, at least
for energies below the TeV scale. ATLAS makes its precise muon
measurement with an air-core toroidal magnet outside the calorimeter
and three groups of tracking chambers to measure the resulting
sagitta. The barrel toroid is made of eight lumped coils, which can be
seen in Figure~\ref{ATLASdet}. Lumped coils are needed to allow
chambers to be placed within the toroid but give a rather complex,
non-uniform field.  The endcap toroid has one group of chambers in
front of it and two groups behind it. The resolution $p_T$ for large
$p_T$ in the central region is 
$$
{\Delta p_T \over p_T} \sim 10\%\left({p_T \over 1\,\TeV}\right)\,;
$$
it remains quite good up to $\eta = 2.5$ because the endcap toroid
gives a $B \propto 1/r$ at small radius, thus increasing the bending
power at large $\eta$.

\begin{table}[t]
\caption{A set of possible Level~1 and Level~2 triggers for
ATLAS\protect\cite{ATLAS} at $\calL = 10^{34}\,\cmsec$. The $\etmiss$
thresholds are not yet known but should be $\sim100\,\GeV$; they will
be set to give the indicated rates.\label{tbl:trig}}
\medskip
\begin{center}
\begin{tabular}{lcc}
\hline\hline
Trigger Requirement				& LVL1 Rate	& LVL2 Rate\\
						& (kHz)		& (kHz)\\
\hline
$\ge1$ muon, $p_T > 20\,\GeV$			& 4		& \\
\quad$\ge1$ isolated $\mu$, $p_T > 20\,\GeV$	& 		& 0.2\\
\quad$\ge1$ $\mu$, $p_T > 40\,\GeV$		& 		& 0.1\\
\hline
$\ge1$ isolated e.m. cluster, $E_T>30\,\GeV$	& 20		& \\
\quad$\ge1$ electron $E_T>30\,\GeV$		& 		& 0.3\\
\quad$\ge1$ isol.\ e.m.\ cluster, $E_T>30\,\GeV$& 		& 0.1\\
\hline
$\ge2$ muons, $p_T>6\,\GeV$			& 1		& \\
\quad$\ge2$ $\mu$, $p_T>10\,\GeV$		&		& 0.1\\ 
\quad$\ge2$ isolated $\mu$, $p_T>6\,\GeV$	&		& 0.01\\ 
\hline
$\ge2$ isolated e.m. clusters, $E_T>20\,\GeV$	& 4		& \\
\quad$\ge2$ $e$ or $\gamma$, $p_T>20\,\GeV$	& 		& 0.2\\
\hline
$\ge1$ jet, $E_T>150\,\GeV$			& 3		& \\
\quad$\ge1$ jet, $E_T>300\,\GeV$		&		& 0.1\\
\quad$\ge3$ jets, $E_T>150\,\GeV$		& 		& 0.04\\
\hline
Missing energy ($\etmiss$)			& 1		& 0.1\\
\hline
Prescaled triggers				& 5		& 0.1\\
\hline                                        
Total						& 38		& 1.4\\
\hline\hline
\end{tabular}
\end{center}
\vskip-12pt
\end{table}

	CMS relies on its central tracking to achieve comparable
resolution for both muons and hadrons, making only a $\sim20\%$
measurement in the external muon system, which utilizes the iron flux
return of the central solenoid for its magnetic field. The CMS muon
system provides triggering and muon identification rather than a
precise measurement.

	{\it Trigger:} The trigger systems are crucial for the LHC
experiments. The $40\,{\rm MHz}$ interaction rate must be reduced by
about a factor of $10^6$ before the events can be saved to tape for
future analysis, and this obviously limits the physics that can be
studied. The trigger is divided into three levels. Level~1 is
hardware-based, synchronous with the beam clock, and deadtimeless. The
data from every detector element must be saved for about $2\,{\rm \mu
s}$ until a Level~1 decision can be made. This decision is based on
fast sums of predefined clusters in the electromagnetic and hadronic
calorimeter and on hits in roads corresponding to stiff tracks in the
muon system.  The thresholds on these are adjusted to reduce the rate
to about $10^4\,{\rm Hz}$. Level~2 refines the selection made at
Level~1 by using the full granularity of the detector and by combining
measurements from more than one subsystem in ``regions of interest''
found by Level~1 trigger. This allows one, e.g., to determine the
$p_T$ of an electron candidate more accurately both by using more
detailed calorimeter information and by comparing it with tracking
information. Finally, at Level~3 the whole detector is read out into a
computer farm, which can run off-line analysis code and save events at
roughly $100\,{\rm Hz}$. A list of possible triggers for ATLAS and
their Level-1 and Level-2 rates is shown in Table~\ref{tbl:trig}.

\subsection{Measuring the Standard Model Quanta\label{sec:detperf}}

	Any SUSY or other new particle will be produced at the LHC
with $p_T \simle M$, so its decay products will be widely distributed
in phase space. It will either decay into quanta of the Standard Model
--- quark or gluon jets, charged leptons, neutrinos, or photons --- or
it will be stable and escape the detector like the lightest SUSY
particle $\lsp$ if $R$ parity is conserved. ATLAS and CMS are designed
to detect such signals, in many cases with redundant measurements.

	{\it Jets:} Jets, the dominant signal at large $p_T$, are
measured as clusters of energy in the electromagnetic and hadronic
calorimeters. There will also be multiple charged tracks connecting
this cluster and the vertex. The jet energy resolution tends to be
dominated more by uncertainties associated with jet clustering and QCD
radiation than by detector performance.

	{\it $b$ Jets:} The most important use of the vertex detector
is to tag $b$ jets. Most of the tracks in an event will point back to
the primary vertex, but those from a $B$ hadron will have a distance
of closest approach characteristic of the $B$ lifetime, i.e., $c\tau
\approx 465\mu{\rm m}$. Tagging is not easier for a highly
relativistic $B$ hadron: while the typical distance traveled by the
$B$ is $\gamma c\tau$, the typical opening angles of the decay tracks
are $\calO(1/\gamma)$, so the typical distance of closest approach to
the primary vertex is independent of $\gamma$ provided $\beta \approx
1$. Calculating the tagging efficiency requires detailed simulation
and obviously involves a tradeoff between efficiency and background
rejection. For a 1\% mistagging rate for light quark jets, the typical
tagging efficiency is 60\%.\cite{ATLAS}

	{\it Photons:} An isolated photon is identified as a cluster
contained in the electromagnetic calorimeter with a radius of order the
radiation length $X_0$ with no hadronic energy behind it and no
high-$p_T$ charged track near it. (Actually, the trackers in ATLAS and
CMS both contain a significant fraction of $1X_0$, so the probability
that the photon converts to an $e^+e^-$ pair in the tracker is not
small.) Isolation criteria are sufficient to reject jets by a factor
of several thousand but still leave some background from those jets in
which one or more $\pi^0$'s carry most of the jet energy. An additional
rejection can be obtained by using a preradiator to count the number
of photons and/or by using detailed shower shape cuts. Thus, because
of the very good ATLAS and CMS electromagnetic calorimeters, a jet
rejection of $\sim10^4$ can be achieved with a photon efficiency of
order 90\%. Hence, the background for $h \to \gamma\gamma$ should be
dominated by the real QCD $\gamma\gamma$ continuum.  Non-isolated
photons of course cannot be separated from the much larger rate for
$\pi^0 \to \gamma\gamma$ in jets.

	{\it Electrons:} An electron gives an electromagnetic shower
like a photon but has a track with $p \approx E$ pointing to it.
Because of this extra constraint, isolated electrons with $p_T \simge
10\,\GeV$ and $|\eta|< 2.5$ can be identified with an efficiency of
order 90\% with a jet rejection of $\sim 10^5$, so that the background
for isolated electrons is dominated in almost all cases by real
electrons. Electrons within jets are much more difficult to identify.

	{\it Taus:} A $\tau$ decaying into a lepton is difficult to
distinguish from a prompt lepton. A $\tau$ decaying into hadrons can
be identified as a narrow hadronic jet with one charged track or three
tracks with a charge $\pm1$ and a mass $<M_\tau$. The background from
QCD jets is significant, and detailed study is required to develop cuts
appropriate for any particular case.

	{\it Muons:} A muon is identified by a charged track in the
central tracker and a matching charged track in the external muon
system. The energy deposition in the calorimeter is small ($\sim
2\,\GeV$) for energies below about $1\,\TeV$. At higher energies
bremsstrahlung and $e^+e^-$ pair production become significant, and
the energy deposited in the calorimeter needs to be considered.

	{\it Missing Energy:} In hadron colliders the total missing
energy is completely dominated by the loss of low-$p_T$ particles in
the beam pipe, so only the missing transverse energy $\etmiss$ can be
used to detect neutrinos or $\lsp$'s. This is measured by summing all
the calorimeters plus any observed muons. The resolution is dominated
by non-Gaussian tails and cracks in the calorimeter; detailed studies
indicate that real neutrinos dominate over the instrumental background
at least for $\etmiss > 100\,\GeV$.

\section{Inclusive SUSY Measurements at LHC\label{sec:reach}}%

	If SUSY is indeed the right new physics at the electroweak
scale, the first task of the LHC will be to detect a deviation from
Standard Model predictions characteristic of SUSY. The ability to do
so is clearly model dependent. For example, if all SUSY particles were
nearly degenerate in mass, then they would decay into very soft jets
or leptons plus an invisible $\lsp$, and nothing would be observable.
Fortunately, such a degenerate spectrum does not occur in any
reasonable model.

\subsection{Simulation of SUSY Signatures\label{sec:reach1}}

	Most recent studies of SUSY signatures at the LHC have assumed
the minimal supergravity (SUGRA) model.\cite{SUGRA} The SUGRA model is
a special case of the Minimal Supersymmetric Standard Model (MSSM),
with two Higgs doublets and a SUSY partner for each Standard Model
one, grand unification at some scale $M_{\rm GUT}$, and soft SUSY
breaking terms added by hand assuming $R$-parity conservation:
\begin{eqnarray*}
-{\cal L}_{\rm soft} &=&A^u h^u \widetilde Q H_u \widetilde u^c
                        + A^d h^d \widetilde Q H_d \widetilde d^c
                        + A^\ell h^\ell \widetilde L H_d \widetilde\ell^c \\
                        &&+B\mu(H_dH_u + h.c.) + M_{H_d}^2 |H_d|^2 \\
                        &&+ M_{H_u}^2 |H_u|^2
                                +M_{\widetilde L}^2 |\widetilde L|^2
                                +M_{\widetilde e}^2 |\widetilde e^c|^2 \\
                        &&+M_{\widetilde Q}^2 |\widetilde Q|^2
                                +M_{\widetilde u}^2 |\widetilde u^c|^2
                                +M_{\widetilde d}^2 |\widetilde d^c|^2 \\
                        &&+\half M_1 \overline{\widetilde B}\widetilde B
                                +\half M_2 \overline{\widetilde W}
                                        \widetilde W
                                +\half M_3 \overline{\widetilde g} \,
                                        \widetilde g\,.
\end{eqnarray*}
In SUGRA the soft breaking terms are assumed to be communicated from
the SUSY breaking sector by gravity and so to be universal at $M_{\rm
GUT}$. The resulting minimal set of parameters is:
\begin{itemize}
\item	$m_0$: the common SUSY-breaking mass of all squarks, sleptons,
and Higgs bosons.
\item	$\mhalf$: the common SUSY-breaking mass of all gauginos.
\item	$A_0$: the common SUSY-breaking trilinear coupling.
\item	$B\mu$: the SUSY-breaking bilinear coupling.
\item	$\mu$: the SUSY-conserving Higgsino mass.
\end{itemize}
All of these parameters, including the SUSY-conserving parameter
$\mu$, should be of order the weak scale. A limitation of the SUGRA
model is the absence of any understanding of why $\mu$ should be of
order the weak scale or why $R$ parity should be conserved. For a more
detailed discussion, see the lectures by Dawson\cite{Sally97} in these
Proceedings and references therein. 

\begin{figure}[t]
\dofig{2.5in}{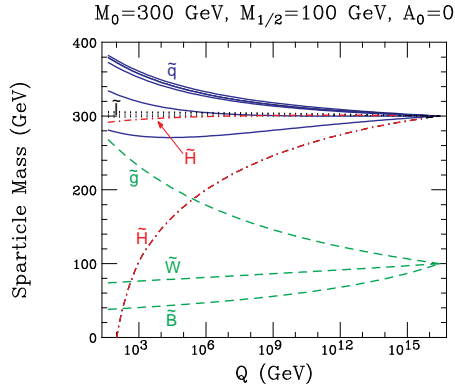}
\pawplot
\caption{Evolution of SUSY masses from the GUT to the electroweak
scale.\protect\cite{Bagger}\label{masses}}
\end{figure}

	The SUGRA model defines the SUSY breaking parameters at the
GUT scale. All of these parameters are essentially couplings and so
obey renormalization group equations (RGE's). The RGE's in the SUGRA
model involve 26 coupled partial differential equations, which have
been studied by various authors;\cite{Ramond94} an example is shown in
Figure~\ref{masses}. ISAJET implements a self-consistent solution of
these RGE's between the weak and the GUT scale. The first step is to
run a truncated set of six equations from $M_Z$ to the GUT scale where
the gauge couplings $g_1$ and $g_2$ meet using approximate SUSY mass
scale:
\begin{eqnarray*}
{dg_1\over dt} &= -{1\over16\pi^2} (-\frac35 - N_f) g_1^3 + 
\hbox{2-loop terms}\\
{dg_2\over dt} &= -{1\over16\pi^2} (5 - N_f) g_2^3 + 
\hbox{2-loop terms}\\
{dg_3\over dt} &= -{1\over16\pi^2} (9 - N_f) g_3^3 + 
\hbox{2-loop terms}\\
\dots\\
{dy_t\over dt} &= {1\over16\pi^2} y_t[6y_t^2 - \frac{13}{15}g_1^2
- 3g_2^2 - \frac{16}{3}g_3^2]\,.
\end{eqnarray*}
Once $M_{\rm GUT}$ is determined, the universal SUGRA boundary
conditions are imposed, and the full set of 26 RGE's are run back to
the weak scale using Runge-Kutta step-by-step integration so that mass
thresholds can be properly taken into account. The Clebsch-Gordon
coefficients in these equations are such that the Higgs mass is driven
negative, breaking electroweak symmetry but not charge or color. The
Higgs effective potential is determined, and the GUT scale parameters
$B$ and $\mu^2$ are determined in terms of the weak scale parameters
$M_Z$ and $\tan\beta=v_2/v_1$. The whole procedure is then iterated
until a self-consistent solution is obtained. The final result is to
express the masses of all 32 SUSY particles plus all the mixing
parameters in terms of just four parameters plus $\sgn\mu=\pm1$:
\begin{itemize}
\item $m_0$: common scalar mass at $M_{\rm GUT}$.
\item $\mhalf$: common gaugino mass at $M_{\rm GUT}$.
\item $A_0$: common trilinear coupling at $M_{\rm GUT}$.
\item $\tan\beta = v_2/v_1$: Ratio of VEV's at $M_Z$.
\item $\sgn\mu = \pm1$.
\end{itemize}

	In the SUGRA model the SUSY masses are mainly determined by
$m_0$ and $\mhalf$, while $\tan\beta$ and $\sgn\mu=\pm1$ mainly affect
the Higgs sector. $A_0$ is not very important for weak-scale physics;
while $A_t$, $A_b$, and $A_\tau$ are important for third-generation
sparticles, they turn out to be only weakly dependent on $A_0$ over
most of the parameter range. Hence it seems to be sufficient to scan
the $m_0$-$\mhalf$ plane for a few values of $\tan\beta$ and
$\sgn\mu=\pm1$ and one value of $A_0$, say $A_0 = 0$. 

	Since the same $m_0$ is used for all scalar particles, it must
be that $m_0^2>0$ so that charge and color are not broken. The SUGRA
model is only possible because the top quark is heavy: it turns out
that the large value of $y_t$ drives the Higgs mass-squared negative,
breaking electroweak symmetry but not color or charge, as illustrated
in Figure~\ref{masses}. SUGRA is surely not the final answer: it sheds
no light on fermion masses, $CP$ violation, etc. But it is a
self-consistent framework representative of a large class of models,
and it might even be close to the truth, so it seems worthy of serious
study. Other models might be easier. In gauge-mediated models, the
lightest SUSY particle is the gravitino; if the next lightest SUSY
particle is the $\lsp$, then $\lsp \to \tilde G \gamma$ decays can be
used to tag SUSY events with two hard photons. In $R$-parity
violating models, the $\lsp$ can decay either into three leptons or
into three quarks; decays into both would lead to weak-scale proton
decay. In the first case the leptons give a good signature. In the
second, there presumably are still leptons from the cascade decays and
it is possible to kinematically reconstruct masses.\cite{BCT97}

\begin{figure}[t]
\dofig{\textwidth}{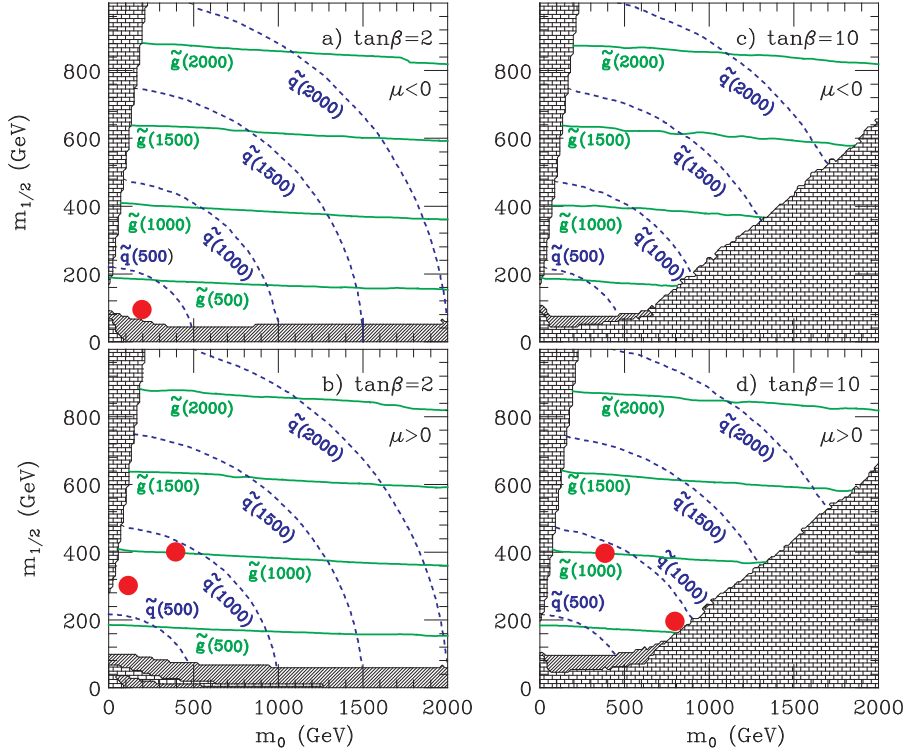}
\caption{Gluino and squark masses\protect\cite{BCPT2} in the minimal
SUGRA model in the $m_0$-$\mhalf$ plane for $\tan\beta=2,10$,
$\sgn\mu=\pm1$, and $A_0=0$. The ``bricked'' regions are excluded
theoretically, while the cross-hatched regions are excluded by
experiment. The dots correspond to the five LHC points described in
the text.\label{fig201}}
\end{figure}

\begin{figure}[t]
\dofig{\textwidth}{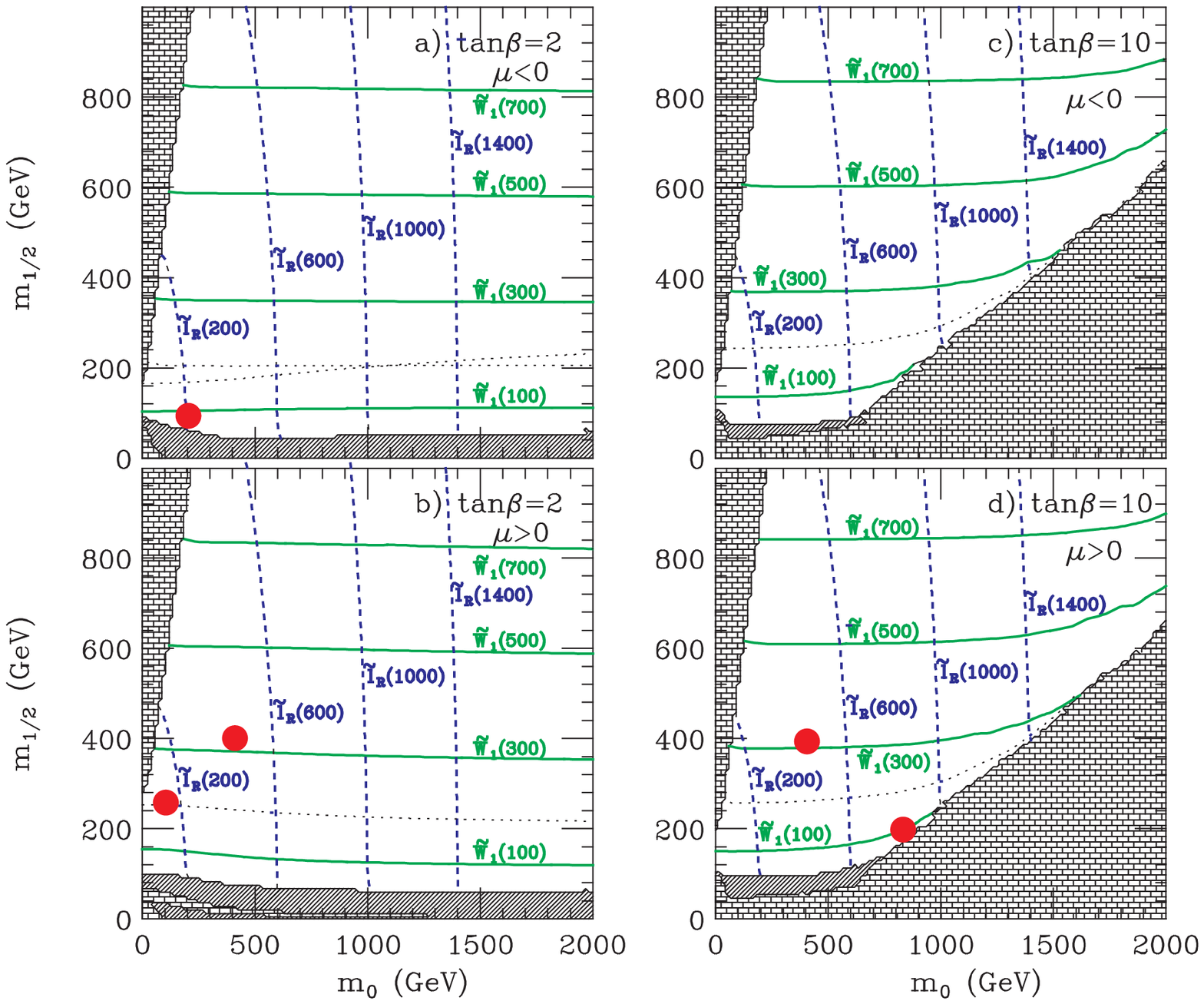}
\caption{Gaugino and slepton masses\protect\cite{BCPT2} in the minimal
SUGRA model in the $m_0$-$\mhalf$ plane for $\tan\beta=2,10$,
$\sgn\mu=\pm1$, and $A_0=0$. The ``bricked'' regions are excluded
theoretically, while the cross-hatched regions are excluded by
experiment. The dots correspond to the five LHC points described in
the text.\label{fig202}}
\end{figure}

	The solution of the renormalization group equations for the
SUGRA model is built into ISAJET.\cite{Isajet} The numerical solution
of these equations uses Runge-Kutta step-by-step integration so that
the thresholds corresponding to the various SUSY masses can be
included in a self-consistent way. First, a truncated set of equations
is used to determine a first estimate of the GUT scale, defined as the
scale at which $\alpha_1$ and $\alpha_2$ meet. The GUT boundary
conditions are then imposed, and the full set of equations is run back
to the weak scale, freezing out mass parameters at their own scales.
The 1-loop Higgs effective potential, including the SUSY masses, is
computed, and the parameters $B$ and $\mu^2$ are eliminated in favor
of $M_Z^2$ and $\tan\beta$. Some optimization of the scale choice is
made; this is equivalent to including some 2-loop contributions. The
equations are then iterated until a self-consistent solution is found.
Once the renormalization equations have been solved, the sfermion and
gaugino mixing matrices are computed, and all the branching ratios for
SUSY particles are computed. In earlier versions mixings in the $\tb$
and $\ttau$ sectors were ignored, limiting the program to
$\tan\beta\simle10$; this restriction has recently been removed. The
branching ratio calculations use the correct matrix elements, but at
present phase space is used in the actual event generation for
technical convenience. Thus, for example, as a squark mass is varied
from just below to just above the gluino mass, the branching ratio
behaves sensibly but the event structure changes discontinuously.

	PYTHIA\cite{Pythia} uses approximate formulas rather than
solving the renormalization group equations, or it can take masses
from an external calculation. It treats the branching ratios in a
similar way.

	Figures~\ref{fig201} and \ref{fig202} show contour plots of
various SUSY masses in the $m_0$--$\mhalf$ plane for $\tan\beta=2,10$,
$\sgn\mu=\pm1$, and $A_0=0$ from ISAJET~7.22.\cite{BCPT2} The
cross-hatched regions are excluded by experiment. The bricked regions
at small $m_0$ are excluded by the requirement that the $\lsp$ rather
than the $\ttau_1$ be the lightest SUSY particle. The bricked regions
small $\mhalf$ and $\tan\beta=10$ were excluded in ISAJET~7.22 by the
absence of electroweak symmetry breaking. It turns out that the size
of this excluded region is very sensitive to the scale at which the
effective potential is minimized; changes in recent versions of ISAJET
intended to make the results more stable for $\tan\beta\gg10$ have
significantly reduced this region.

	Figures~\ref{fig201} and \ref{fig202} illustrate a number of
general features of the SUGRA mass spectrum:
\begin{itemize}
\item	Gluino and gaugino mass depend mainly on $\mhalf$.
\item	Slepton masses depend mainly on $m_0$.
\item	Squark masses depend mainly on $\sqrt{m_0^2 + 4\mhalf^2}$.
\item	$M_\tq \simge 0.9M_\tg$.
\item	$M_\lsp \approx 0.5M_{\tchi_2^0} \approx 0.5M_{\tchi_1^\pm}
\sim 0.5\times0.3 M_\tg$.
\end{itemize}
\noindent The last point is more general than the SUGRA model. It
means that there is a large energy release at each step in the cascade
decays of SUSY particles. If all the SUSY particles were nearly
degenerate, they would be much more difficult to detect.

\subsection{Reach of SUSY Signatures\label{sec:reach2}}

	Recall that a $\tg$ or $\tq$ is produced at the LHC with $p_T
\sim M$ and decays into jets, possible leptons, and a $\lsp$, which is
neutral and weakly interacting and so escapes the detector. Thus the
most generic prediction of ($R$-parity conserving weak scale) SUSY is
an excess of events with multiple jets plus missing energy compared
with the Standard Model sources, i.e., $W$, $Z$, and heavy quark
production and mismeasurement of QCD jet events. The first task is to
determine whether such an excess exists.

	The standard requirement for discovery of a new phenomenon
is a significance of at least $5\sigma$. That is, the probability that
the background fluctuates up to the observed signal should be less
than the tail of a Gaussian distribution beyond $5\sigma$, i.e.,
$5.7\times 10^{-5}$. This may seem overly conservative but is
essential because one always looks at many different distributions with
different cuts, and one of them is likely to have an unlikely
fluctuation. For large numbers of events, the $5\sigma$ requirement is
equivalent to
$$
S/\sqrt B > 5
$$
where $S$ and $B$ are the number of signal and background events
respectively. For small numbers, Poisson probabilities should be used.
Of course the $S/B$ ratio, or more properly the error on determining
what the background should be, must also be considered. Ruling out the
existence of a signal is less demanding, and limits are generally
quoted for 90\% or 95\% confidence.

	The approach\cite{BCPT1,BCPT2} for determining the LHC reach
in the SUGRA parameter space is to scan the $m_0-\mhalf$ plane for
selected values of the other parameters, generating a sample of SUSY
events for each choice of parameters. These consist mainly of $\tg$
and $\tq$ production, but all processes are included. Since one event
typically takes about $0.5\,{\rm s}$ on an HP-735, the feasible data
samples correspond to $\sim10\,\fb^{-1}$ for detailed studies but much
less for such a scan. It is clearly not possible to generate a
representative sample of the Standard Model total cross section.
Instead, high-$p_T$ events which potentially can give large $\etmiss$,
namely
\begin{itemize}
\item	$W+\jets$, $W \to \ell\nu,\tau\nu$
\item	$Z+\jets$, $Z \to \nu\bar\nu, \tau\tau$.
\item	$t \bar t$.
\item	QCD jets, including $g \to b\bar b, c\bar c$ branching and
decay.
\end{itemize}
\noindent These samples are generated using several approximately
equal intervals of $\log p_T$ for the primary hard scattering. The
event generator of course produces not just the hard scattering but
also parton showers, hadronization of the partons into jets, and beam
jets. The studies described here generally have assumed low
luminosity, $\calL=10^{33}\,\cmsec$, and have therefore neglected
pileup from overlapping events.

\begin{figure}[t]
\dofig{5in}{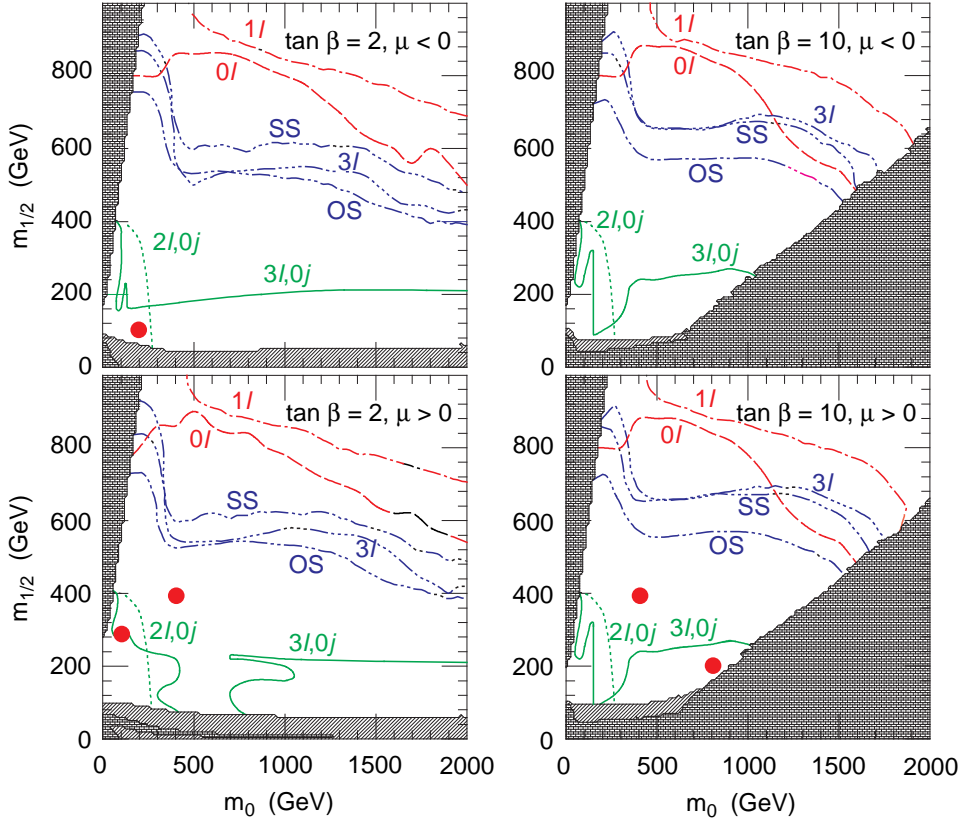}
\caption{SUGRA reach in the $m_0$-$\mhalf$ plane for $\tan\beta=2,10$,
$\sgn\mu=\pm1$, and $A_0=0$ for $10\,\fb^{-1}$ luminosity at the
LHC.\protect\cite{BCPT2} The ``bricked'' regions are excluded
theoretically, while the cross-hatched regions are excluded by
experiment. The dots correspond to the points selected by the
LHCC.\label{figrea}}
\end{figure}

	All events are passed through a toy detector simulation. This
takes into account the overall coverage and Gaussian resolutions but
not cracks, resolution tails, multiple scattering, or many other
effects. It is possible to take all these effects into account, but
the detector simulation then requires more than $1\,{\rm hr}$ per
event. The toy simulation is not adequate to determine the $\etmiss$
background from mismeasured QCD jets. More detailed studies show that
this background is less than that from real Standard Model neutrinos
for $\etmiss\simge100\,\GeV$, and this cut will generally be made
whenever $\etmiss$ is used.

	Jets are found using a simple fixed-cone algorithm. That is,
the highest remaining unused cell of the calorimeter is found, and the
total $E_T$ is summed in a cone in
$$
R = \sqrt{(\Delta\eta)^2 + (\Delta\phi)^2}\,,
$$
which is equivalent to a polar angle at $\eta=0$ but is $z$-boost
invariant and so behaves properly in the forward direction. Generally
$R=0.4$ is used for complex multijet events as a compromise between
identifying nearby jets and containing all of the jet energy. Electrons
and muons are treated equivalently, requiring isolation $E_T <
5\,\GeV$ in cone $R=0.3$ both to suppress the $b,c \to \ell X$
background and to permit $e$ identification by track/shower matching.
The latter requires a more detailed simulation to implement, so
generator information is used for lepton identification. 

	Figure~\ref{figrea} summarizes the reach of the LHC to observe
SUSY in various channels in the $m_0-\mhalf$ plane for two
representative values of $\tan\beta$, $\sgn\mu=\pm1$, and $A_0=0$. The
reach limits are based on a $5\sigma$ signal after $10\,fb^{-1}$,
corresponding to one year at low luminosity.

	$0\ell$, $1\ell$: These curves show the reach in the basic
channel, multiple jets plus missing energy. The $0\ell$ curve includes
a veto on muons and isolated electrons, while the $1\ell$ curve
requires a lepton. The lepton veto improves the $S/B$ ratio for low
masses, but at high masses so many leptons are produced that requiring
a lepton improves the reach. For both sets of curves the following
cuts are made to reject the Standard Model background and to enhance
the acceptance for heavy particles produced with $p_T \sim M$:
\begin{itemize}
\item	$N_\jets \ge 2$ with $E_T>100\,\GeV$, $|\eta|<3$.
\item	Closest jet $j_c$ has $30^\circ < \Delta\phi(\etmiss,j_c) <
90^\circ$.
\item	$S_T>0.2$
\item	$\etmiss > E_T^c$, $E_T(j_1), E_T(j_2) > E_T^c$.
\end{itemize}
Here $E_T^c$ is a cut which is adjusted at each point to optimize
$S/\sqrt{B}$, and the optimum value is used to define the reach. The
variable $S_T$ is the transverse sphericity or circularity, which is
defined as
$$
S_T = {2\lambda_2 \over \lambda_1+\lambda_2}\,,
$$
where $\lambda_1>\lambda_2$ are the eigenvalues of the transverse
sphericity tensor
$$
S_{ij} = \sum_n p_{n,i}p_{n,j}\,,\qquad i,j=1,2\,.
$$
This cut selects ``round'' events characteristic of heavy particle
production, but it is highly correlated with the multijet and other
cuts.

	$SS$: This curve shows the reach in the like-sign dilepton
channel, $\ell^\pm\ell^\pm$, $\ell=e,\mu$. The leptons are required to
have $p_T>20\,\GeV$ and $|\eta|<2.5$, and to satisfy the isolation
criterion $E_T<5\,\GeV$ in a cone $R=0.3$. Since the gluino is a
Majorana fermion, it is its own antiparticle and so has equal
branching ratios into $\ell^+ X$ and $\ell^- X$, e.g., through
$\tilde\chi_1^\pm$ cascade decays. (There may be other SUSY sources of
like-sign dileptons.) The dominant Standard Model isolated dilepton
backgrounds, $t \bar t$, Drell-Yan, and $W^+W^-$, only give
opposite-sign dileptons. There are Standard Model like-sign dilepton
backgrounds, e.g., from $t \bar t$ production with $t \to
\ell^+ X$ and $\bar t \to \bar b \to \ell^+ X$, but these will
normally fail the isolation test. 

	$OS$: SUSY can also give opposite-sign dileptons
$\ell^+\ell^-$, $\ell=e,\mu$.  e.g., from $\tchi_2^0 \to \lsp
\ell^+\ell^-$. The same lepton cuts are made as for the $SS$ curves,
and a $Z$ mass cut is also made for identical flavor leptons.
Opposite-sign dilepton decays are enhanced at low $m_0$, for which the
sleptons are light and the decay can proceed via $\tchi_2^0 \to \tell
\ell \to \lsp \ell^+\ell^-$ with substantial branching ratios. While
the Standard Model backgrounds are larger in this channel than for
$SS$, the statistical reach is comparable; the $OS$ channel also
provides important independent information.

	$3\ell$: SUSY can produce trilepton events from a variety of
sources, including the decay of one gluino or squark via $\tchi_1^\pm
\to \lsp \ell^\pm\nu$ and the other through $\tchi_2^0 \to \lsp
\ell^+\ell^-$. The Standard Model background for three isolated
leptons is fairly small, so the reach in this channel is comparable to
the dilepton channels even though the total branching ratio is
smaller. 

	$3\ell,0j$, $2\ell,0j$: These channels require two or three
leptons with the same cuts as before, $\etmiss$, and no jets with
$p_T>25\,\GeV$ and $|\eta|<3$.  The jet veto is designed to select the
direct production of gaugino or slepton pairs. These channels are the
best way of searching for SUSY at the Tevatron, where the limited
energy suppresses the production of the heavier gluinos and squarks.
The search range is limited by competition from $\tchi_2^0 \to \lsp Z,
\lsp h$ once $\mhalf$ is large enough that these are kinematically
allowed.\cite{BCPT2} Even at smaller $\mhalf$ the branching ratio can
be suppressed by interference between the virtual $Z$ and slepton
exchange graphs, leading to the holes in the reach seen in
Figure~\ref{figrea}. Nevertheless, these channels would provide useful
additional information should they be observed.

	By comparing Figure~\ref{figrea} and the mass contours in
Figure~\ref{fig201}, one can see that the LHC can search the whole
SUGRA parameter space at the $5\sigma$ level for gluino and squark
masses up to about $2\,\TeV$ with only $10\,\fb^{-1}$ of luminosity.
Similar conclusions have been found by the ATLAS\cite{ATLAS} and
CMS\cite{CMS} Collaborations using the more general MSSM model and a
more realistic parameterization of the detectors. In addition, various
multilepton signatures can be observed for gluino and squark masses up
to about $1\,\TeV$, i.e., over the whole range favored by fine-tuning
arguments. The multilepton signatures with a jet veto are limited to
relatively small values of $\mhalf$ or $m_0$. Thus, if SUSY exists at
the weak scale, ATLAS and CMS should observe characteristic deviations
from the Standard Model after one year of operation at only 10\% of
design luminosity. The LHC should either find SUSY or exclude it. Only
experiment will decide whether SUSY at the weak scale is a crucial
element in physics or an interesting exercise in mathematics.

\subsection{Introduction to Precision
Measurements\label{sec:preciseintro}}%

	While observing signatures characteristic of SUSY at the LHC
would be one of the most exciting developments in particle physics of
all time, it is important to be able to determine the masses and other
parameters of SUSY particles and thus to get a handle on the
underlying dynamics. If $R$ parity is conserved, however, then every
SUSY event is missing two $\lsp$'s, and there are not enough kinematic
constraints to determine their momenta. 

	It may be useful to compare the SUSY case with the production
of $t \bar t$ at the Tevatron:
$$
q + \bar q \to t + \bar t \to \ell^+\nu b + \bar q' q'' \bar b\,.
$$
In these events there is one missing $\nu$ and hence three unknown
kinematic variables $\vec p_\nu$. To determine these, there are two
measured components of $\etmiss$ and two additional constraints
expressed as quadratic equations in the components of $\vec p_\nu$:
\begin{eqnarray*}
(p_e+p_\nu)^2		&=& M_W^2\,, \\
(p_e+p_\nu+p_b)^2 	&=& (p_{q'}+p_{q''}+p_{\bar b})^2\,.
\end{eqnarray*}
Thus there is one more constraint than unknown; this is known in
ancient bubble chamber terminology as a 1C fit. Using all the
constraints, one can fully reconstruct the event despite the missing
neutrino. If top were produced singly, then one would have only one
quadratic constraint, a 0C fit; in this case $\vec p_\nu$ could still
be reconstructed, but there would be a 2-fold ambiguity. Of course for
$t \bar t$ production one can also reconstruct the 3-jet mass
directly. 

\begin{table}[t]
\caption{Parameters for the five LHC SUGRA points.\label{Points}}
\medskip
\begin{center}
\begin{tabular}{cccccc}
\hline\hline
Point & $m_0$ & $m_{1/2}$ & $A_0$ & $\tan\beta$ & $\sgn{\mu}$ \\
      & (GeV) & (GeV)   & (GeV)   &             &             \\
\hline
1 & 400 & 400 &   0 & \phantom{0}2.0 & $+$\\
2 & 400 & 400 &   0 & 10.0 & $+$\\
3 & 200 & 100 &   0 & \phantom{0}2.0 & $-$\\
4 & 800 & 200 &   0 & 10.0 & $+$ \\
5 & 100 & 300 & 300 & \phantom{0}2.1 & $+$\\
\hline\hline
\end{tabular}
\end{center}
\end{table}

\begin{table}[t]
\caption{Masses of the superpartners, in GeV, at the five LHC SUGRA
points from ISAJET~7.22. The first and second generation squarks and
sleptons are degenerate.\label{Masses}}
\medskip
\begin{center}
\begin{tabular}{cccccc}
\hline\hline
Point & 1 & 2& 3 & 4& 5 \\
\hline
$\widetilde g$                 & 1004 & 1009  & 298 & 582  &  767\\
$\widetilde \chi_1^\pm$        &  325 &  321  &  96 & 147  &  232\\
$\widetilde \chi_2^\pm$        &  764 &  537  & 272 & 315  &  518\\
$\widetilde \chi_1^0$          &  168 &  168  & 45  & 80   &  122\\
$\widetilde \chi_2^0$          &  326 &  321  & 97  & 148  &  233\\
$\widetilde \chi_3^0$          &  750 &  519  & 257 & 290  &  497\\
$\widetilde \chi_4^0$          &  766 &  538  & 273 & 315  &  521\\
$\widetilde u_L$               &  957 &  963  & 317 & 918  &  687\\
$\widetilde u_R$               &  925 &  933  & 313 & 910  &  664\\
$\widetilde d_L$               &  959 &  966  & 323 & 921  &  690\\
$\widetilde d_R$               &  921 &  930  & 314 & 910  &  662\\
$\widetilde t_1$               &  643 &  710  & 264 & 594  &  489\\
$\widetilde t_2$               &  924 &  933  & 329 & 805  &  717\\
$\widetilde b_1$               &  854 &  871  & 278 & 774  &  633\\
$\widetilde b_2$               &  922 &  930  & 314 & 903  &  663\\
$\widetilde e_L$               &  490 &  491  & 216 & 814  &  239\\
$\widetilde e_R$               &  430 &  431  & 207 & 805  &  157\\
$\widetilde \nu_e$             &  486 &  485  & 207 & 810  &  230\\
$\widetilde \tau_1$            &  430 &  425  & 206 & 797  &  157\\
$\widetilde \tau_2$            &  490 &  491  & 216 & 811  &  239\\
$\widetilde \nu_\tau$          &  486 &  483  & 207 & 806  &  230\\
$h^0$                          &  111 &  125  & 68  & 117  &  104\\
$H^0$                          & 1046 &  737  & 379 & 858  &  638\\
$A^0$                          & 1044 &  737  & 371 & 859  &  634\\
$H^\pm$                        & 1046 &  741  & 378 & 862  &  638\\
\hline\hline
\end{tabular}
\end{center}
\end{table}

	For SUSY, there are two missing $\lsp$'s and so six unknown
momentum components in addition to the $\lsp$ mass. The SUSY signal
contains many different processes; there is no simple constraints like
the $W$ mass in the top case, so there are only two constraints on the
six variables from the two components of $\etmiss$. Hence it is not
possible to reconstruct the events in general. It is possible,
however, to use the kinematic endpoints of various distributions to
make a precise determination of combinations of masses. This is
simplest in a $e^+e^-$ (or $\mu^+\mu^-$) collider, where the SUSY
particles are produced with a known beam energy, giving extra
constraints, but it is also possible at the LHC, as will be seen in
Section~\ref{sec:precise} below.

	While the possibility of making such precision measurements is
quite general, which ones can be made depends on the assumed masses
and branching ratios and so can be determined only by simulating in
detail events for specific choices of the SUSY parameters. The LHC
Committee (LHCC), the CERN committee overseeing the LHC experiments,
selected the five SUGRA points listed in Table~\ref{Points} for
detailed study by the ATLAS and CMS collaborations. Point~3 is the
``comparison point,'' selected so that every existing or proposed
accelerator could discover something. At this point, the Tevatron
would discover winos and zinos, the LHC would discover gluinos and
squarks, the NLC would discover sleptons, and LEP would have recently
announced the discovery of a light Higgs boson with a mass of
$68\,\GeV$.  Points~1 and 2 have gluino and squark masses of about
$1\,\TeV$ and so test the reach of the LHC for such masses. Point~4
has large $m_0$, so that sleptons and squarks are much heavier than
gauginos and gluinos. It was also close to the boundary of the allowed
electroweak symmetry breaking region with ISAJET~7.22, so that $\mu$
was quite small and there was large mixing between the gauginos and
higgsinos. More recent versions of ISAJET find that Point~4 further
from this boundary, so that $\mu$ is larger and the mixing of gauginos
and higgsinos is smaller. Finally, Point~5 was chosen to be in the
center of the region giving the right amount of cold dark matter for
cosmology.\cite{CDM} Heavy stable $\lsp$'s tend to overclose the
universe; getting the right amount of cold dark matter generally
requires enhancing the $\lsp$ annihilation cross section and hence
having relatively light sleptons. 

	The masses of the SUSY particles for these five points as
calculated with ISAJET~7.22 are listed in Table~\ref{Masses}. These
masses and the corresponding branching ratios are used in all the
analyses described below.

\subsection{Effective Mass Analysis\label{sec:effmass}}

	The SUSY reach limits discussed in Section~\ref{sec:reach} are
based on just counting the number of events with some specified set of
cuts. Because QCD corrections to hadronic cross sections are large,
the signal expected in the Standard Model is somewhat uncertain. It is
therefore desirable to measure for some variable a distribution
which agrees with the Standard Model in some range and then deviates
from it, thus giving a more convincing signal and also providing an
estimate of the SUSY mass scale.

\begin{figure}[t]
\dofig{2.5in}{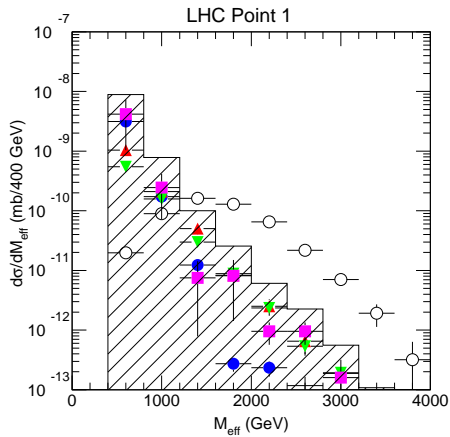}
\pawplot
\caption{LHC Point~1 signal and Standard Model
backgrounds.\protect\cite{HPSSY} Open circles: SUSY signal.  Solid
circles: $t\bar t$. Triangles:  $W\to\ell\nu$, $\tau\nu$.  Downward
triangles:  $Z\to\nu\bar\nu$, $\tau\tau$.  Squares: QCD jets.
Histogram: sum of all backgrounds.\label{lhc1-147}}
\end{figure}

\begin{figure}[t]
\dofig{2.5in}{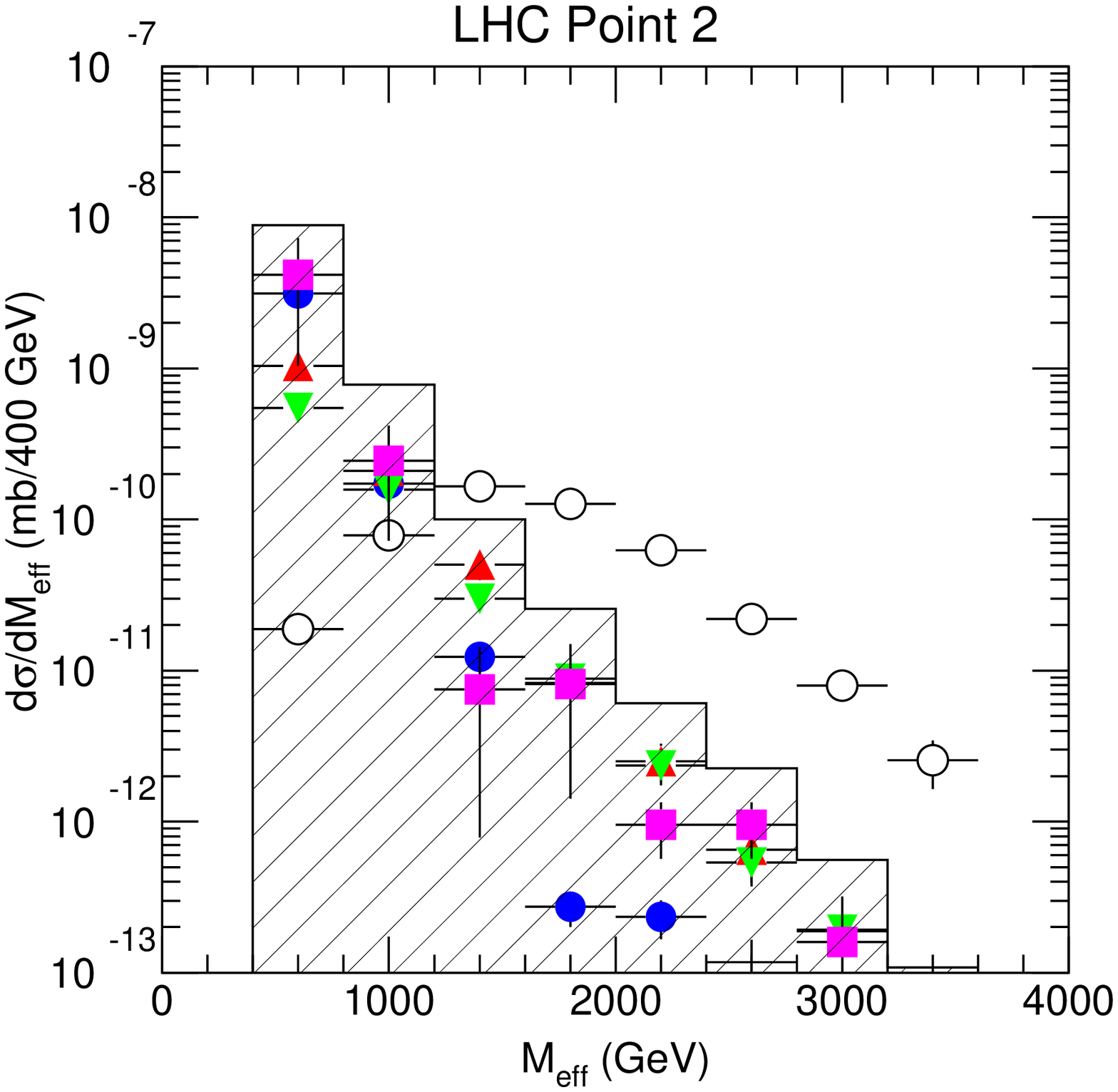}
\pawplot
\caption{LHC Point~2 signal and Standard Model
backgrounds.\protect\cite{HPSSY} See Figure~\protect\ref{lhc1-147} for
definitions of the symbols.\label{lhc2-147}}
\end{figure}

\begin{figure}[t]
\dofig{2.5in}{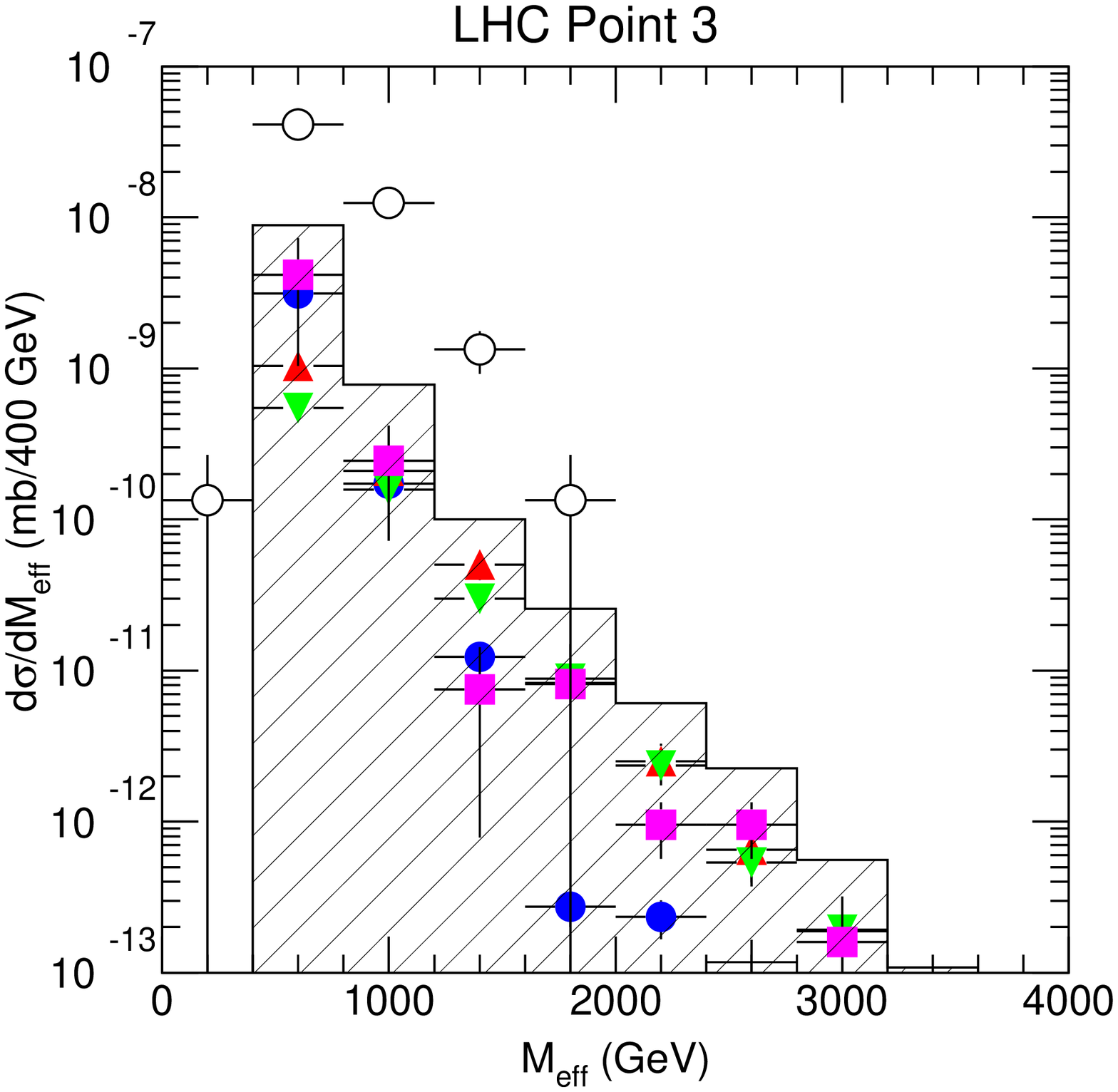}
\pawplot
\caption{LHC Point~3 signal and Standard Model
backgrounds.\protect\cite{HPSSY} See Figure~\protect\ref{lhc1-147} for
definitions of the symbols.\label{lhc3-147}}
\end{figure}

\begin{figure}[t]
\dofig{2.5in}{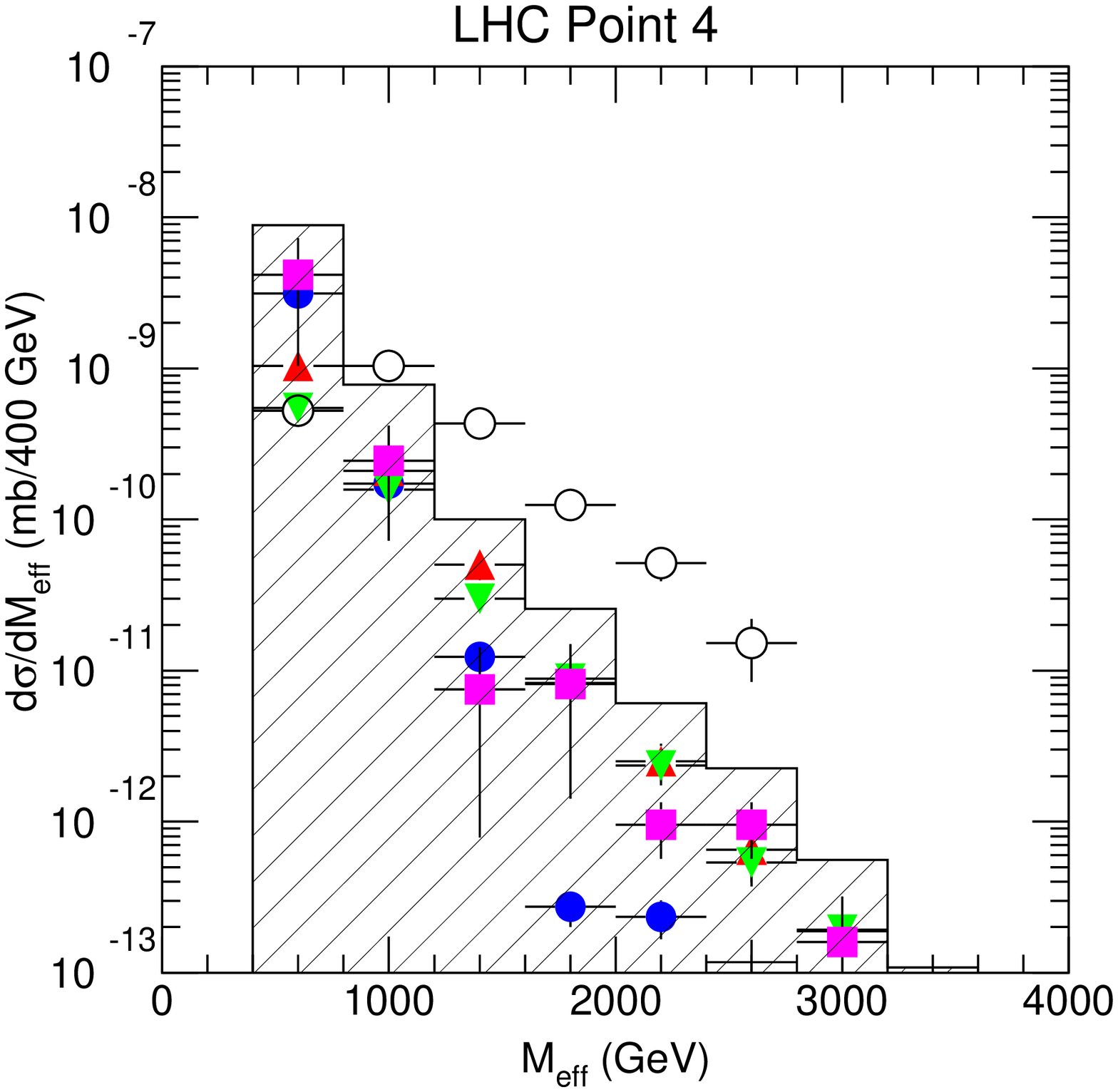}
\pawplot
\caption{LHC Point~4 signal and Standard Model
backgrounds.\protect\cite{HPSSY} See Figure~\protect\ref{lhc1-147} for
definitions of the symbols.\label{lhc4-147}}
\end{figure}

\begin{figure}[t]
\dofig{2.5in}{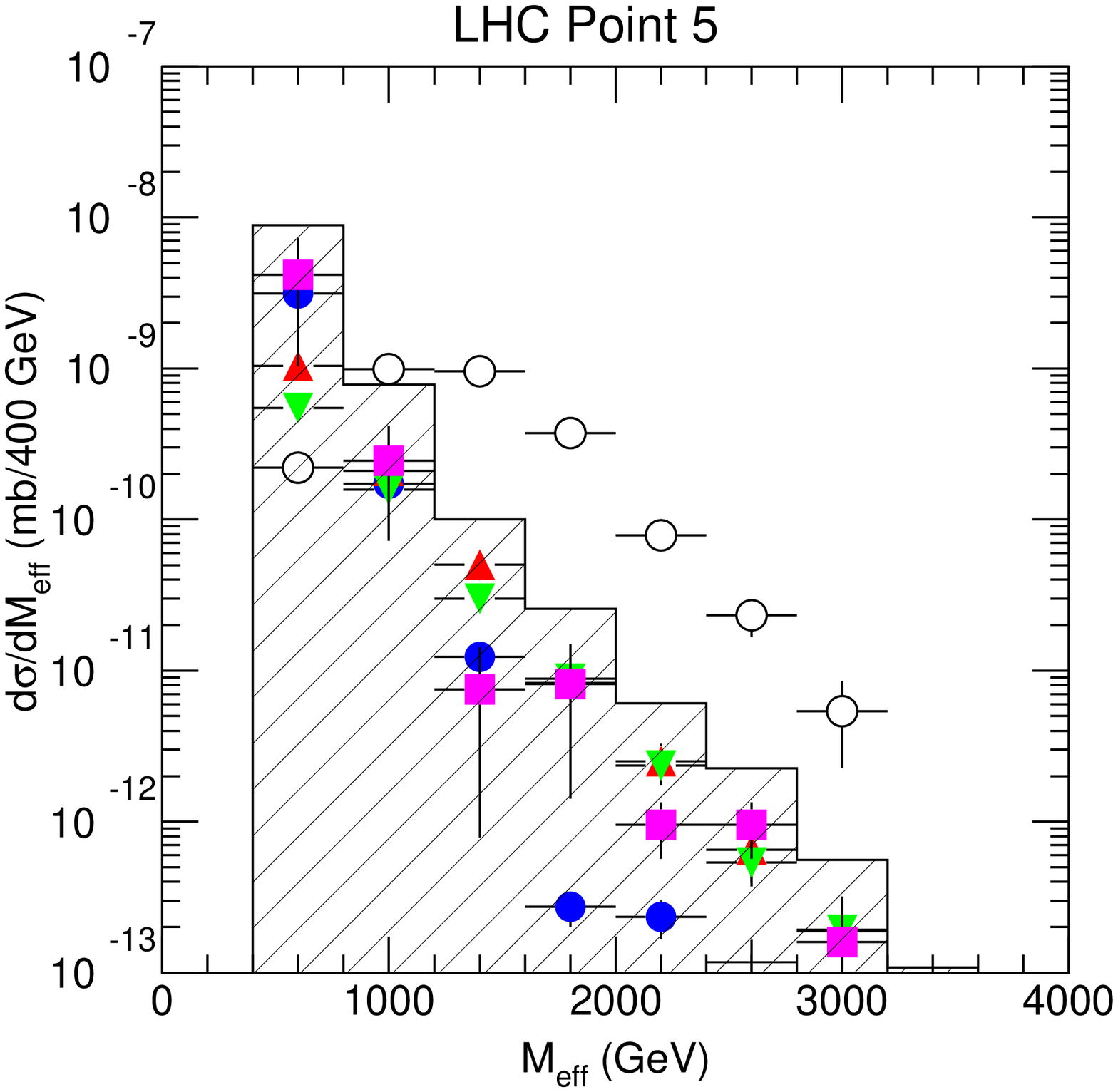}
\pawplot
\caption{LHC Point~5 signal and Standard Model
backgrounds.\protect\cite{HPSSY} See Figure~\protect\ref{lhc1-147} for
definitions of the symbols.\label{lhc5-147}}
\end{figure}

	What properties should such a variable have? At least in the
SUGRA model, the squarks are never much lighter than the gluino, so
gluino production, which is enhanced by color and spin factors, is
always important. If the squarks are heavier than the gluino, then the
dominant gluino decays will be $\tg \to \tchi q \bar q$, giving a
minimum of four jets plus missing energy. If the gluino is heavier,
then the decay chain $\tg \to \tq \bar q$, $\tq \to \tchi q$ will
dominate. In either case there will be at least four jets plus missing
energy, more if one or more of the gauginos in the process decay
hadronically. QCD cross sections fall rapidly with momentum transfer
--- the jet cross section at $\sqrt{s}=14\,\TeV$ falls over the
relevant range of $p_T$ roughly like
$$
{d\sigma\over dp_T^2} \sim p_T^{-6}
$$
--- so it is clearly important to compare signal and background at
comparable $p_T$ scales. The invariant mass of the produced system is
not the best measure of this because it is too much influenced by
possible soft jets at large rapidity. The scalar sum of the $p_T$'s of
the four hardest jets and the $\etmiss$ works well and will be called
the ``effective mass''\cite{HPSSY}
$$
\Meff = p_{T,1} + p_{T,2} + p_{T,3} + p_{T,4} + \etmiss\,.
$$
Backgrounds from QCD processes with multiple jets and neutrinos from
heavy quarks generally have missing energy small compared to the $Q^2$
scale of the event. To avoid such backgrounds the $\etmiss$ cut is
made proportional to $\Meff$,
$$
\etmiss > 0.2\Meff\,,
$$
where the coefficient $0.2$ was chosen after studying the SUSY and
Standard Model Monte Carlo distributions.

\begin{table}[t]
\caption{The value of $\Meff$ for which $S = B$ compared to $\Msusy$,
the lighter of the gluino and squark ($\tilde{u_R}$) masses. Note that
Point~3 is strongly influenced by the $\etmiss$ and jet $p_T$
cuts.\label{tbl:meff}}
\medskip
\begin{center}
\begin{tabular}{cccc}
\hline\hline
LHC Point& $\Meff\,(\GeV)$& $M_{\rm SUSY}\,(\GeV)$& Ratio\\
\hline
1 &      1360           &   926 &    1.47 \\
2 &      1420           &   928 &    1.53 \\
3 &      \phantom{0}470 &   300 &    1.58 \\
4 &      \phantom{0}980 &   586 &    1.67 \\
5 &      \phantom{0}980 &   663 &    1.48 \\
\hline\hline
\end{tabular}
\vskip-10pt
\end{center}
\end{table}

	Several additional cuts were made for technical reasons: a
missing energy cut $\etmiss>100\,\GeV$ to ensure that the Standard
Model $\etmiss$ background is dominated by neutrinos rather than
mismeasured jets; a jet cut $p_{T,j}>50\,\GeV$ to ensure that the jets
were well identified and measured, and a cut $p_{T,1}>100\,\GeV$ on
the hardest jet to limit the range of QCD background that had to be
generated. These cuts require $\Meff>350\,\GeV$ and so limit the
sensitivity to very light SUSY particles. In addition, there is a cut
on transverse sphericity $S_T>0.2$ to select ``round'' events
characteristic of SUSY, although this is highly correlated with the
previous cuts. Finally, there is a veto on muons or isolated electrons
with $p_T>15\,\GeV$; this minimizes the background for SUSY masses
comparable to the top mass but reduces the sensitivity for high
masses. 

	After all these cuts, the Standard Model background dominates
the $\Meff$ distribution for low $\Meff$, but the SUSY signal
dominates by a factor of 5--10 for large $\Meff$ for all of the LHC
points except the comparison point, Point 3, as can be seen from
Figures~\ref{lhc1-147}--\ref{lhc5-147}. For Point~3,
Figure~\ref{lhc3-147}, the SUSY signal is larger than the Standard
Model background for all values of $\Meff$ allowed by the technical
cuts described above. Observing such a change in shape from a curve
dominated by Standard Model physics to one a factor of 5--10 larger
would be convincing evidence for new physics.

	The signal curves in Figures~\ref{lhc1-147}--\ref{lhc5-147}
clearly shift with the SUSY masses. Since the SUSY cross section is
dominated by $\tg$ and $\tq$ production, it is natural to use
$$
\Msusy \equiv \min(M_{\tg}, M_{\tilde u_R})
$$ 
as a measure of the mass scale. Table~\ref{tbl:meff} shows the points
at which the signal and background points cross
Figures~\ref{lhc1-147}--\ref{lhc5-147}. Clearly these points scale
quite well with $\Msusy$.

\begin{figure}[t]
\dofig{2.5in}{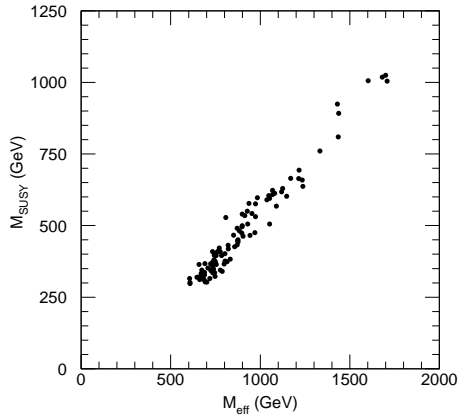}
\pawplot
\caption{Scatterplot of $\Msusy = \min(M_{\tg}, M_{\tilde u})$
vs.\ $\Meff$ for randomly chosen SUGRA models having the same light
Higgs mass within $\pm3\,\GeV$ as LHC Point~5.\protect\cite{HPSSY}
\label{newscan1}}
\end{figure}

\begin{figure}[t]
\dofig{2.5in}{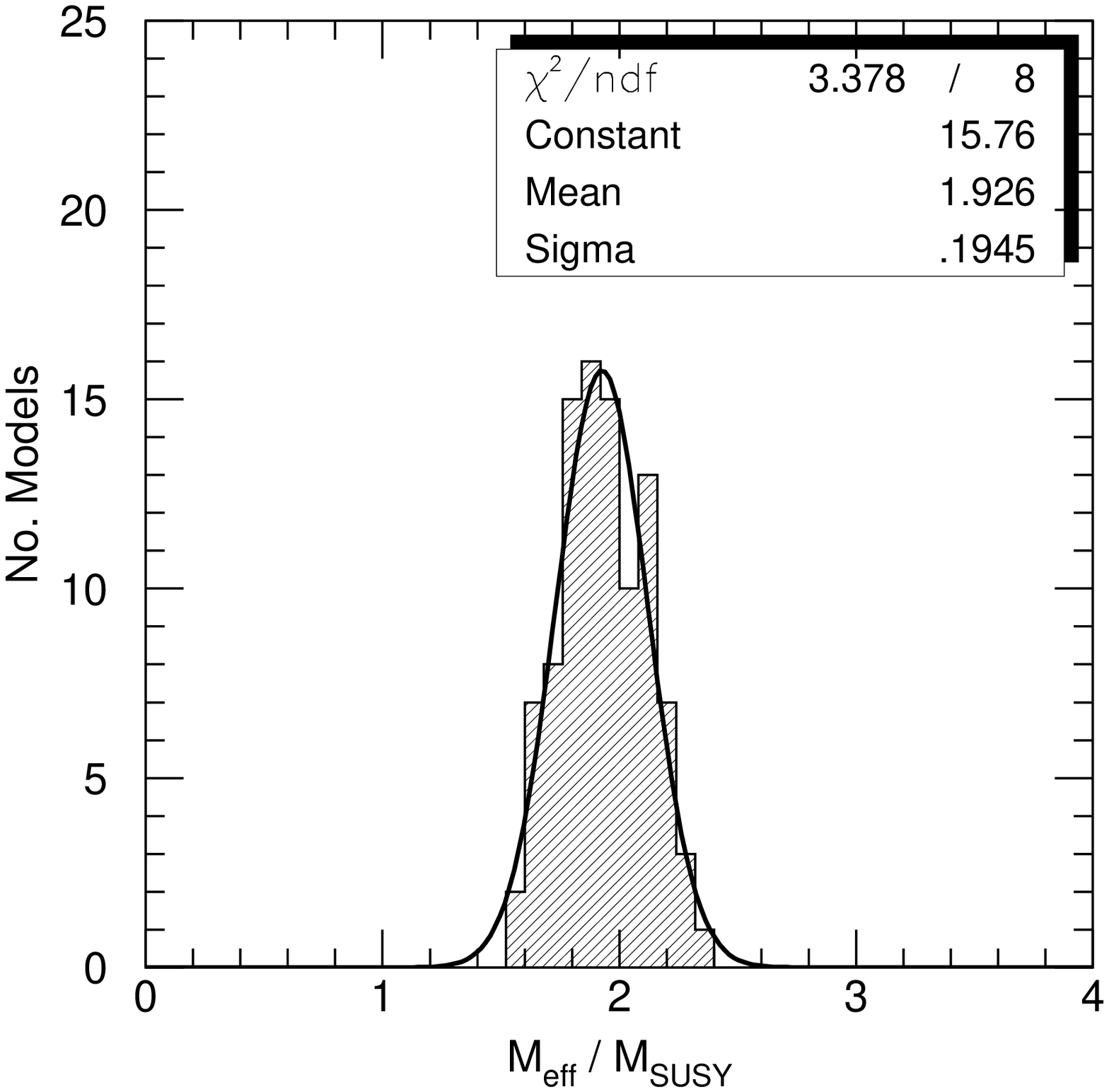}
\pawplot
\caption{Ratio $\Meff/\Msusy$ from
Figure~\protect\ref{newscan1}.\protect\cite{HPSSY}\label{newscan3}}
\end{figure}

	To see whether the scaling in Table~\ref{tbl:meff} might be
accidental, a comparison of $\Meff$ was made\cite{HPSSY} between 100
random SUGRA models and Point~5. The models were generated with
parameters uniformly distributed in the intervals $100 < m_0 <
500\,\GeV$, $100 < m_{1/2} < 500\,\GeV$, $-500 < A_0 < 500\,\GeV$,
$1.8 < \tan\beta < 12$, and $\sgn\mu=\pm1$. All 100 models were
selected to have $M_h$ within an assumed theoretical uncertainty of
$\pm3\,\GeV$ from the $M_h$ mass for Point~5. The value of $\Meff$ for
each model was determined not by the intersection with the Standard
Model background but by the peak of the signal, which is somewhat
higher. (It is not at all obvious that this is the optimal procedure,
but it is what has been done.) The scatter plot of the peak $\Meff$
vs.\ $\Msusy$ for each model is shown in Figure~\ref{newscan1}, and
the projection is shown in Figure~\ref{newscan3}. Evidently the
scaling of $\Meff$ vs.\ $\Msusy$ works remarkably well for this random
selection as well as for the five LHC points. While the scaling is
physically plausible, it is not known how well it works for arbitrary
SUSY models.

\section{Precision Measurements with Exclusive Final
States\label{sec:precise}}%

	While $\Meff$ seems to work quite well as a measure of the
SUSY mass scale, it clearly averages over many final states and
branching ratios, so it can only be a rough approximation. To do
better, one needs to reconstruct specific final states. If $R$ parity
is conserved, then every SUSY event is missing two $\lsp$'s, so no
masses can be reconstructed directly. It is possible, however, to
determine precisely combinations of masses by finding endpoints of
kinematic distributions in specific final states, starting at the
bottom of the decay chains for the SUSY particles and working
up.\cite{HPSSY,Snow} For simple SUSY models such a SUGRA with only a
few parameters, this approach can determine the model parameters with
good accuracy, at least in favorable cases. Even for more complicated
models it is a good starting point. This Section describes a number of
such precision measurements\cite{HPSSY,Snow} for the five LHC SUGRA
points.

\subsection{Measurement of $M(\tchi_2^0)-M(\lsp)$}

	First consider Point~3. This point has unusual branching
ratios because $M_\tb < M_\tg$ but $M_\tq > M_\tb$, so that $\tg \to
\tb \bar b$ is very much enhanced: 
\begin{eqnarray*}
B(\tg \to \tb_1 \bar b + {\rm h.c.}) &= 89\%\,, \\
B(\tb_1 \to \tchi_2^0 b) &= 86\%\,, \\
B(\tchi_2^0 \to \lsp \ell^+\ell^-) &= 33\%\,.
\end{eqnarray*}
While these branching ratios are not typical, it is common for heavy
flavors in SUSY decays to be comparable to light ones or even
enhanced.

	The SUSY particles at this point are relatively light and so
give $\etmiss$ in the range for which detector effects are not
negligible. Hence, $\etmiss$ is not used in the event selection at
this point. Instead, events are selected by requiring
\begin{itemize}
\item $\ell^+\ell^-$ pair with $p_{T,\ell}>10\,\GeV$, $\eta<2.5$.
\item $\ge 2$ jets tagged as $b$ quarks with $p_T>15\,\GeV$ and
$\eta<2$.
\item No $\etmiss$ cut.
\end{itemize}
making use of the large $\tchi_2^0 \to \lsp \ell^+\ell^-$ branching
ratio. The details of the selection are certainly specific to this
point, but it should be possible in general to use leptonic modes to
observe SUSY particles in this low mass range.

\begin{figure}[t]
\dofig{2.5in}{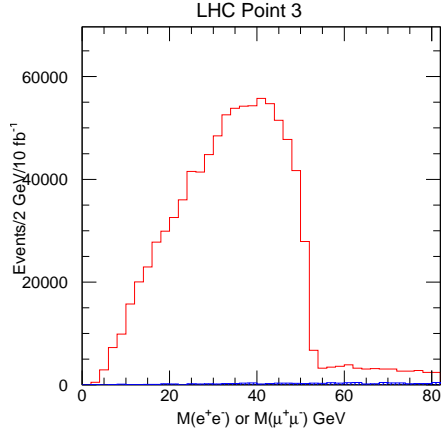}
\pawplot
\caption{$\ell^+\ell^-$ mass for Point~3 (solid) and Standard Model
background (shaded, nearly invisible).\protect\cite{HPSSY}
\label{c3-edge}}
\end{figure}

	SUSY signal and Standard Model background events were
simulated as described above, and events were selected with the
criteria just listed. Then the $\ell^+\ell^-$ mass distribution was
plotted, including a 60\% tagging efficiency for $b$'s and a 90\%
efficiency for electrons and muons. This mass distribution shows a
spectacular edge at the $M(\tchi_2^0)-M(\lsp)$ endpoint,
Figure~\ref{c3-edge}. This distribution reflects the strong signal
production, the large branching ratios, and the distinctive signature,
resulting in almost no Standard Model background. The signal has huge
statistics, and measuring the position of the edge is clearly easier
than measuring the $W$ mass at the Tevatron, since only the lepton
resolution and not the global $\etmiss$ resolution enters. Since the
latter has already achieved an error of about $40\,\MeV$, it seems
conservative to estimate that one could measure the position of this
edge and determine $M(\tchi_2^0)-M(\lsp)$ to $\sim 50\,\MeV$.

\begin{figure}[t]
\dofig{2.5in}{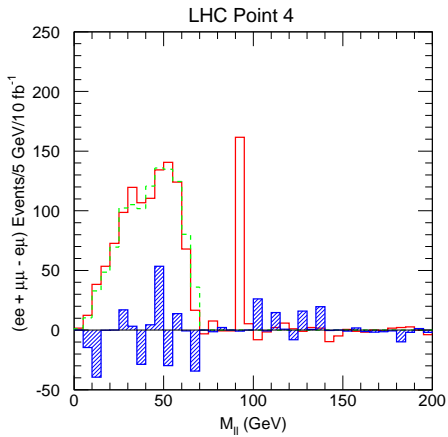}
\pawplot
\caption{$M_{\ell^+\ell^-}$ distribution at Point~4 for opposite-sign,
same-flavor dileptons (solid), opposite-sign, opposite-flavor
dileptons (dashed) and Standard Model background
(shaded).\protect\cite{HPSSY}\label{c4-mllosdif}}
\end{figure}

	The huge statistics from the low masses and the extra handle
of the $b$ tag make this measurement unusually easy.  But there is a
similar edge at Point~4 plus a $Z$ peak coming from decays of the
heavier gauginos. The dominant background is from two independent
$\tchi_1^\pm$ decays in SUSY events. This and the Standard Model
backgrounds, e.g., from top decays, contribute equally to all
combinations of leptons and so vanishes up to statistical fluctuations
in the combination $e^+e^- + \mu^+\mu^- - e^+\mu^- - e^-\mu^+$.
Figure~\ref{c4-mllosdif} shows a plot of this difference for Point~4.
The fluctuations in the background reflect the limited Monte Carlo
statistics. Given the number of signal events, one could measure
$\Delta\left(M(\tchi_2^0)-M(\tchi_1^0)\right) = \pm1\,\GeV$ with
$10\,\fb^{-1}$ at this point.

	This method should work for any point for which the direct
$\tchi_2^0 \to \lsp \ell^+\ell^-$ branching ratio is not too small.
This is generally the case for $M(\tchi_2^0) \simle 200\,\GeV$ so that
$\tchi_2^0 \not\to \lsp Z$, $\lsp h$. But there is a region of $m_0$
for $\sgn\mu=+1$ where the interference of the $Z$ and slepton
exchange graphs makes the branching ratio small. This is the same
effect that produces the holes in the $3\ell,0j$ reach curves in
Figure~\ref{figrea}.

\subsection{Reconstruction of $\tg$ and $\tb_1$}%

\begin{figure}[t]
\dofig{2.5in}{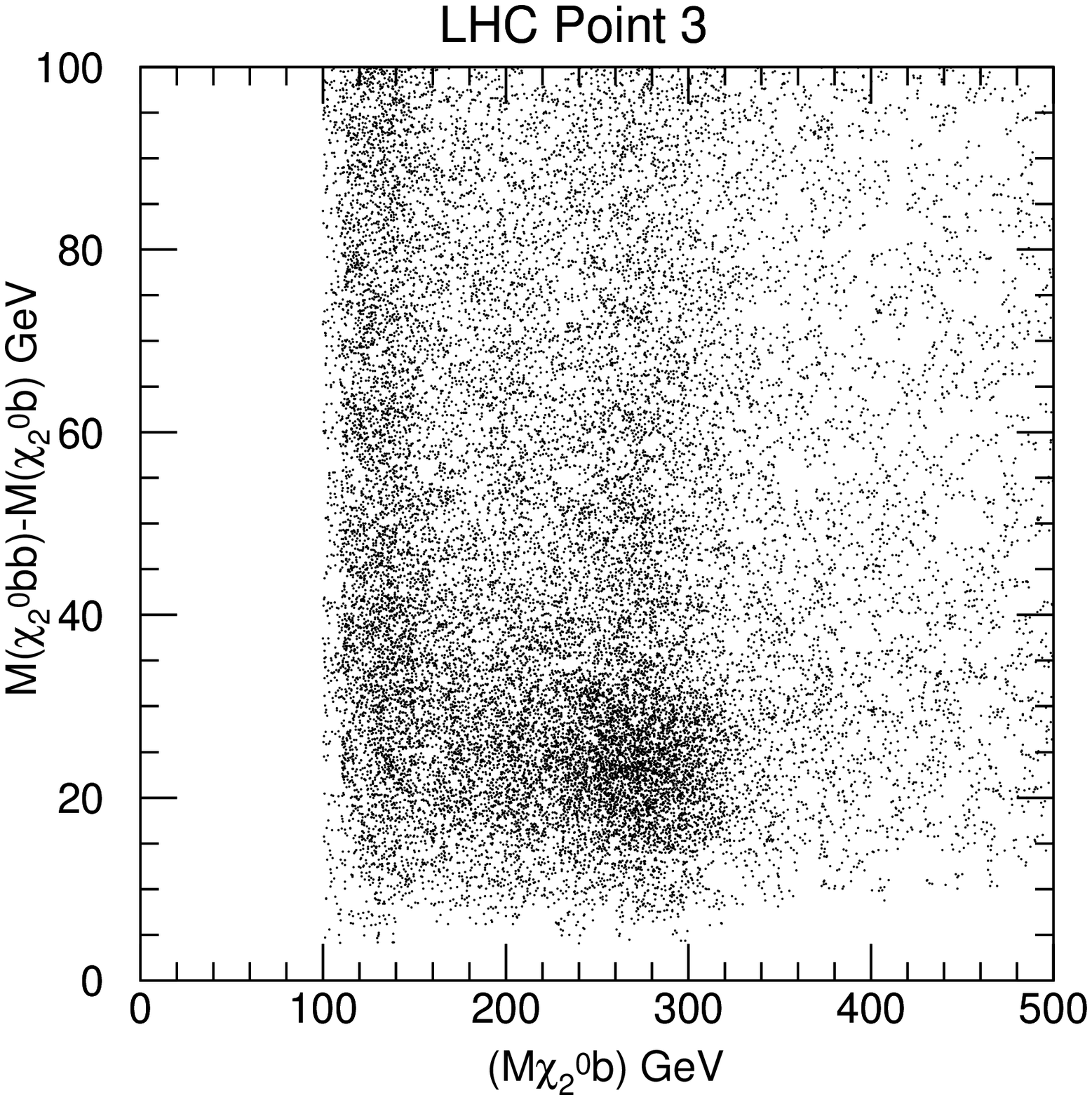}
\pawplot
\caption{Scatter plot of $M(b+\tchi_2)$ vs.\
$M(bb\tchi_2)-M(b\tchi_2)$ assuming $M_\lsp$ known.\protect\cite{HPSSY}
\label{c3-scatter}}
\end{figure}

\begin{figure}[t]
\dofigs{2.5in}{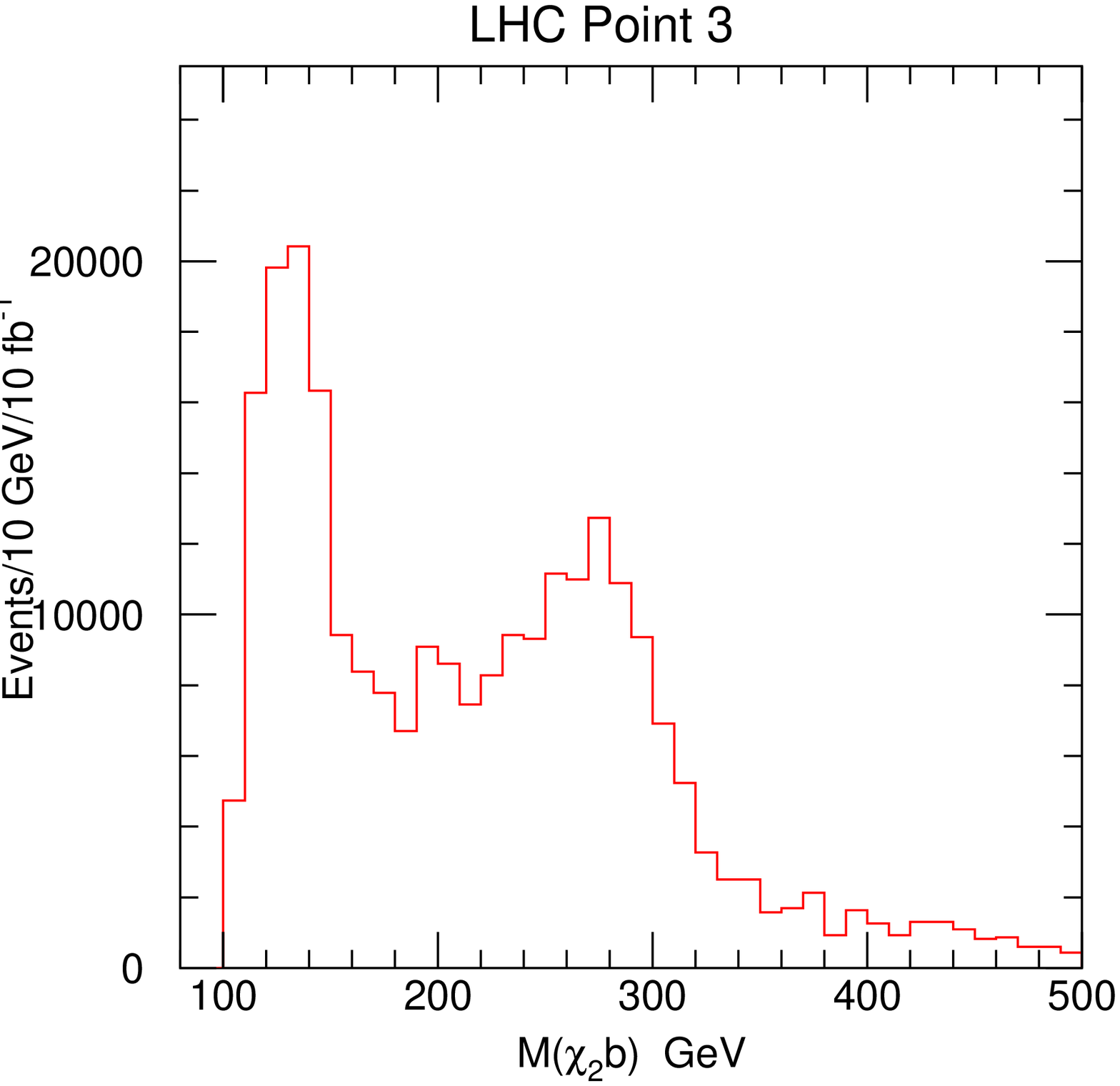}{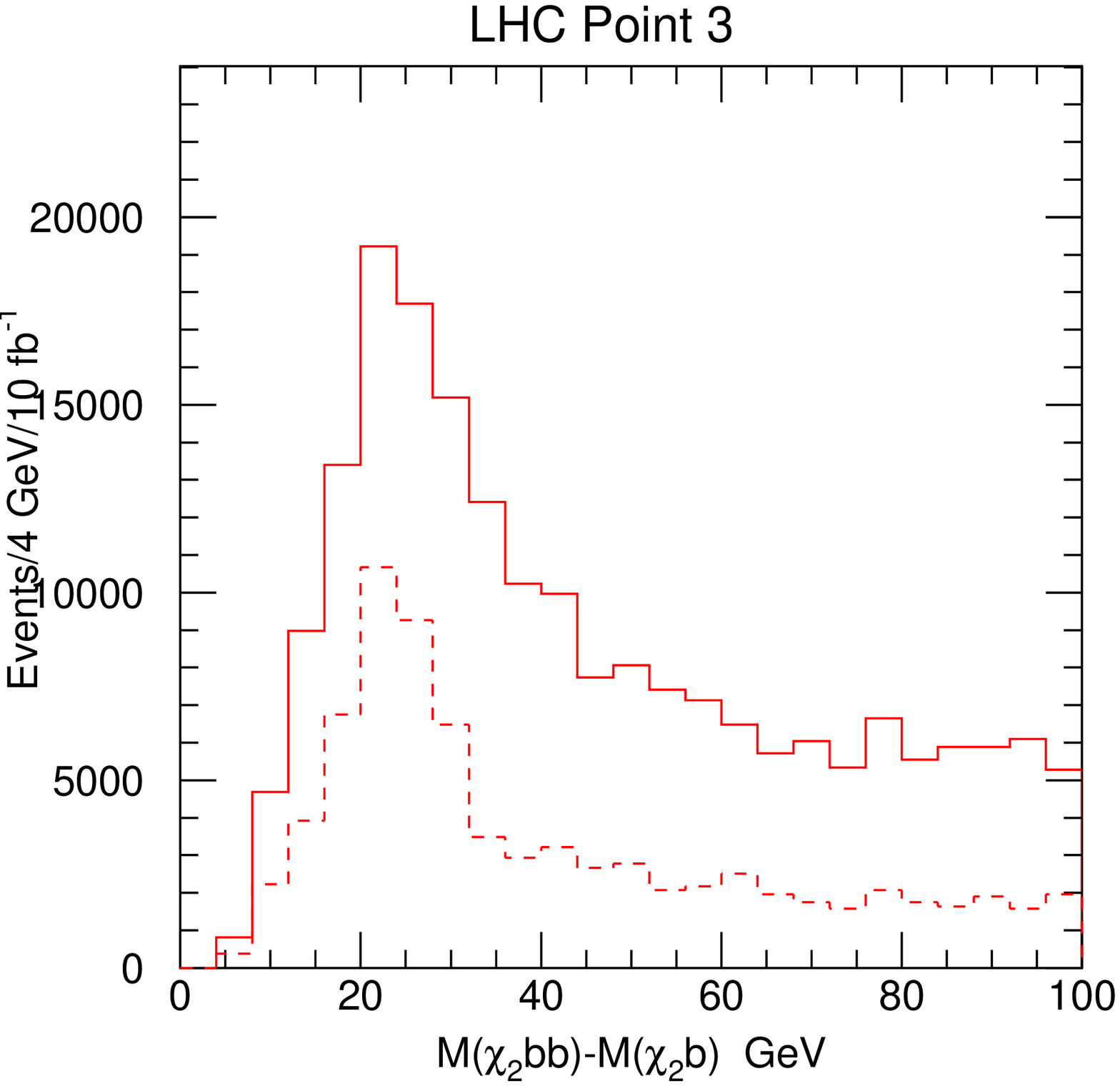}
\pawplot
\caption{Projections of
Figure~\protect\ref{c3-scatter}.\protect\cite{HPSSY} The dashed curve
shows the mass difference for events in the $\tb$ mass
peak.\label{c3-proj}}
\end{figure}

	The $\tchi_2^0 \to \lsp \ell^+\ell^-$ decays at Point~3 can be
combined with $b$ jets to reconstruct the $\tg \to \tb_1 \bar b$,
$\tb_1 \to \tchi_2^0 b$ decay chain. The trick is to select an
$\ell^+\ell^-$ pair near endpoint of the distribution. Then the $\lsp$
must be soft in the $\ell^+\ell^-$ rest frame, so that
$$
\vec p(\tchi_2^0) \approx \left(1 + {M(\lsp) \over M(\ell\ell)}\right) 
\vec p(\ell\ell)\,,
$$
where $M_\lsp$ must be determined from the model. (A first
approximation is $M(\tchi_2^0) = 2M(\lsp)$, so that the $\ell^+\ell^-$
endpoint directly gives $M(\lsp)$.)  The following cuts are made for
this analysis:
\begin{itemize}
\item $\ge 2$ jets tagged as $b$ jets with $p_T>15\,\GeV$, $\eta<2$;
\item $\ell^+\ell^-$ pair with $45<M(\ell\ell)<55\,\GeV$.
\end{itemize}
Again, no use is made of $\etmiss$. Then the inferred $\tchi_2^0$
momentum is combined with one $b$ jet to make $M(\tb_1)$, and a second
$b$ jet is added to make $M(\tg)$. A scatter plot including all the
possible combinations for each event is shown in
Figure~\ref{c3-scatter}. The projections of this scatter plot are
shown in Figure~\ref{c3-proj}.

	Figure~\ref{c3-scatter} was made assuming the correct $\lsp$
mass, which is not directly measurable but must be determined from a
global fit as described in Section~\ref{sec:fits} below. If $M(\lsp)$
is varied about its nominal value, then the value of $M(\tb_1)$
extracted from Figure~\ref{c3-proj} shifts linearly, while the value
of $M(\tg)-M(\tb_1)$ remains constant. Thus the errors on these masses
are estimated to be
\begin{eqnarray*}
&\Delta M(\tb_1) = \pm 1.5\Delta M(\lsp) \pm 3\,\GeV\,,&\\
&\Delta(M(\tg) - M(\tb_1)) = \pm 2\,\GeV\,.&
\end{eqnarray*}
The fact that the difference $\Delta(M(\tg) - M(\tb_1))$ is
insensitive to the assumed mass is a simple consequence the kinematics
and the fact that one $b$ jet is soft. Exactly the same effect is
familiar from $D^* \to D \pi$ decays. In addition to the uncertainty
from the $\lsp$ mass, we must worry about the calibration of the
hadronic energy scale. This can be studied using Standard Model
events. For example, one could use $Z \to e^+e^-$ events to calibrate
the electromagnetic calorimeter and then use $p_T$ balance in
$\gamma+\jets$ and $Z+\jets$ events to transfer this calibration to
the hadronic calorimeter. This is an important issue for the detector
collaborations but need not concern us here.

\subsection{Reconstruction of $h\to b \bar b$}%

\begin{figure}[t]
\dofig{2.5in}{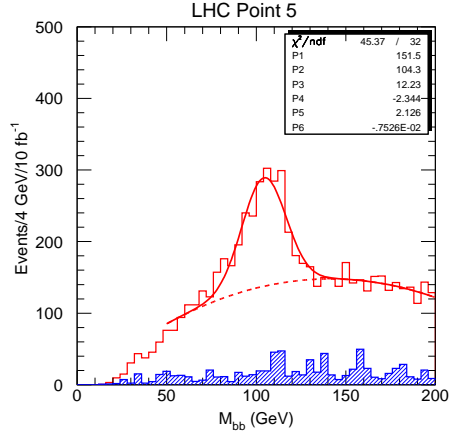}
\pawplot
\caption{$M_{bb}$ distribution for Point~5 (solid), a Gaussian fit to
the peak, and the Standard Model background (shaded).\protect\cite{HPSSY}
\label{c5-mbb}}
\end{figure}

	For Point~5, the decay $\tchi_2^0 \to \lsp h$ has a large
branching ratio. A set of cuts similar to those made in the $\Meff$
analysis is used to select the SUSY events and reject the Standard
Model background:
\begin{itemize}
\item $\ge4$ jets with $p_T>50\,\GeV$, $p_{T,1}>100\,\GeV$;
\item Transverse sphericity $S_T>0.2$;
\item $\Meff>800\,\GeV$;
\item $\etmiss > \max(100\,\GeV, 0.2\Meff)$.
\end{itemize}
After these basic cuts, at least two jets are required to be tagged as
$b$'s with the vertex detector and to have $p_{T,b}>25\,\GeV$ and
$\eta_b<2$. The distribution of masses of all pairs of $b$ jets is
shown in Figure~\ref{c5-mbb}. The width of the peak at the Higgs mass
is set by clustering effects and resolution. The dominant background is
other $b$ jets from SUSY events, not Standard Model background. This
peak is much easier to detect than $h \to \gamma\gamma$ and would be
the discovery mode of the Higgs boson at this point. 

\begin{figure}[t]
\dofigs{2.5in}{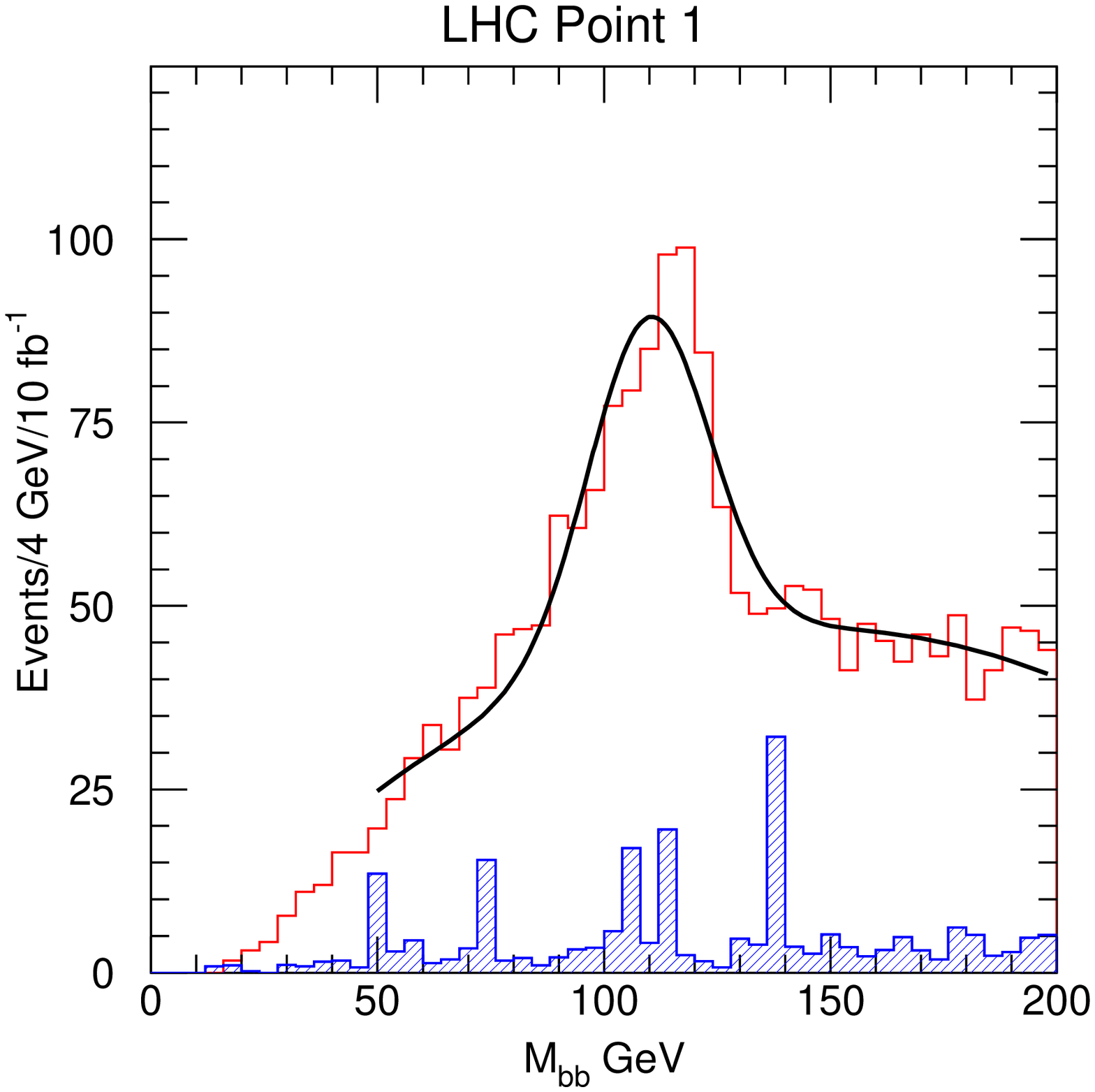}{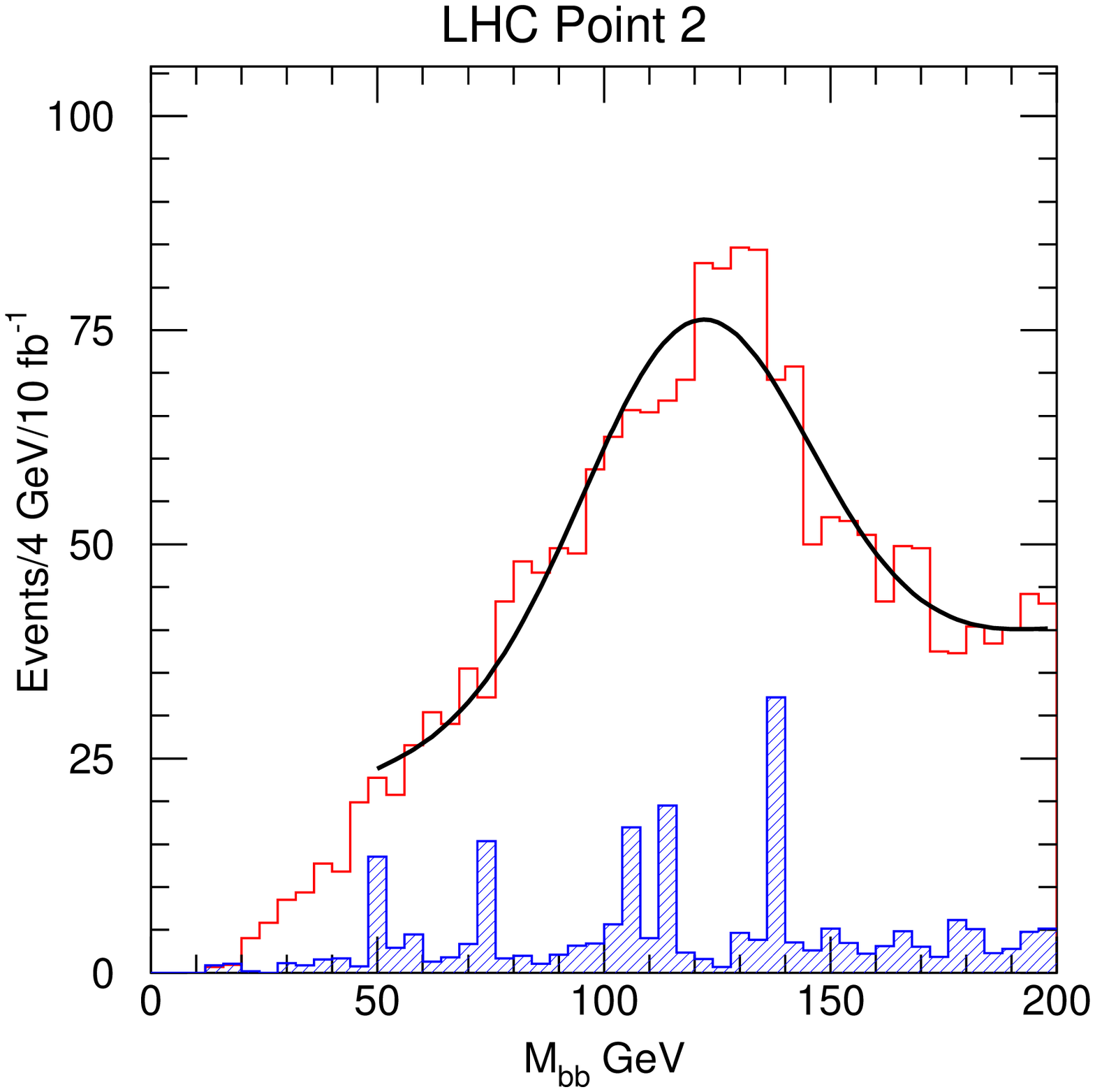}
\pawplot
\caption{$M_{bb}$ distributions for Points~1 and 2.\protect\cite{HPSSY}
See Figure~\protect\ref{c5-mbb}.\label{c12-mbb}}
\end{figure}

	With the clustering algorithm described in
Section~\ref{sec:reach2} the Higgs peak actually came out several GeV
below the Higgs mass. This is due partly to missing neutrinos and
partly to leakage of energy out of the jet cone. A correction,
$$
E_b^{\rm true} = 1.08\left(E_b^{\rm observed} + 2.9\,\GeV\right)\,,
$$
was derived from the Monte Carlo data and applied to make
Figure~\ref{c5-mbb}. In the real experiment, one would need to use the
precision mass from $h \to \gamma\gamma$ to determine this correction.

	The branching ratio for $\tchi_2^0 \to \lsp h$ is generally
substantial if it is kinematically allowed, i.e., for $\mhalf \simge
250\,\GeV$. There are similar Higgs mass peaks for the two other LHC
points in this range; see Figure~\ref{c12-mbb}. One can hope to
observe either $\tchi_2^0 \to \lsp h$ or $\tchi_2^0 \to \lsp
\ell^+\ell^-$ over most of the parameter space.

	It is also possible to reconstruct $W \to q \bar q$ at
Point~5. Since there is no $b$ tag to select the relevant jets, the
Standard Model backgrounds are large, and very hard cuts are
needed.\cite{HPSSY} 

\subsection{Reconstruction of $\tg + \tg \to \tq_L q + \tq_R q$}

\begin{figure}[t]
\dofig{2.5in}{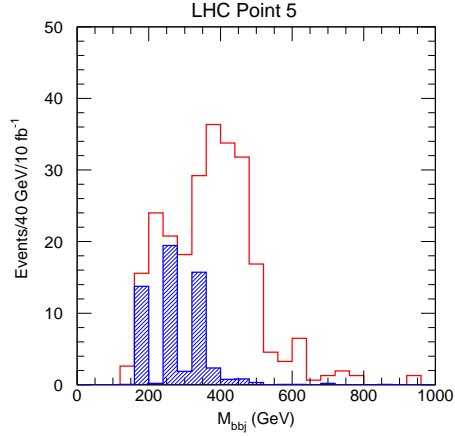}
\pawplot
\caption{Smaller of two $b \bar b j$ masses at Point~5 (solid) and
Standard Model background (shaded).\protect\cite{HPSSY} The background
statistics are limited by the Monte Carlo event sample.\label{c5-mbbj}}
\end{figure}

	Just as $\tchi_2^0 \to \lsp \ell^+\ell^-$ was used as the
starting point to reconstruct partially $\tb_1$ and $\tg$ at Point~3,
one can use $h \to b \bar b$ as the starting point for more complex
analyses. For example, at Point~5, the gluino is somewhat heavier than
the squarks but has a larger cross section because of color and spin
factors, producing as one possible signal
\begin{eqnarray*}
&\tg + \tg \to \tq_L q + \tq_R q\,,& \\
&\tq_L \to \tchi_2^0 q \to \lsp h q, \quad \tq_R \to \lsp q\,.&
\end{eqnarray*} 
The quark jets from the $\tq$ decays are hard, while the other jets
are softer. These events can be selected by requiring in addition to
the two $b$ jets two and only two additional jets with
$p_T>75\,\GeV$.  Then one of the two $q b \bar b$ combinations should
come from a $\tq$ decay, so the smaller of the two $q b \bar b$ masses
should have an endpoint at the kinematic limit for $\tq \to \tchi_2^0
q \to \lsp h q$.  This limit is readily found from two-body kinematics
to be
\begin{eqnarray*}
(M_{hq}^{\rm max})^2 &=& M_h^2
+ \left(M_{\tq}^2 - M_{\tchi_2^0}^2\right) \\
&&\times\left[{M_{\tchi_2^0}^2 + M_h^2 - M_{\lsp}^2 +
\sqrt{(M_{\tchi_2^0}^2 - M_h^2 - M_{\lsp}^2)^2 -4M_h^2
M_{\lsp}^2}} \over 2 M_{\tchi_2^0}^2\right]\,.
\end{eqnarray*}
The average of the $u_L$ and $d_L$ masses gives $M_{hq}^{\rm max} =
506\,\GeV$, which is consistent with the edge seen in
Figure~\ref{c5-mbbj}. The estimated error on this combination of
masses is about one bin or $40\,\GeV$ for $10\,\fb^{-1}$. It is
statistics limited and so should improve with more luminosity.

\subsection{$\ell^+\ell^-$ Distribution at Point~5\label{ll5}}%

\begin{figure}[t]
\dofig{2.5in}{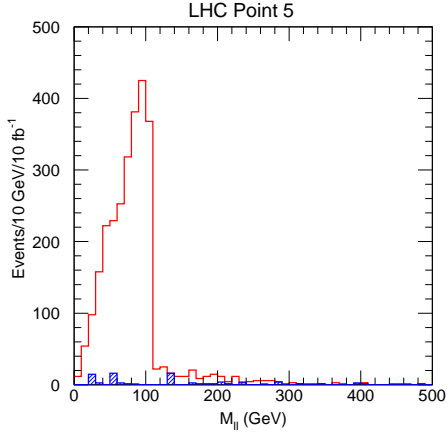}
\pawplot
\caption{$M(\ell^+\ell^-)$ distribution for Point~5 (solid) and
Standard Model background (shaded).\protect\cite{HPSSY}\label{c5-mll}}
\end{figure}

	The $\ell^+\ell^-$ mass distribution has already been used to
measure $M(\tchi_2^0) - M(\lsp)$ at Points~3 and 4. Consider now this
mass distribution for Point~5, one of the points for which a
$\tchi_2^0 \to \lsp h$ signal has been observed. The event selection
combines the same basic cuts as before plus the requirement of a
lepton pair:
\begin{itemize}
\item $\Meff > 800\,\GeV$;
\item $\etmiss > 0.2\Meff$;
\item $\ge 1$ jet with $p_T > 100\,\GeV$;
\item Isolated $\ell^+\ell^-$ pair with $p_T>10\,\GeV$, $\eta<2$;
\item Transverse sphericity $S_T > 0.2$.
\end{itemize}
\noindent The opposite-sign, same-flavor dilepton distribution after
these cuts is shown in Figure~\ref{c5-mll} and obviously has a
dramatic edge with very little background.

	It might at first seem that one should interpret this edge as
a measure of $M(\tchi_2^0)-M(\lsp)$ from the kinematic limit of
$\tchi_2^0 \to \lsp \ell^+\ell^-$. But this interpretation is very
implausible given the observation of $h \to b \bar b$ peak at this
point, since this would require a 3-body decay to compete with a
2-body one, whereas 3-body phase space is much smaller than 2-body
phase space. A much more plausible explanation is that another
two-body decay, $\tchi_2^0 \to \tell\bar\ell$, gives rise to the
dilepton endpoint. In SUGRA one generally has $M(\tell_R) <
M(\tell_L)$, and in fact the observed endpoint in this case comes from
$$
\tchi_2^0 \to \tell^\pm \ell^\mp \to \lsp \ell^\pm \ell^\mp\,.
$$
Simple two-body kinematics for the $\tchi_2^0 \to \tell_R\bar\ell$ and
$\tell_R \to \lsp \ell$ decays gives for the $M(\ell^+\ell^-)$ endpoint
$$
M_{\rm max}(\ell\ell) = M(\tchi_2^0)
\sqrt{1-{M_{\tilde\ell}^2 \over M_{\tilde\chi_2^0}^2}}
\sqrt{1-{M_{\lsp}^2 \over M_{\tilde\ell}^2}}\,.
$$
The estimated error on this combination of masses is about $1\,\GeV$
for $10\,\fb^{-1}$. It is statistics limited and so should improve
with more luminosity.

	This example illustrates that the interpretation of edges of
kinematic distributions at the LHC may be ambiguous. If a signal for
$\tchi_2^0 \to \lsp h$, $h \to b \bar b$, had not already been
observed at this point, it would not be clear whether the edge
measured $M(\tchi_2^0)-M(\lsp)$ from direct decays or the above
combination of three masses from decays through sleptons. Thus it is
important to measure as many combinations of masses and distributions
as possible so as to overconstrain models.

\begin{figure}[t]
\dofig{2.5in}{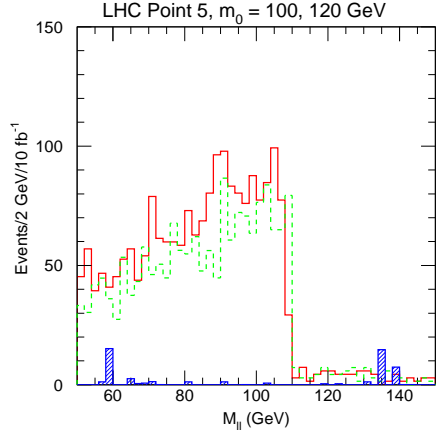}
\pawplot
\caption{Same as Figure~\protect\ref{c5-mll} but for Point~5 (solid)
and $m_0=120\,\GeV$ (dashed), and Standard Model background
(shaded).\protect\cite{HPSSY}\label{c5p20-mll}}
\end{figure}

\begin{figure}[t]
\dofig{2.5in}{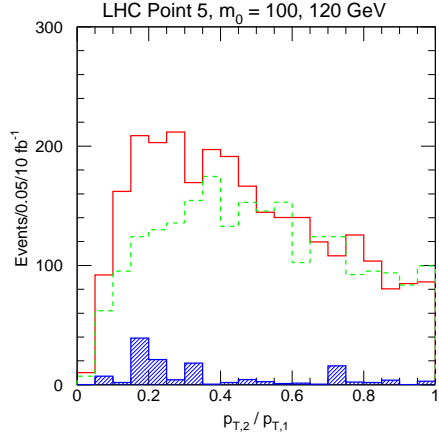}
\pawplot
\caption{Ratio $p_T(\ell_2)/p_T(\ell_1)$ for Point~5 
(solid), $m_0=120\,\GeV$ (dashed), and Standard Model background
(shaded).\protect\cite{HPSSY}\label{c5p20-pt2pt1}}
\end{figure}

	To get the maximum information from the dilepton distribution
at this point, one should fit at least the $M(\ell\ell)$,
$p_T(\ell\ell)$, $p_T(h)$, and $p_T(\ell_2)/p_T(\ell_1)$ distributions
varying $M(\tchi_2^0)$, $M(\tilde\ell_R)$, $M(\lsp)$, and the $p_T$
distribution of $\tchi_2^0$. This is certainly what one would do if the
distribution in Figure~\ref{c5-mll} were real experimental data, but
it requires more effort than seems warranted now. Instead, one
additional sample of events was generated with $m_0 = 120\,\GeV$
rather than $100\,\GeV$; this change increases $M(\ell_R)$ by
$13\,\GeV$ while having small effects on all the other relevant
masses. As a result there is a $2\,\GeV$ change in the location of the
edge, as seen in Figure~\ref{c5p20-mll}. The changes in the $p_T$
distributions are small. The most significant change is in the
distribution of the variable $p_T(\ell_2)/p_T(\ell_1)$, which is shown
in Figure~\ref{c5p20-pt2pt1}.  This makes physical sense:  increasing
the slepton mass reduces the phase space and hence the $p_T$ for the
first lepton, while increasing the phase space and the $p_T$ for the
second lepton from the slepton. It is clear that a change of this
magnitude is easily detected; the estimated error on the slepton mass
from such measurement is $\sim5\,\GeV$ on $m_0$ and $\sim3\,\GeV$ on
the slepton mass. Changing the slepton mass also changes the
$\tchi_2^0 \to \tell\bar\ell$ branching ratio, providing additional
information.

\subsection{Measurement of $M(\tg)-M(\tchi_2^0)$ and
$M(\tg)-M(\tchi_1^\pm)$ at Point~4}%

	Gluino production dominates at Point~4 since $m_0 \gg \mhalf$,
so that the squarks are much heavier than the gluino. An analysis
described previously for this point found a $\tchi_2^0 \to \lsp
\ell^+\ell^-$ edge and hence a measure of $M(\tchi_2^0)-M(\lsp)$. The
jet multiplicity from gluino decay is large, so there is a lot of
combinatorial background, and reconstructing a gluino signal is not
trivial. The strategy for this analysis\cite{Fabiola} is to select
$$
\tg + \tg \to \tchi_2^0 q \bar q + \tchi_1^\pm q \bar q
$$
using leptonic decays of both gauginos to identify them and so to
reduce the combinatorial background. Then the right combination of
jet-jet masses has a common endpoint since $M(\tchi_2^0) \approx
M(\tchi_1^\pm)$.

\begin{figure}[t]
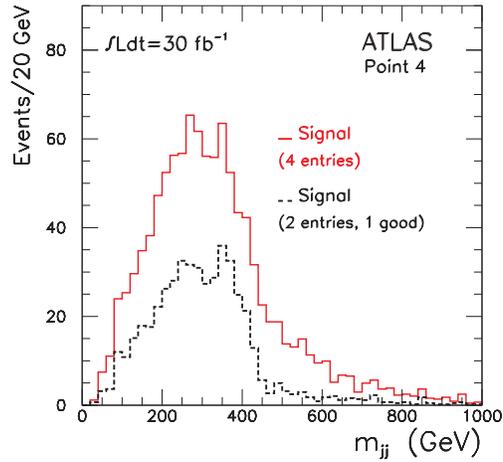

\dofig{2.5in}{fabiola16.ai}
\caption{Jet-jet mass distribution for Point~4 with three leptons and
four jets (solid) and for correct pairing
(dashed).\protect\cite{Fabiola}
\label{fabiola16}}
\end{figure}

\begin{figure}[t]
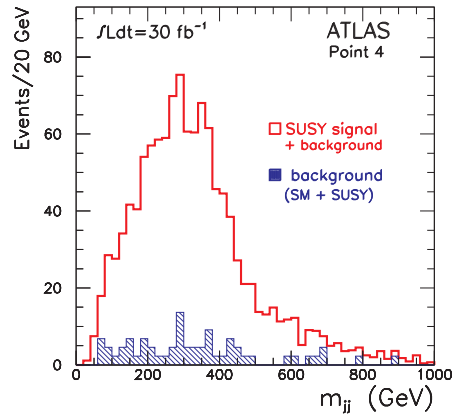

\dofig{2.5in}{fabiola17.ai}
\pawplot
\caption{Jet-jet mass signal (solid) and background (shaded) for
Point~4 with three leptons and four
jets.\protect\cite{Fabiola}\label{fabiola17}}
\end{figure}

	The event selection for this analysis imposes the following
set of cuts:
\begin{itemize}
\item	3 isolated leptons with $p_T > 20$, 10, $10\,\GeV$ and
$|\eta|<2.5$, with one $OS$, $SF$ lepton pair.
\item	$M<72\,\GeV$, the edge in Figure~\ref{c4-mllosdif}, for the
$OS$, $SF$ pair, consistent with $\tchi20$ decay and not with possible
$Z$ backgrounds.
\item	4 jets with $p_T > 150$, 120, 70, $40\,\GeV$ and $|\eta|<3.2$.
\item	No additional jets with $p_T>40\,\GeV$ and $|\eta|<5$ to
minimize combinatorial backgrounds.
\item	No $\etmiss$ cut.
\end{itemize}
\noindent After these cuts there 250 signal events with 30 $\tg\tq$
background and 18 other SUSY and Standard Model background for
$30\,\fb^{-1}$ of integrated luminosity. Thus the signal purity is
quite good, and the signal efficiency, $\sim35\%$, is also good. 

	There are three ways of pairing the jets for event. The one
that pairs the two highest and the two lowest $p_T$ jets is usually
wrong since the $\tg$ are mainly at low $p_T$ and hence is
removed.\cite{BCDPT} The distribution for the two remaining
combinations, Figure~\ref{fabiola16}, shows drop a near the endpoint
for the correct combination of jets, shown as a dashed curve. The
error on the mass was calculated by varying $M_\tg$, generating a new
distribution, and using a Komogorov-Smirnov test to find whether the
two distributions could be distinguished. The result of this analysis
gives an estimated error
$$
M_\tg -M_{\tchi_1^\pm}/M{\tchi_2^0} = 434 {+5.0\atop-16}
\pm4.5\,\GeV\,.
$$
(The Komogorov-Smirnov test is a standard statistical test to see
whether two distributions are identical. For a derivation and
discussion, see Knuth.\cite{Knuth})

\section{Global Fits to Determine SUSY Parameters\label{sec:fits}}%

	Section~\ref{sec:precise} has described a number of precision
measurements that can be made at the LHC by relating features of
kinematic distributions to combinations of SUSY masses. Quite a few
other such measurements have been developed for these points, and many
more would surely be found with the incentive of real data indicating
the discovery of SUSY. It will also be possible to measure many other
distributions that cannot be related to combinations of masses in such
a precise way but which can be used to constrain the SUSY model. Given
experimental observation of such signals, one would certainly use all
the available data in a global fit to determine the SUSY model, just
as the parameters of the Standard Model have been determined by global
fits at LEP and SLC.

	Making such a fit requires generating large samples of many
possible SUSY signals, combining them with the Standard Model
background, and comparing the results with the distributions for the
assumed SUSY signal. This requires far more effort than can be
justified at present. Instead, a much simpler approach has been
adopted. Samples of signal events have been generated for each of the
five LHC points. For each point, precision measurements of
combinations of masses have been studied, and the errors on these
combinations have been estimated. Then, SUGRA models have been
generated choosing the parameters at random, the masses have been
calculated, and the distributions of the model parameters have been
determined for those models giving combinations of masses consistent
with the precision measurements.

\begin{figure}[t]
\dofig{2.5in}{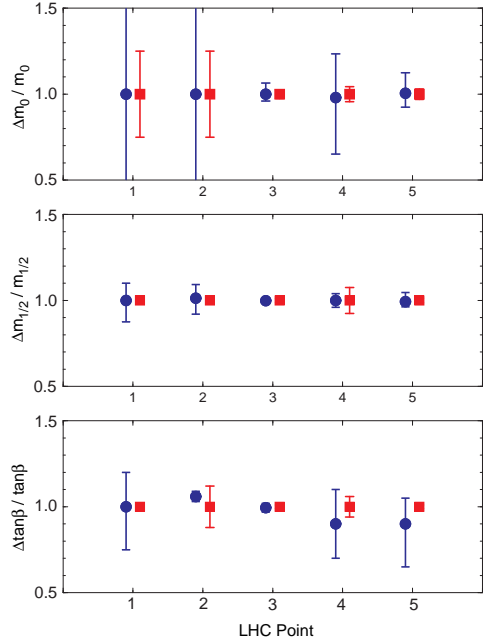}
\caption{Errors on SUGRA parameters from the Hinchliffe et
al.{}\protect\cite{HPSSY} (circles) and Froidevaux\protect\cite{Froid}
(squares) fits.\label{errors}}
\end{figure}

	Two such restricted global fits have been performed. The first
fit is based on the set of precision measurements studied by
Hinchliffe et al.{}\cite{HPSSY} and uses statistical errors
appropriate to $10\,\fb^{-1}$, plus an assumed $3\,\GeV$ theoretical
error on the light Higgs mass. The second fit, done by
Froidevaux\cite{Froid} as part of the SUSY studies of the ATLAS
Collaboration, is based on all of the Hinchliffe et al.{} measurements
plus a number of additional ones and uses experimental errors scaled
to three years at the full LHC design integrated luminosity,
$300\,\fb^{-1}$.  Froidevaux also assumes an error on the light Higgs
mass set by the experimental error of $0.2\,\GeV$. That is, the
Hinchliffe et al.{} fit is conservative both in regards to what is
included and in regards to the integrated luminosity. The Froidevaux
fit uses more distributions and the full LHC luminosity and it assumes
a theoretical Higgs mass error much smaller than that currently
available, so it requires a substantial improvement on the calculation
of the Higgs effective potential in terms of the underlying SUGRA
parameters. Nevertheless, even the results of Froidevaux might be
improved by additional work and by studying actual data.

	The results of these fits are shown in Figure~\ref{errors},
which displays the fractional errors for the Hinchliffe et al.\ and
Froidevaux fits on each of the SUGRA parameters except $A_0$ at each
of the five LHC points. Both fits do very well on $\mhalf$, the
gaugino mass that sets the overall SUSY mass scale. The Hinchliffe, et
al., fits do quite poorly on $m_0$ at Points~1 and 2 because the
squark masses are insensitive to it and the slepton masses are too
heavy to enter into any of the cascade decays. The Froidevaux fits do
rather better because they include more information. The fits to
$\tan\beta$ and $\sgn\mu$ uses $M_h$ heavily and so are sensitive to
the assumed theoretical errors, which dominate over the experimental
ones. Finally, neither fit gives a significant constraint on $A_0$. It
is possible to constrain the weak-scale $A_t$, $A_b$, and $A_\tau$, but
the renormalization group equations imply that these are only weakly
related to $A_0$ and instead flow to quasi-fixed points.

	Figure~\ref{errors} shows that the LHC has the potential to
determine the parameters of the SUSY model with good precision, at
least for the cases studied. It is important to realize, however, that
the use of a simple model is important in obtaining these results.
Most of the measurements depend mainly on the gluino and squarks and
on their main decay products, the lighter gauginos, $\tchi_1^0$,
$\tchi_2^0$, or $\tchi_1^\pm$, and the light Higgs boson $h$. It would
be a useful exercise to vary the SUGRA model, e.g., by allowing
splittings among the scalar masses, and to see how well these could be
constrained.

\section{Example Point with Large $\tan\beta$\label{sec:large}}%

	When the five LHC points were selected, ISAJET and PYTHIA
still neglected some mixing effects due to the $y_b$ and $y_\tau$
couplings. Hence it was necessary to choose points with
$\tan\beta\le10$, for which these effects are small, since
$$
\displaylines
{
y_t={g m_t\over\sqrt{2}M_W\sin\beta}\,, \\
y_b={g m_b\over\sqrt{2}M_W\cos\beta}\,,
\quad y_{\tau}={g m_{\tau}\over\sqrt{2}M_W\cos\beta}\,.
}
$$
More recent versions of ISAJET have incorporated these mixing effects
and so can be used for all values of $\tan\beta$. Large $\tan\beta$
has significant consequences for the phenomenology of the minimal
SUGRA model:\cite{BCDPT}
\begin{itemize}
\item The renormalization group equations produce smaller $\tb_{L,R}$
and $\ttau_{L,R}$ soft masses. 
\item The squark and slepton masses have larger off-diagonal terms,
increasing the mixing and so reducing the $\tb_1$ and $\ttau_1$
masses. 
\item The combination of smaller masses and larger Yukawa couplings
enhance $b$ and $\tau$ decays.
\item $M_A$ and related masses are also reduced.
\end{itemize}
\noindent Perhaps the most important consequence of these changes is
that it is possible to have $M(\ttau_1) < M(\tchi_2^0),
M(\tchi_1^\pm)$ but $M(\tell) > M(\tchi_2^0), M(\tchi_1^\pm)$, leading
to dominant $\tau$ decays of the gauginos.

\begin{figure}[t]
\dofig{2.5in}{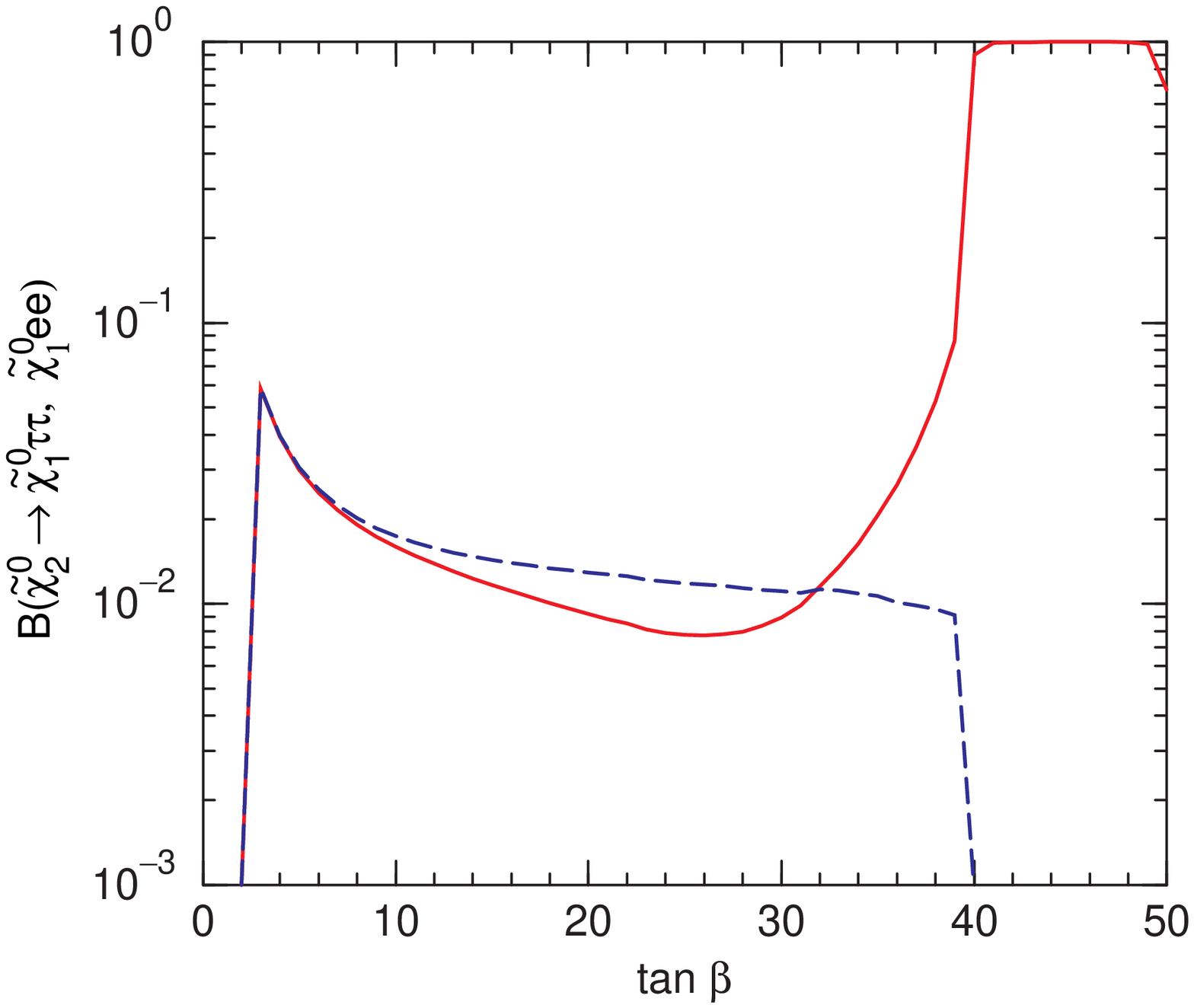}
\caption{Branching ratios for $\tchi_2^0 \to \lsp e^+e^-$ (dashed) and
$\lsp \tau^+\tau^-$ including $\ttau_1^\pm \tau^\mp$
(solid).\label{betascan}}
\end{figure}

        ``Point 6'' is a minimal SUGRA point with $m_0 = \mhalf =
200\,\GeV$, $A_0=0$, $\tan\beta=45$, $\mu<0$. For this choice of $m_0$
and $\mhalf$, $W$, $Z$ and $h$ decays of the $\tchi_2^0$ and
$\tchi_1^\pm$ are kinematically forbidden. Choosing $\tan\beta=45$
makes the $\ttau_1$ light enough that $\tau$ decays of these gauginos
are dominant, so that the $\tchi_2^0 \to \ttau_1 \bar\tau$ and
$\tchi^\pm \to \ttau_1^\pm \nu_\tau$ decays are dominant, as can be
seen from Figure~\ref{betascan}.

	The dominant decays of $\tau$'s are into leptons or into one
or three charged hadrons:\cite{PDG}
\begin{eqnarray*}
\tau^- \to	& e^- \nu_e \nu_\tau\hfill 	&\quad 17.8\% \\
		& \pi^- \nu_\tau\hfill		&\quad 11.3\% \\
		& \rho^- \nu_\tau\hfill		&\quad 25.2\% \\
		& a_1^-(1260)\hfill		&\quad 16.3\%\,.
\end{eqnarray*}
The $a_1(1260)$ is an isovector, axial vector meson which dominantly
decays into $\rho\pi$. Thus $\tau$'s can be identified experimentally
as hadronic jets with a mass less than $M(\tau)=1.777\,\GeV$ and with
either one charged track or three charged tracks with net charge
$\pm1$.

\begin{figure}[t]
\dofig{2.5in}{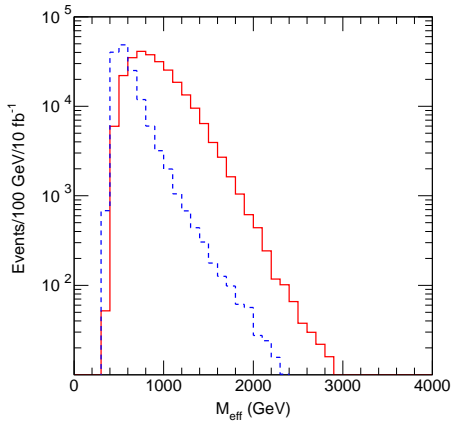}
\pawplot
\caption{Effective mass $\Meff$ for Point~6 signal (solid) and
Standard Model background (dashed).\label{c6-meffcut}} 
\end{figure}

        As is normally the case at the LHC, gluino and squark
production dominate at this point. The same basic cuts are made as for
the points discussed previously:
\begin{itemize}
\item $\ge4$ jets with $p_{T,1}>100\,\GeV$, $p_{T,2,3,4}>50\,\GeV$.
\item Missing energy $\etmiss>100\,\GeV$.
\item Transverse sphericity $S_T>0.2$.
\item $\Meff=\etmiss+p_{T,1}+p_{T,2}+p_{T,3}+p_{T,4}>500\,\GeV$.
\end{itemize}
\noindent After these cuts, discovery is straightforward via multiple
jets plus $\etmiss$. Figure~\ref{c6-meffcut} shows the distribution of
the variable $\Meff$ defined previously for the signal and the sum of
all Standard Model backgrounds. Evidently the signal dominates for
large $\Meff$ without using any $\tau$ identification cuts.

\begin{figure}[t]
\dofig{2.5in}{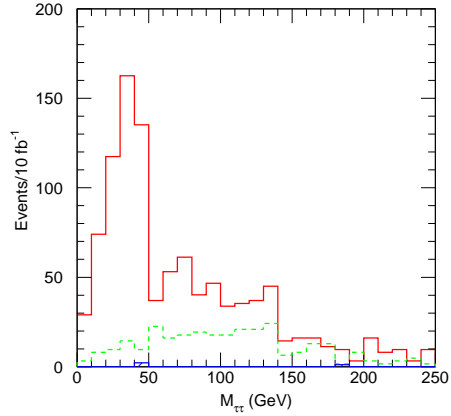}
\pawplot
\caption{Visible $\tau^+\tau^-$ (solid) and $\tau^\pm\tau^\pm$ mass
(dashed) for Point~6 and Standard Model background
(shaded).\label{c6-mtautau3p}}
\end{figure}

\begin{figure}[t]
\dofig{2.5in}{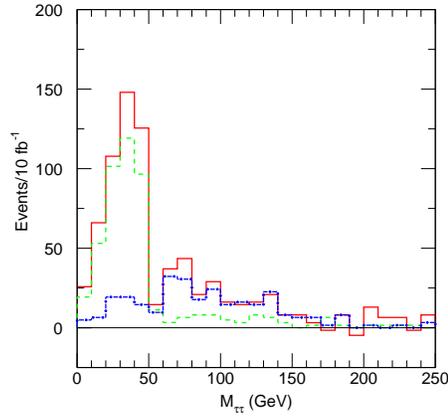}
\pawplot\pawplot
\caption{Visible $\tau^+\tau^- - \tau^\pm\tau^\pm$ mass for Point~6
signal (solid), $\tchi_2^0$ decay contribution (dashed), and heavy
gaugino contribution (dash-dotted).\label{c6-mtautaudif}}
\end{figure}

	After the above cuts, the signal is selected by requiring
$\Meff > 500\,\GeV$, two hadronic $\tau$ decays, and no other isolated
leptons. The $\tau$'s were selected using information from the event
generator, since the actual detector cuts require rather a detailed
analysis.\footnote{This is in progress. It is expected that
backgrounds from misidentified jets are not very important since the
SUSY signal dominates before any $\tau$ cuts are made.} Background
from two independent decays, either from other SUSY channels or from
Standard Model background, were removed by subtracting same-sign
events from opposite-sign ones.  Finally, only $\tau$ decays into
three charged particles (mainly $a_1\nu_\tau$ and the $3\pi\nu_\tau$)
were used. Since the 3-prong branching ratio is only about 15\%, this
involves a substantial loss in statistics, but the visible 3-prong
momentum is closer to the true $\tau$ momentum.  The opposite-sign and
same-sign visible mass distributions after these cuts for the Point~6
signal and Standard Model background are shown in
Figure~\ref{c6-mtautau3p}. The difference of the opposite-sign and
same-sign signals is shown in Figure~\ref{c6-mtautaudif}. This figure
also shows the contributions from $\tchi_2^0 \to \ttau_1 \bar\tau \to
\lsp \tau^+\tau^-$ and from events involving heavier gauginos. The
former is responsible for the clear edge in the experimental
distribution, while the latter dominates at larger masses. 

	Point~6 and similar points at which $\tau$ decays dominate
will clearly be more difficult to study than points with decays into
electrons and muons. More work is needed both to optimize the strategy
and to understand the detector implications. Nevertheless, the results
presented here are not unencouraging. Preliminary results indicate,
however, that it will be difficult to detect any signals for such
points at the Tevatron.

\section{Higgs Bosons at LHC\label{sec:higgs}}%

	Since the main reason for introducing SUSY at the weak scale
is to cancel the quadratically divergent corrections to the Higgs
masses, observing Higgs bosons at the LHC is just as important as
observing SUSY particles. It is also a good deal more difficult,
partly because the couplings to light quarks and hence the production
cross sections are small, and partly because only some rare decay
modes can be observed over the Standard Model backgrounds. Indeed, the
ability to search for Higgs bosons has been the main design criterion
for the detectors.\cite{ATLAS,CMS}

	In the Minimal Supersymmetric Standard Model, there are five
Higgs bosons,\cite{HHG} $h$, $H$, $A$, and $H^\pm$. These are
described at tree level by just two parameters, generally taken to be
the mass $M_A$ of the pseudoscalar Higgs boson and $\tan\beta$, the
ratio of vacuum expectation values. The renormalization group
equations and radiative electroweak symmetry breaking relate $M_A$ to
the SUGRA parameters.  The other masses are then given by
\begin{eqnarray*}
M_{H^\pm}^2 &=& M_A^2 + M_W^2 \\
M_{h,H}^2 &=& \half\left[{M_A^2 + M_Z^2 \pm \sqrt{(M_A^2 + M_Z^2)^2 -
4 M_A^2M_Z^2 \cos^22\beta}}\right]\,.
\end{eqnarray*}
The mixing angle $\alpha$ between $h$ and $H$ is
\begin{eqnarray*}
\cos2\alpha &= - \cos2\beta \left({\displaystyle {M_A^2 - M_Z^2 \over 
M_H^2 - M_h^2}}\right)\,, \\
\sin2\alpha &= - \sin2\beta \left({\displaystyle {M_H^2 + M_h^2 \over 
M_H^2 - M_h^2}}\right)\,.
\end{eqnarray*}
The couplings of the MSSM Higgs bosons relative to the Standard Model
ones can all be expressed in terms of $\alpha$ and $\beta$, as shown
in Table~\ref{tbl:hcouple}.  Note that for $M_A \gg M_Z$, $\alpha
\approx \beta - \half\pi$, so that the light Higgs $h$ has couplings
identical to those of a Standard Model Higgs of the same mass.

\begin{table}[t]
\caption{MSSM Higgs couplings relative to a Standard Model Higgs of
the same mass.\label{tbl:hcouple}}
\medskip
\begin{center}
\begin{tabular}{cccc}
\hline\hline
\strut	& $d \bar d$, $\ell^+\ell^-$	& $u \bar u$	& $WW$\\
\hline
$h$	& $-\sin\alpha/\cos\beta$	& $\cos\alpha/\sin\beta$ &
$\sin(\beta-\alpha)$\\
$H$	& $\cos\alpha/\cos\beta$	& $\sin\alpha/\sin\beta$ &
$\cos(\beta-\alpha)$\\
$A$	& $-i\gamma_5\tan\beta$		& $-i\gamma_5\cot\beta$ &
$0$\\
\hline\hline
\end{tabular}
\end{center}
\end{table}

\subsection{Observing Standard Model Higgs Bosons}%

	The Standard Model Higgs is simpler than the MSSM Higgs sector
and has been studied in more detail by ATLAS and CMS, so it makes
sense to discuss it first. The dominant production is by $gg$ fusion
through a $t$-quark loop, although $WW$ and $ZZ$ fusion also
contribute for large masses. The same $5\sigma$ discovery limit
discussed in Section~\ref{sec:reach2} will be used here. The search
for the Standard Model Higgs naturally divides into three mass
regions\cite{HHG,ATLASnote48}:

\subsubsection{$M_H > 2M_Z$}%

	The dominant Higgs decay modes in this mass region are $WW$
and $ZZ$, with $t \bar t$ at most $\sim10\%$ and all other modes very
small. The signal for $H \to ZZ \to \ellell \ellell$ is robust, with
the only important background being the $ZZ$ continuum. It may also be
possible to use the $H \to ZZ \to \ellell \nu\bar\nu$ and $H \to ZZ
\to \ellell q \bar q$ modes, although the backgrounds in these
channels are much more significant. The only problem in this region is
that a very heavy Standard Model Higgs is strongly coupled to
longitudinal $WW$ and $ZZ$ pairs, so that
$$
\Gamma_H \approx {3 G_F M_H^3 \over 16\pi\sqrt{2}}, \quad M_H \gg
M_Z\,.
$$
Thus at very large masses, $M_H > 600$--$800\,\GeV$, the Higgs becomes
so broad that it ceases to be a well defined particle and vanishes
into the $WW$ or $ZZ$ continuum. Recent precision electroweak data
from LEP indicates that the Standard Model Higgs is lighter than
$420\,\GeV$ with 95\% confidence.\cite{Ward97} In this mass region,
detection of a Higgs decaying into four leptons should be easy.

\subsubsection{$130\,\GeV \simle M_H < 2M_Z$}%

\begin{figure}[t]
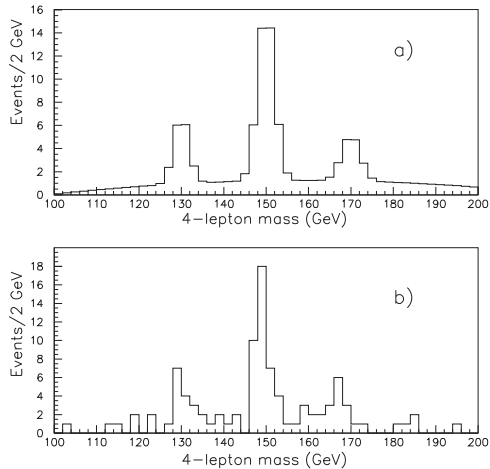

\dofig{2.5in}{fig11_14.ai}
\caption{Standard Model $H \to ZZ^*$ signals for $M_H = 130$, 150, and
$170\,\GeV$, showing both the average signal (top) and the signal for
a typical experiment (bottom).\protect\cite{ATLAS}\label{h4l}}
\end{figure}

	In this mass region, the dominant Higgs decay mode is $H \to b
\bar b$, but the background from QCD $b \bar b$ production is
overwhelming. The best mode for detecting such a Higgs is $H \to ZZ^*
\to \ellell \ellell$. While the rates are small, isolation cuts can be
used to reject the $Z b\bar b$ background. Then, after the cuts
$M_{12} \approx M_Z$ and $M_{34}>10\,\GeV$ to reject the $Z\gamma^*$
continuum, the $S/B$ is very good, as can be seen in Figure~\ref{h4l},
although the signal is quite small. It seems clear that one can
eventually detect a Higgs in this range, although it may take more
than one year for $M_H \approx 170\,\GeV$, for which the $W^+W^-$ mode
dominates.

\begin{table}
\caption{$H \to ZZ^*$ event numbers.\label{tbl:h4l}}
\medskip
\begin{center}
\begin{tabular}{cccccc}
\hline\hline
\hfill Mass (GeV) & 120 & 130 & 150 & 170 & 180\\
\hline
Signal ($3\times 10^4\,\pb^{-1}$) & 2.6 & 12.5 & 34.9 & 10.2 & 26.5\\
\hline
$t \bar t$	& $<0.1$ & $<0.1$ & $<0.1$ & $<0.1$ & $<0.1$\\
$Z b \bar b$	& 0.1 & 0.2 & 0.2 & 0.2 & 0.2\\
$ZZ^*$, $Z\gamma^*$	& 1.6 & 2.8 & 3.3 & 3.5 & 3.3\\
Background ($3\times 10^4\,\pb^{-1}$) & 1.7 & 3.0 & 3.6 & 3.7 & 3.6\\
\hline
$S/\sqrt{B}$ ($3\times 10^4\,\pb^{-1}$) & 2.0 & 7.2 & 18.3 & 5.3 & 14\\
\hline
Signal ($10^5\,\pb^{-1}$)       & 5.2 & 24.5 & 68.5 & 19.9 & 51.9\\
Background ($10^5\,\pb^{-1}$)   & 4.7 & 8.2 & 10.0 & 9.5 & 9.0\\
$S/\sqrt{B}$ ($10^5\,\pb^{-1}$) & 2.4 & 8.5 & 21.7 & 6.5 & 17.3\\
\hline\hline
\end{tabular}
\end{center}
\end{table}

	For low statistics processes such as $H \to ZZ^*$, it is
important to treat the statistics correctly. The right question to ask
is, ``What is the probability for the background to fluctuate up to
the expected signal?'' For large numbers of events, the answer is
given by using Gaussian statistics to calculate the significance,
$S/\sqrt{B}$, in standard deviations. For low numbers of events,
however, it is important to use Poisson statistics and then to
translate the probability to an equivalent Gaussian significance.
Note, for example, that the Poisson probability for a background of 1
event to fluctuate up to 6 events is
$$
P(5+1|1) = {1^6 e^{-1} \over 6!} = 5.1 \times 10^{-4}\,,
$$
compared to $5\sigma$ Gaussian probability of $5.7 \times 10^{-5}$.
The probably that an expected signal $S+B=6$ fluctuates down to the
background $B\le1$ is much larger:
$$
P(0|6) + P(1|6) = e^{-6}[1 + {6 \over 1!}] = 0.017\,.
$$
Since there are many possible fake signals but presumably only one
real one, this is acceptable.

	The Gaussian significances shown in Table~\ref{tbl:h4l} are
thus slightly too optimistic, but nevertheless they show that it is
possible to discover a Higgs boson decaying through $ZZ^*$ into four
leptons for $M_H \ge 130\,\GeV$. The maximum luminosity is needed for
$M_H \sim 170\,\GeV$ because the $WW$ channel is open, reducing the
$ZZ^*$ branching ratio. The acceptance used in this table may be
somewhat optimistic, particularly for muons, but this can be
compensated by more running time.

\begin{figure}[t]
\dofig{2.5in}{fig11_6.ai}
\pawplot
\caption{$120\,\GeV$ Standard Model Higgs signal and $\gamma\gamma$
background with the ATLAS detector.\protect\cite{ATLAS}\label{fig11-6}}
\end{figure}

\begin{figure}[t]
\dofig{0.9\textwidth}{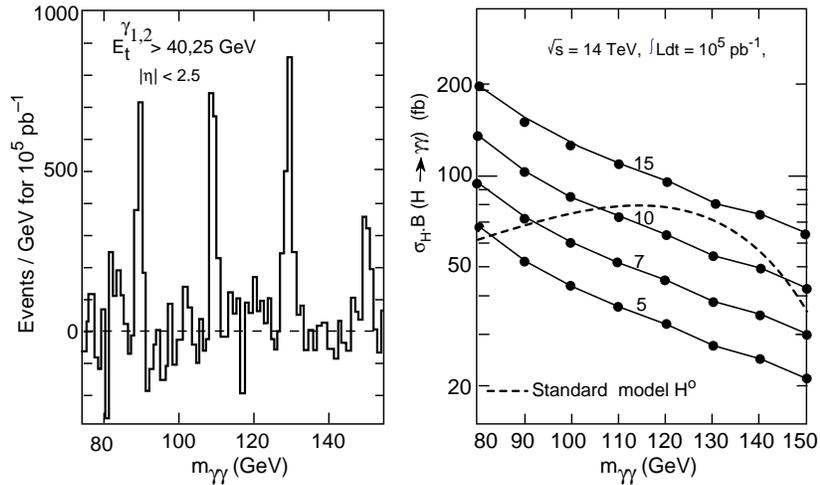}
\caption{Background-subtracted signals for $H\to\gamma\gamma$
with the CMS detector and cross sections required for 5, 7, 10, and
$15\sigma$ significance.\protect\cite{CMS}\label{hgg}}
\end{figure}
\relax

\subsubsection{$M_H \simle 130\,\GeV$}%

	Below about $130\,\GeV$ the $H \to ZZ^* \to \ellell \ellell$
branching ratio becomes too small to be observable, and the dominant
$b \bar b$ and $\tau\tau$ modes are swamped by background. Thus in
this mass range the only way to observe the Higgs is via the
$\gamma\gamma$ mode. This mass region is particularly important for
SUSY, since the light Higgs boson is always in this range and
generally is almost identical to a Standard Model Higgs of the same
mass. There are large backgrounds from the QCD $\gamma$-$\gamma$
continuum which can only be overcome with excellent $\gamma$-$\gamma$
mass resolution.  There are also huge potential backgrounds from the
$\gamma$-jet and jet-jet backgrounds. Detection of $H \to
\gamma\gamma$ is an extraordinary challenge, and the need to do so has
driven the design of the electromagnetic calorimeter --- and, to a
lesser extent, of the whole detector --- for both ATLAS and CMS.

	The first problem in detecting $H \to \gamma\gamma$ is to
overcome the $\gamma\gamma$ continuum. The lowest order QCD process is
$q \bar q \to \gamma\gamma$, but it is important to include the (gauge
invariant) higher order process $gg \to \gamma\gamma$ through a box
graph because $g(x)$ is very large at small $x$. The background
calculation also includes $\gamma$ bremsstrahlung from quarks in the
leading-log approximation. The following cuts are made in the ATLAS
analysis to reduce this real $\gamma\gamma$ background compared to the
signal:
\begin{itemize}
\item	$p_{T,1}>40\,\GeV$, $p_{T,2}>25\,\GeV$, $|\eta|<2.5$.
\item	$p_{T,1}/(p_{T,1}+p_{T,2})<0.7$ to minimize bremsstrahlung
backgrounds.
\item	Exclude regions of the calorimeter with degraded resolution.
\item	Choose $\Delta M$ for each mass to maximize $S/\sqrt{B}$.
\end{itemize}
\noindent With these cuts and with no fake $\gamma\gamma$ background,
even the most favorable mass, $M_H=120\,\GeV$ is hard, as can be seen
from Figure~\ref{fig11-6}. Lower masses, $M_H < 100\,\GeV$, are the
most difficult, as can be seen from Figure~\ref{hgg}. For a Standard
Model Higgs in this mass range, one or more years at full luminosity
are required to reach a $5\sigma$ significance. For all masses the
$S/B$ ratio is poor, but because the peak is narrow, the side bands
can be used to determine the background. Thus, the search limits are
set by statistics and not by the understanding of the background.

	The potential jet-jet and $\gamma$-jet backgrounds are very
large; a $\gamma/\jet$ rejection of $\sim 10^4$ is essential to reduce
these to a fraction of the QCD $\gamma\gamma$ continuum. Very detailed
simulations of the isolation cuts and of the rejection of single and
multiple $\pi^0$ using the detailed properties of the electromagnetic
showers indicates that the required rejection can be achieved. An
efficient trigger on two photons is also possible. But detecting this
signal is probably the most difficult challenge for ATLAS and CMS.

\begin{figure}[t]
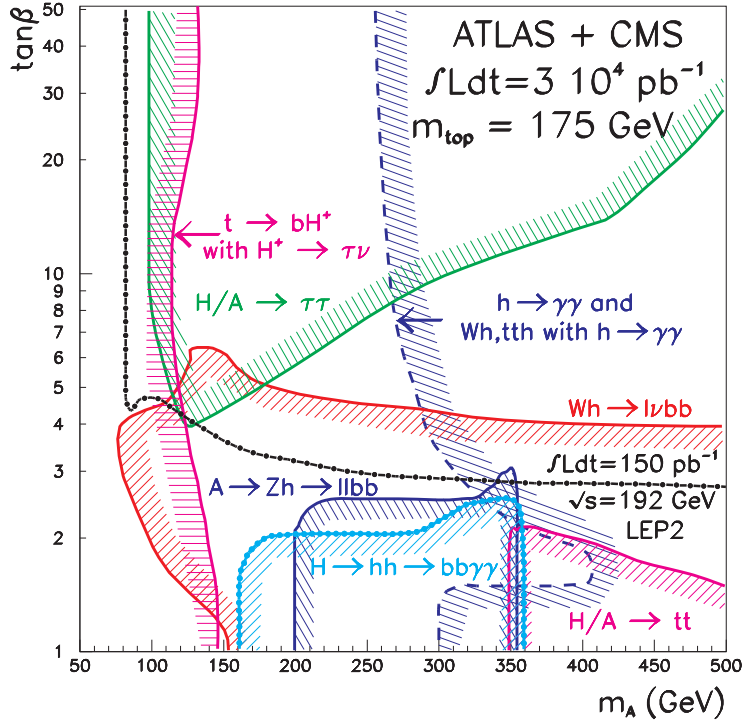

\dofig{0.9\textwidth}{lhc30top175_mA.ai}
\pawplot
\caption{Possible Higgs $5\sigma$ discovery contours for
$30\,\fb^{-1}$ in various modes assuming all SUSY masses are
$\sim1\,\TeV$.\protect\cite{ATLASnote74} Each signal is observable on
the shaded side of the corresponding curve.\label{h30}} 
\end{figure}

\subsection{Observing SUSY Higgs Bosons}%

\begin{figure}[t]
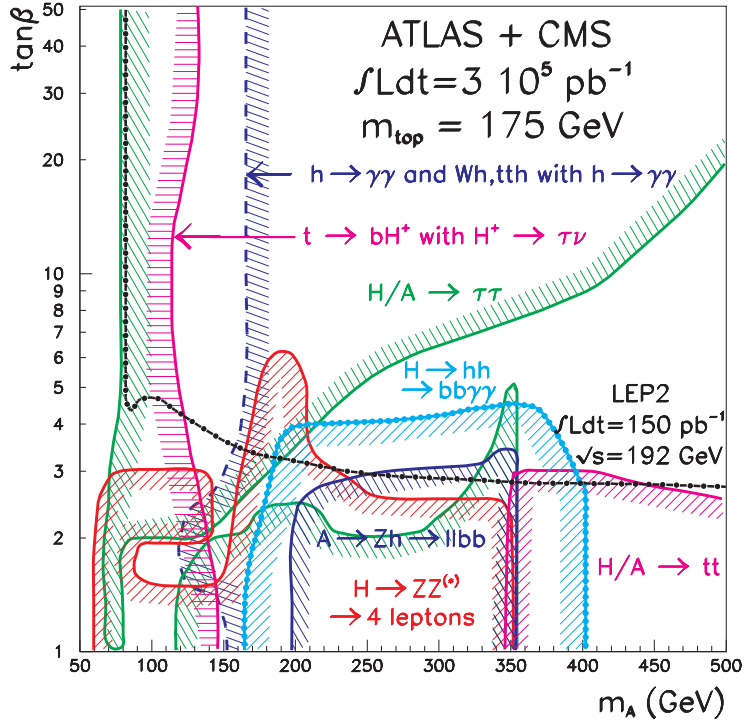

\dofig{0.9\textwidth}{lhc300top175_mA.ai}
\pawplot
\caption{Possible Higgs $5\sigma$ discovery contours for
$300\,\fb^{-1}$ in various modes assuming all SUSY masses are
$\sim1\,\TeV$.\protect\cite{ATLASnote74} Each signal is observable on
the shaded side of the corresponding curve.\label{h300}}
\end{figure}

	In the MSSM the Higgs sector at tree level depends only on two
parameters, $M_A$ and $\tan\beta$. In the SUGRA model it is usually
true that $M_A \gg M_Z$, so that the light Higgs $h$ is very similar
to a Standard Model Higgs of the same mass, while the rest of the
Higgs bosons are heavy, $M_H \approx M_{H^\pm} \approx M_A$. Many
modes might be used to search for the SUSY Higgs bosons.
Figures~\ref{h30} and \ref{h300} show the $5\sigma$ reach
contours\cite{ATLASnote74} for various signatures in the
$M_A$-$\tan\beta$ plane, calculated assuming that all SUSY masses are
$\sim1\,\TeV$ so that SUSY decay modes do not
contribute.\cite{ATLASnote74} Each mode has a statistical $5\sigma$
significance on the shaded side of the corresponding curve. The plot
does not show constraints such as the rate for $b \to s \gamma$ that
limit the Higgs parameter space for the assumed SUSY masses. Perhaps
the only thing clear from these figures is that many modes can be used
to search for SUSY Higgs bosons. By making use of all of the modes, it
appears to be possible to discover at least one Higgs boson over the
entire $M_A$--$\tan\beta$ plane by combining LEP at $195\,\GeV$ and
ATLAS and CMS with $300\,\fb^{-1}$ each.\cite{ATLASnote74} Over much
of the parameter space, more than one Higgs can be observed, but it is
generally not possible to observe all of them.

	All of the Standard Model Higgs search modes will be used to
search for SUSY Higgs. In particular, in the SUGRA model the light
Higgs $h$ is almost always in the range covered by the $\gamma\gamma$
mode because $M_A$ is large, so that the $h$ couplings are similar to
the standard model ones. Hence the search for the $h$ is similar to
that for a Standard Model Higgs in the same mass range. The $5\sigma$
discovery limits are indicated in Figures~\ref{h30} and \ref{h300}.

	For the heavy Higgs bosons $H^0$ and $A^0$, the $ZZ$ modes are
suppressed in the MSSM compared to the Standard Model. Because of $CP$
invariance, $A \not\to ZZ$, and the $H$ coupling to $WW$ and $ZZ$ is
proportional to $\cos(\beta-\alpha)$, which vanished for large $M_A$.
Hence the four-lepton modes are of limited value for SUSY Higgs,
although there are regions indicated in Figures~\ref{h30} and
\ref{h300} where they can be observed.

	Standard Model Higgs decays into $\tau\tau$ are probably not
possible to observe. The $H$ and $A$ decays to $\tau\tau$ are enhanced
for large $\tan\beta$, as can be seen from Table~\ref{tbl:hcouple},
and the competing $WW$ and $ZZ$ modes are reduced or eliminated. Since
the $\tau$ always decays to a missing $\nu_\tau$, it is not possible
to reconstruct the $\tau\tau$ mass in general. But if the Higgs is
produced with high $p_T$, $p_T \simge M_H$, then the two $\tau$'s are
not back to back and it is possible to resolve the missing energy
$\etmiss$ along the observed $\tau$ directions, reconstructing the two
$\tau$ momenta and hence the $\tau\tau$ mass. The mass resolution for
such events has been studied with full simulation.\cite{ATLASnote74}
The dominant $\tau\tau$ background is $Z^*/\gamma^* \to \tau\tau$, and
this can be determined using $Z \to ee$, $\mu\mu$, and the known
$\tau$ decay modes. Hence, the statistical errors should be dominant,
and the $5\sigma$ discovery limits indicated in Figures~\ref{h30} and
\ref{h300} should be reliable.

	The process $Wh \to \ell\nu b \bar b$ is another one that is
difficult to observe for Standard Model Higgs, even though the lepton
tag greatly reduces the Standard Model $b \bar b$ background. This
signal is enhanced for large $\tan\beta$ in the MSSM and is at least
statistically observable in the regions shown in Figures~\ref{h30} and
\ref{h300}. But it must be noted that there is a large $t \bar t$
background: e.g., there are 475 signal events on 12700 background for
a $100\,\GeV$ Standard Model Higgs. While this gives a statistical
significance $S/\sqrt B = 4.2$, it is necessary to know the background
to better than 1\% in order to be sure that the signal is real. While
the $S/B$ for $H \to \gamma\gamma$ is also poor, the signal peak in
that case is very narrow, so the sidebands can be used to determine
the background. For $b \bar b$ decays the mass resolution will be of
order 10\%, and the $t \bar t$ background can produce background that
varies significantly over this range. While it may be possible to
observe this signal, the small value of $S/B$ implies that it will be
very difficult.

\section{Conclusions\label{sec:last}}%

	If SUSY exists at the electroweak scale, i.e., with squark and
gluino masses less than 1--$2\,\TeV$, it should be straightforward to
find signals for it at the LHC in the $\jets + \etmiss$ channel and
perhaps in many other channels.  Discovery of a deviation from the
Standard Model should be possible with an integrated luminosity of
$10\,\fb^{-1}$ or even less for masses below $1\,\TeV$. In many cases,
it should be possible to determine combinations of masses from
features of kinematic distributions, giving precision measurements of
these mass combinations. If the SUSY model turns out to be simple, it
will also be possible to determine its parameters from such precision
measurements. Shortly after the LHC starts operation, either SUSY will
become a central part of particle physics and of every subsequent TASI
Summer School, or it will be relegated to an obscure corner of
mathematical physics.

	The LHC will mainly produce gluinos, squarks, and their main
decay products, the light gauginos, $\lsp$, $\tchi_2^0$, and
$\tchi_1^\pm$. The dominant backgrounds for SUSY signatures come not
from Standard Model processes but from other SUSY processes.  For some
choices of the SUSY model, it will also be possible to detect other
SUSY particles, including some or all of the sleptons and the heavier
gauginos. However, it is generally not possible to detect the whole
SUSY and heavy Higgs spectrum. Thus, some of the conclusions from any
LHC SUSY analysis will probably be model dependent.

	The Next Linear Collider --- or the Next Lepton Collider if a
muon collider should turn out to be preferable --- probably will be
completed much later than the LHC. An NLC generally can detect all
SUSY particles which are kinematically allowed. Using the fact that
SUSY particles are pair produced at the beam energy allows one to
reconstruct the $\lsp$ mass as well as other masses.\cite{TFMYO,NFT}
Such a machine may have to measure gauginos, Higgs's, and perhaps
sleptons with masses $\sim M_\tg$ This would require $\sqrt{s} >
2M_\tg$.  While this scale is not now known, it could be $2\,\TeV$. Of
course, if the SUSY model turns out to be complicated, a lower energy
machine capable of studying the sleptons and light gauginos could
prove essential.  There are exciting experimental prospects for SUSY
not just at the LHC but at future machines.

	I wish that I could be as optimistic about the future of
particle physics in the United States. 

\bigskip

	It is a pleasure to thank my collaborators, including Howie
Baer, Chih-Hao Chen, Manuel Drees, Daniel Froidevaux, Fabiola
Gianotti, Ian Hinchliffe, Serban Protopopescu, Marjorie Shapiro, Jesper
S\"oderqvist, Xerxes Tata, and Weiming Yao. 

	This work was supported in part by the United States
Department of Energy under Contract DE-AC02-76CH00016.

\end{document}